\documentclass{emulateapj}
\usepackage{graphicx}
\usepackage{latexsym}
\usepackage{amssymb}
\usepackage[applemac]{inputenc}
\usepackage{float}
\usepackage{fancyhdr}
\usepackage{amsfonts}
\usepackage{amsmath}
\usepackage{color}
\usepackage{mathrsfs}
\usepackage{bm}
\usepackage{epsfig}
\usepackage{array}
\usepackage{multirow}
\newcommand{\beq}{\begin{equation}}
\newcommand{\eeq}{\end{equation}}
\newcommand{\beqa}{\begin{eqnarray}}
\newcommand{\eeqa}{\end{eqnarray}}
\newcommand{\lsim}{\mbox{\raisebox{-.6ex}{~$\stackrel{<}{\sim}$~}}}
\newcommand{\gsim}{\mbox{\raisebox{-.6ex}{~$\stackrel{>}{\sim}$~}}}

\newcommand{\s}{\,{\rm s}}
\newcommand{\km}{\,{\rm km}}
\newcommand{\kmps}{\km\s^{-1}}

\newcommand{\kpc}{\,{\rm kpc}}

\newcommand{\rsun}{R_0}
\newcommand{\vsun}{V_0}
\newcommand{\Rsun}{R_0}
\newcommand{\Rsuntilde}{\tilde{R}_0}
\newcommand{\Vsun}{V_0}
\newcommand{\Vsuntilde}{\tilde{V}_0}
\newcommand{\GCset}{\left[\frac{\Rsun}{\kpc}\,, 
\frac{\Vsun}{\kmps}\right]}

\newcommand{\los}{{\it los}}
\newcommand{\vc}{V_c}
\newcommand{\vctilde}{\tilde{V}_c}

\newcommand{\rh}{r_{\rm h}}
\newcommand{\Rt}{R_t}

\newcommand{\vh}{v_{\rm h}}

\newcommand{\vlsr}{v_{\rm LSR}}
\newcommand{\vgsr}{v_{\rm GSR}}
\newcommand{\vt}{v_t}
\newcommand{\vtlsr}{v_{t, {\rm LSR}}}

\newcommand{\Uodot}{U_\odot}
\newcommand{\Vodot}{V_\odot}
\newcommand{\Wodot}{W_\odot}

\newcommand{\ntracer}{n_{\rm tr}}

\newcommand{\sigmagsr}{\sigma_{\rm GSR}}
\newcommand{\sigmar}{\sigma_r}
\newcommand{\sigmat}{\sigma_t}

\newcommand{\sigmazerokmps}{\frac{\sigma_0}{\kmps}}

\newcommand{\Msun}{M_\odot}

\begin{document}
%
\title{Rotation Curve of the Milky Way out to $\sim$ 200 kpc}
\author{Pijushpani Bhattacharjee\altaffilmark{1,2,3}, 
Soumini Chaudhury\altaffilmark{2,4}, and 
Susmita Kundu\altaffilmark{2,5} 
}
\altaffiltext{1}{McDonnell Center for the Space Sciences \& 
Department 
of Physics, Washington University in St. Louis, Campus Box 1105, 
One Brookings Drive, St. Louis, MO 63130. USA.}
\altaffiltext{2}{AstroParticle Physics \& Cosmology Division 
and Centre 
for AstroParticle Physics, 
Saha Institute of Nuclear Physics, 1/AF 
Bidhannagar, Kolkata~700064.~India.}
\altaffiltext{3}{E-mail: pijush.bhattacharjee@saha.ac.in}
\altaffiltext{4}{E-mail: soumini.chaudhury@saha.ac.in}
\altaffiltext{5}{E-mail: susmita.kundu@saha.ac.in}
%
\begin{abstract}
{The rotation curve (RC) of our Galaxy, the Milky Way, is 
constructed starting from its very inner regions (few hundred pc) out 
to a large galactocentric distance of $\sim 200\kpc$ using kinematical 
data on a variety of tracer objects moving in the gravitational 
potential of the Galaxy, without assuming any theoretical models of the 
visible and dark matter components of the Galaxy. 
We study the effect on the RC due to the 
uncertainties in the values of the Galactic Constants (GCs) $\rsun$ and 
$\vsun$ (these being the sun's distance from and 
circular rotation speed around the Galactic center, respectively) and 
the velocity anisotropy parameter $\beta$ of the halo tracer objects 
used for deriving the RC at large galactocentric distances. The 
resulting RC in the disk region is found to depend significantly on the  
choice of the GCs, while the dominant uncertainty in the RC at large 
distances beyond the stellar disk comes from the uncertainty in the 
value of $\beta$. In general we find that the mean RC steadily declines 
at distances beyond $\sim 60\kpc$, independently of the value of 
$\beta$. Also, at a given radius, the circular 
speed is lower for larger values of $\beta$ (i.e., for more radially 
biased velocity anisotropy). Considering that the largest possible value 
of $\beta$ is unity, which corresponds to stellar orbits being purely 
radial, our results for the case of $\beta=1$ give a lower limit to the 
total mass of the Galaxy within $\sim 200\kpc$, $M(200\kpc)\gsim 
(6.8\pm4.1)\times10^{11}\Msun$, independently of any model of the dark 
matter halo of the Galaxy.} 
\end{abstract}

\keywords{Galaxy: rotation curve -- Galaxy: dark matter -- 
Galaxy: mass}


\section{Introduction}\label{sec:intro}
The circular velocity, $\vc(r)= \sqrt{GM(r)/r}$, of a test particle 
at a radial distance $r$ from the center of a mass distribution gives a 
direct measure of the total gravitational mass, $M(r)$, contained 
within that 
radius. A measured profile of $\vc$ as a function of $r$ for a spiral 
galaxy --- often simply called its Rotation Curve (RC) --- is therefore a 
direct probe of the spatial distribution of the total gravitating mass 
inside the galaxy including its dark matter (DM) content;  
see, e.g., \citet{Sofue_Rubin_RC_rev_2001}, 
\citet{Trimble_rev_1987} for reviews. Recent comprehensive discussions 
of the RC and mass models for our Galaxy, the Milky way, can be found, 
e.g., in 
\citet{Weber_deBoer_2010,Sofue_GrandRC_2012,Nesti_Salucci_2013}. 

Recently, it has been shown that the RC of the Milky Way can be directly 
used to derive not only the local density of DM, but also the velocity 
distribution of the DM particles in the Galaxy~\citep{BCKM_PRD_2013}, 
which are crucial for analyzing the results of both direct as well as 
indirect DM search experiments \citep{Jungman_etal_PhysRep_1996}; see 
also 
\citet{CRBM_NewA_2007,CBC_King_JCAP_2010,KB_King_PRD_2012,Burch_Cowsik_2013}. 
For this purpose, it is essential to derive the RC of the Galaxy to as 
large a galactocentric distance as possible without referring to any 
specific model of the DM halo of the Galaxy. In this paper we derive the 
RC of the Galaxy spanning a large range of galactocentric distances 
starting from its inner regions ($\sim 0.2\kpc$) out to $\sim 200\kpc$ 
using kinematical data on a variety of tracer objects moving in the 
gravitational potential of the Galaxy, without assuming any model of the 
DM halo of the Galaxy.  

The circular velocity of a test particle in the Galaxy is, of course, 
not a directly measured quantity. The RC of the Galaxy has to be 
derived from the kinematical as well as positional data 
for an appropriate set of tracer objects moving in the gravitational 
field of the Galaxy. Except in few cases, the full 3-D velocity 
information of the tracers is not available, and the RC has to be 
reconstructed from only the measured line-of-sight (\los) velocity and 
positional information of various tracer objects in the Galaxy. 

For deriving the RC in the disk region of the Galaxy, one usually makes 
the reasonable assumption that the disk tracer objects move in circular 
orbits around the Galactic center. From the observed 
heliocentric \los\  velocities, $\vh$, of the tracers and their position 
coordinates in the Galaxy, and with an assumed set of 
values of the Galactic Constants (GCs), [$\rsun, \vsun$], where $\rsun$ 
and $\vsun$ are the sun's distance from and circular rotation speed 
around the Galactic center, respectively, that define the Local 
Standard of Rest (LSR) frame, and applying corrections for 
the peculiar motion of the sun with respect to the LSR, one can 
obtain the circular velocities around the Galactic center, $\vc$, in a 
fairly straightforward manner~\citep{Binney_Merrifield_1998}. 
Observations on  
a variety of tracers such as HI regions, CO emission associated with HII 
regions, compact objects like Carbon stars (C stars), Cepheids, 
planetary nebulae (PNe), masers, and so on, have been used to derive the 
RC of the Galaxy in the disk region. Some recent compilations of RC data 
for the disk region of the Galaxy can be found, e.g., in 
\citet{Sofue_etal_2009} and \citet{Burch_Cowsik_2013}. 

To derive the RC in the outer regions of the Galaxy beyond the Galactic 
disk, one has to rely on distant tracers like Blue Horizontal Branch 
(BHB) stars, K Giant (KG) stars and relatively rare tracer objects like 
Globular Clusters (GCl), dwarf spheroidal (dSph) galaxies and so forth  
which populate the Milky Way's extended DM halo out to 
galactocentric distances of several hundreds of kpc. Unlike the disk 
tracers, these non-disk tracers do 
not exhibit any systematic motion, and move about in the Galaxy along 
various different orbits. The standard approach then is to assume that 
the tracer population under consideration is isotropically 
distributed in the halo of the Galaxy and then use the Jeans 
equation~\citep{Binney_Tremaine_2008} for spherical systems relating the 
circular velocity $\vc$ at radius $r$ to the number density and 
galactocentric radial as well as transverse velocity dispersions of the 
tracers at that radius. Of course, in absence of full 3-D velocity 
information, with only the observed radial velocity dispersion 
available, the RC constructed using Jeans equation depends on the 
unknown velocity anisotropy parameter $\beta\equiv 
1-\sigmat^2/2\sigmar^2$ ($\sigmar$ and $\sigmat$ being the radial and 
transverse velocity dispersions of the tracers, respectively; see 
section \ref{sec:nondisk} below). 

The Jeans equation approach has been used in several recent studies to 
extend the RC of the Galaxy to distances beyond the extent of the 
Galaxy's stellar disk. Accurate measurements of \los\ velocities of a 
sample of 2401 BHB stars drawn from SDSS DR6 
\citep{SDSS-DR6} were used by \citet{Xue_etal_2008} to derive the RC 
of the Galaxy to $\sim 60\kpc$ for two constant ($r$-independent) values 
of $\beta$, namely $\beta=0$ (isotropic velocity distribution) and 
$\beta=0.37$, the 
latter derived from results of numerical simulations. More recently, the 
Jeans equation has also been employed, together with certain analytical 
models of the phase-space distribution 
function of the tracer population, to construct the RC of the Galaxy to 
various distances of $\sim$ 25 to $\sim$ 80 kpc 
\citep{Gnedin_etal_2010,Deason_etal_2012a,Kafle_etal_2012}. 

A crucial ingredient in the derivation of the distant RC using 
Jeans equation is the measured radial velocity dispersion of the tracers 
as a function of their galactocentric distance $r$. An important finding 
in this regard is the result, first shown by 
\citet{Battaglia_etal_2005_2006}, that the radial 
velocity dispersion remains almost constant at a value of $\sim 
120\kmps$ out to $\sim 30\kpc$ and then steadily {\it declines} down to 
a value of $\sim 50\kmps$ at $r\sim120\kpc$. In 
their work \citet{Battaglia_etal_2005_2006} used a heterogeneous sample 
of about 240 halo objects consisting of field blue horizontal branch 
stars, red giant stars, globular clusters and distant 
satellite galaxies. Similar trend of the radial 
velocity dispersion profile has been found in several subsequent studies 
using different samples of tracers, e.g., by  
\citet{Xue_etal_2008,Brown_etal_2010,Gnedin_etal_2010,Deason_etal_2012a,Deason_etal_coldveil_2012b}, 
and most recently in large cosmological simulations by 
\citet{Rashkov_etal_eris_2013}.    

In this paper we consider a combination of currently available 
largest samples of a variety of both disk and non-disk tracers to 
construct the RC of the Galaxy from $\sim 0.2\kpc$ to $\sim200\kpc$. We 
perform detailed analysis of the dependence of the RC on 
the choice of the GCs and also the dependence on the anisotropy parameter 
$\beta$ of the non-disk tracers. We find that, while the RC in the disk 
region is significantly influenced by the choice of the GCs, the dominant 
uncertainty in the RC at large distances beyond the stellar disk comes 
from the uncertainty in the value of $\beta$. Since currently not much 
reliable observational information on $\beta$ is available, in this 
paper we calculate the circular velocities using Jeans equation 
with the velocity anisotropy $\beta$ of the tracers taken as (a) a 
radially constant free parameter varying over a possible range of values 
from $\beta=0$ (corresponding to complete isotropy of the tracers' 
orbits) to $\beta=1$ (corresponding to completely radial orbits of the 
tracers), (b) a radially varying $\beta$ of the Osipkov-Merritt (OM) form  
\citep[see][p.297-298]{Binney_Tremaine_2008} given by 
$\beta(r)=(1+r_a^2/r^2)^{-1}$, with $r_a$ the ``anisotropy radius", and 
(c) a radial profile of $\beta$ 
obtained from a recent large high resolution 
hydrodynamical simulations of formation of late-type spirals like our 
Galaxy~\citep{Rashkov_etal_eris_2013}. 

We find that, irrespective of the value of $\beta$, the mean RC steadily 
declines with $r$ beyond $r\sim 60\kpc$. The circular speed 
at a given radius decreases as $\beta$ is increased (i.e., as the 
tracers' orbits are made more radially biased). Thus, the lowest value 
of the rotation speed at any $r$ obtains for the case of complete radial 
anisotropy ($\beta=1$) of the non-disk tracers. This fact allows us to 
set a lower limit on the total mass of the Galaxy, $M(r)$, within 
a radius $r$, giving $M(200\kpc)\geq 
(6.8\pm4.1)\times10^{11}\Msun$. In this context, it may be noted 
that the recent numerical simulation study of 
\citet{Rashkov_etal_eris_2013} indicates 
an increasingly radially biased velocity ellipsoid of the Galaxy's 
stellar population at large distances, with stellar orbits tending to be 
purely 
radial ($\beta\to 1$) beyond $\sim 100\kpc$. If this behavior of 
$\beta$ is confirmed by future observational data, then the above lower 
limit on the Galaxy's mass (obtained from our results with $\beta=1$) may 
in fact be a good estimate of the actual mass of the Galaxy out to 
$\sim 200\kpc$. 

The rest of this paper is arranged as follows. In Section
\ref{sec:disk} we derive the RC on the disk of the Galaxy up to a 
distance of $\sim20\kpc$ from the Galactic center. We specify the 
various tracer samples used in our derivation of the RC and study 
the dependence of the RC on the chosen set of values of the GCs, 
[$\rsun, \vsun$]. In Section \ref{sec:nondisk} we extend the RC to 
larger  
distances (up to $\sim 200 \kpc$) by an extensive analysis of various 
non-disk tracer samples discussed there in details. Finally, in 
Section \ref{sec:combined_vc}, we present our unified RC and our 
estimates of the total mass of the Galaxy within $\sim 200\kpc$ 
and conclude by summarizing our main results in Section 
\ref{sec:summary}.

\section{Rotation curve from disk tracers}\label{sec:disk}
\begin{figure*}[!htb]
\begin{tabular}{cc}
\epsfig{file=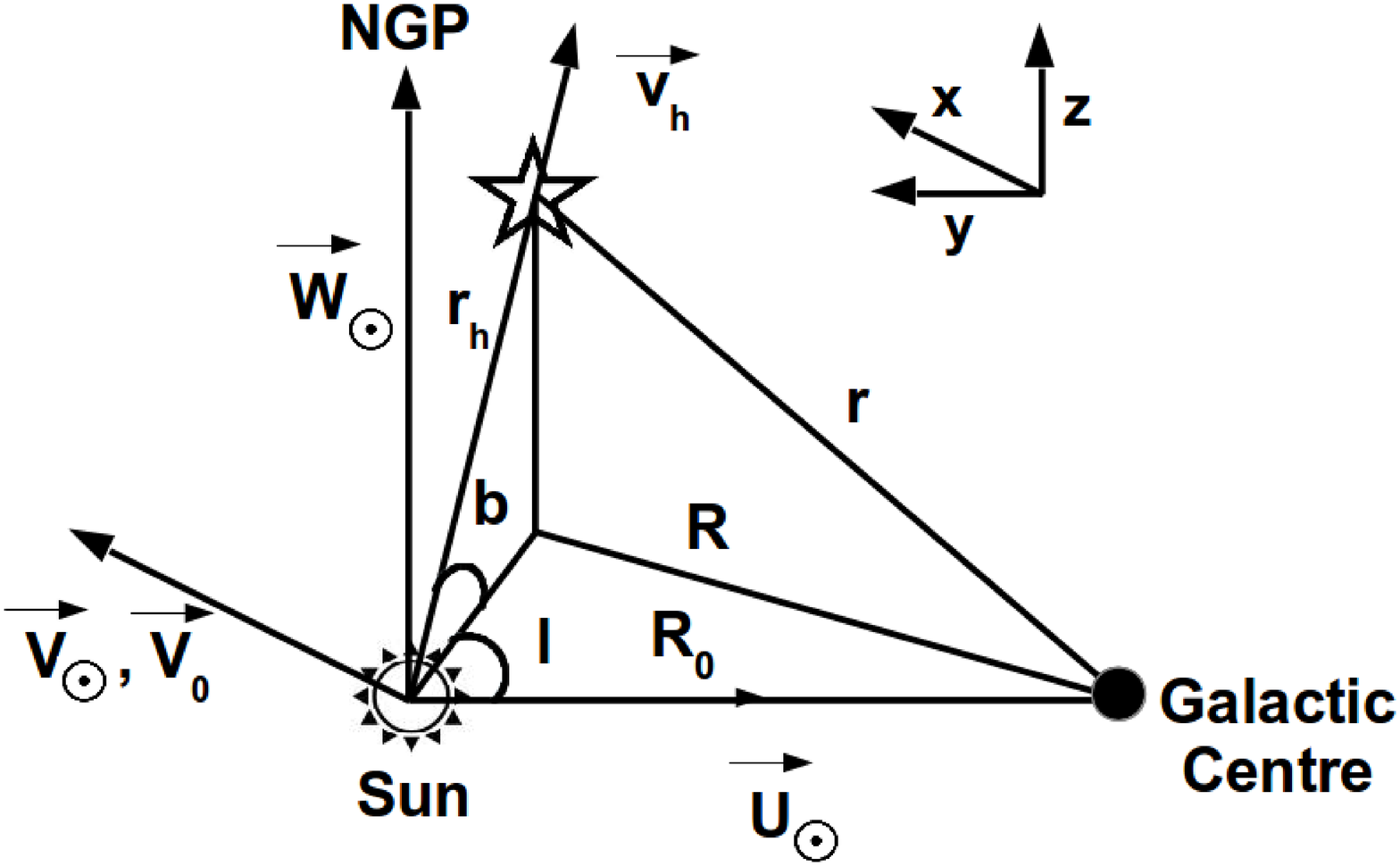,angle=0,width=\columnwidth}&
\epsfig{file=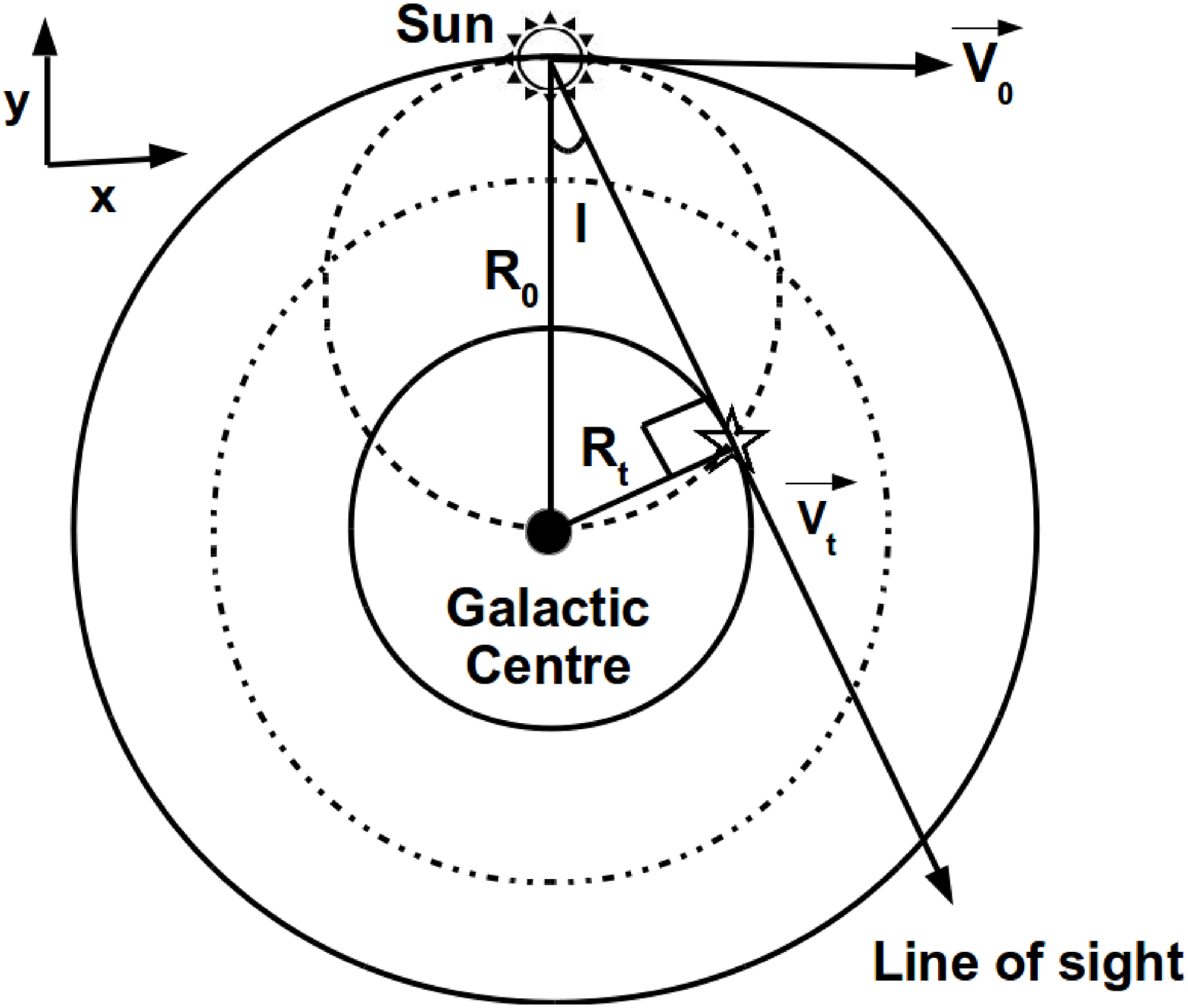,angle=0,width=\columnwidth}
\end{tabular}
\caption{Left: Schematic diagram showing the coordinate system,
velocity and distance notations used in this work. Right: Illustration
of the tangent point method for deriving the circular speeds
for distances $R<\rsun$ on the disk.}
\label{Fig:Fig_coordinates_a_axes_b_TPM_clr}
\end{figure*}

Let us consider a tracer object with Galactic coordinates ($l,~b$) at a 
heliocentric distance $\rh$ and observed heliocentric \los\ velocity 
$\vh$ (see Figure \ref{Fig:Fig_coordinates_a_axes_b_TPM_clr}). 

We shall assume that the tracer follows a nearly circular orbit about 
the Galactic center. The velocity of the tracer as would be measured by 
an observer stationary with respect to the LSR, $\vlsr$, can be obtained 
from the measured $\vh$ through the relation 
\begin{equation}
\vlsr=\vh+\Uodot\cos b \cos l + \Vodot\cos b \sin l + \Wodot \sin b\,,
\label{eq:vh2vlsr}
\end{equation}
where $(\Uodot, \Vodot, \Wodot)$ denote the peculiar motion of the 
sun with respect to LSR; see Figure \ref{Fig:Fig_coordinates_a_axes_b_TPM_clr}. 
In our calculations below we shall take $(\Uodot, \Vodot, \Wodot) = 
(11.1,~12.24,~7.25)$ ($\kmps$) \citep{Schonrich_etal_pecvel_2010}. 
Simple algebraic steps then allow us to relate the desired circular
velocity with respect to Galactic center rest frame, $\vc$, to $\vlsr$  
as \citep{Binney_Merrifield_1998} 
\begin{equation}
\vc(R)=\frac{R}{\rsun}~\left[\frac{\vlsr}{\sin l \cos b} + 
\vsun\right]\,, 
\label{eq:vlsr2vc}
\end{equation}
where, $R$ is the projection of the galactocentric distance $r$ onto the
equatorial plane,
\begin{equation}
R=\sqrt{\rsun^2 + \rh^2 \cos^2 b - 2 \rsun ~\rh \cos b \cos l}\,.  
\label{eq:rh2R}
\end{equation}
For a given set of GCs, $[\rsun,\vsun]$, the Cartesian coordinates of the 
tracer are given by 
\begin{eqnarray}
x & = & \rh ~\cos b ~\sin l \,,\nonumber \\ 
y & = &\rsun-\rh~\cos b~\cos l\,, \label{xyz_general}
\\ 
z &=  &\rh~\sin b\,,\nonumber 
\end{eqnarray}
with Galactic center at the origin and sun lying on the Galactic 
mid-plane ($z=0$) with coordinates $(x,y,z) = (0,\rsun, 0)$ as 
illustrated in the left panel of Figure 
\ref{Fig:Fig_coordinates_a_axes_b_TPM_clr}.  
Hence, for known ($l,~b,~\rh,~\vh$) one can
solve for $\vc$ from Equation (\ref{eq:vlsr2vc}) for a given set of GCs. 

{\it Tangent Point Method} (TPM) : For $R<\Rsun$, one can calculate 
$\vc$ by the simple tangent point method \citep{Binney_Merrifield_1998} 
as follows: Along a given \los,\ the maximum \los\ velocity will occur 
for the tracer closest to the Galactic center, with the \los\ tangent to 
the circular orbit of the tracer at that point (see right panel of
Figure \ref{Fig:Fig_coordinates_a_axes_b_TPM_clr}). This maximum \los\ 
velocity, called the terminal velocity ($\vt$), is easily seen to be 
related to $\vc$ through the relation 
\begin{equation}
\vc(\Rt)=  \left|\vtlsr(\Rt)+\Vsun \sin l\right|\,, ~~~(b=0)\,,
\label{eq:vc_tangent}
\end{equation}
where  
\begin{equation}
\Rt = |\Rsun~\sin l|
\label{eq:R_tangent}
\end{equation}
is the distance of the tangent point from the Galactic center, and 
$\vtlsr$ is the $\vt$ corrected for the sun's peculiar motion as in 
Equation (\ref{eq:vh2vlsr}).

For non zero galactic latitude ($b$), Equation (\ref{eq:vc_tangent}) 
generalizes to:
\begin{equation}
\vc(\Rt)= \left|\frac{\vtlsr(\Rt)}{\cos b} + \Vsun~\sin l\right|\,,
\label{eq:vc_tangent_b}
\end{equation}
and in this case the Cartesian coordinates of the tracer are given by  
\begin{eqnarray}
x &=&\rsun~\sin l~\cos l \,, \nonumber \\
y &=& \rsun~\sin^2 l\,, \label{eq:vc_tangent_xyz}
\\
z &=& \rsun ~\cos l ~\tan b\,.\nonumber
\end{eqnarray}
Hence the circular velocity $\vc$ can be calculated 
directly from the measured terminal velocity by using 
Equation (\ref{eq:vc_tangent_b}).

The details of the disk tracer samples used in this paper along with 
references to 
the corresponding data sources for each tracer genre are given in 
the Appendix (Table \ref{Table:Disk_samples}). 
The cuts on $l$ and $b$ are adopted from 
the published source papers. Towards the Galactic center ($l\to 
0\textdegree$) or 
anti-center ($l\to180\textdegree$), we expect $\vlsr$ to approach zero 
to prevent unphysical $\vc$ values there [see 
Equation (\ref{eq:vlsr2vc})]. However, $\vlsr$ observations in practice 
have finite values due to contamination from non circular motions 
dominant there. Therefore, additional restrictions have been applied on 
$l$ ranges so as to ensure that we avoid observations too close to 
Galactic center (anti-center) regions. We further impose a cut to 
keep only the tracers whose $|z| \leq 2 \kpc$ and $ R \leq 25 \kpc$ so 
as to ensure that the selected tracers `belong' to the stellar disk of 
the Galaxy. The $x$--$y$ and $l$--$z$ scatter plots for the selected 
disk tracers listed in Table \ref{Table:Disk_samples} are shown in 
the Appendix (Figures \ref{Fig:Fig_disk_xy_scatter_clr} and
\ref{Fig:Fig_disk_lz_scatter_clr}, respectively).  

It is clear from Equations (\ref{eq:vlsr2vc}) -- 
(\ref{eq:vc_tangent_xyz}) 
that the RC depends on the set of values of the GCs  
([$\Rsun, \Vsun$]) adopted in the calculation. Values of $\Rsun$ in the 
range $\sim (7 - 9) 
\kpc$ and $\Vsun$ in the range $\sim (180 - 250)\kmps$ exist in 
literature \citep[see, 
e.g.,][]{Reid_1993_r0_8,Olling_M_1998_GC_range,Ghez_etal_2008_r0,Reid_etal_2009_VLBI_maser,MB_2010,Sofue_etal_2011_GC,Brunthaler_etal_2011_r0_v0,Schonrich_2012_r0_v0}. 
Actually, the ratio $\Vsun/\Rsun = (A - B)$, $A$ and $B$ being the 
Oort constants \citep[see, e.g.,][]{Binney_Merrifield_1998}, is 
considerably better constrained. Maser observations and 
measurements of stellar orbits around SgrA* near the Galactic center
report values of $(A-B)$ in the range from about 29 to 32 
$\kmps\kpc^{-1}$ 
\citep{Reid_Brunthaler_2004_GC_ratio_sgA_pm,Reid_etal_2009_VLBI_maser,MB_2010}.  
RCs have been traditionally presented with the IAU 
recommended set of values, $\GCset_{\rm IAU} = [8.5, 220]$, for which, 
however, the ratio $\Vsun/\Rsun = 25.9$ is outside the range of values 
of this ratio mentioned above. A recently suggested set of 
values of $[\Rsun, \Vsun]$, consistent with observations of 
masers and stellar orbits around SgrA* mentioned above, is $\GCset = 
[8.3, 244]$ \citep[see, e.g.,][]{Bovy_etal_2009,Gillessen_etal_2009}. 

In general, as easily seen from Equation (\ref{eq:vlsr2vc}), given a RC, 
$\vc(R)$, for a certain set of values of $[\Rsun, \Vsun]$, one can 
obtain the new RC, $\vctilde(R)$, for another set of values of the GCs 
denoted by $[\Rsuntilde, \Vsuntilde]$ through the relation 
\begin{equation} 
\vctilde(R) = 
\frac{\Rsun}{\Rsuntilde} \left[\vc(R) - 
\frac{R}{\Rsun}\left(\Vsun - \Vsuntilde\right)\right]\,.
\label{eq:v_rescale} 
\end{equation} 
In order to illustrate the dependence of the RC on the 
choice of the GCs, in this paper we shall calculate RCs with three 
different sets 
of values of $\GCset$, namely the set [8.3, 244] mentioned above as well 
as two other sets, the IAU recommended set [8.5, 220] and 
the set [8.0, 200] \citep{Sofue_GrandRC_2012}.   

\begin{figure*}[!htb]
\begin{tabular}{cc}
\epsfig{file=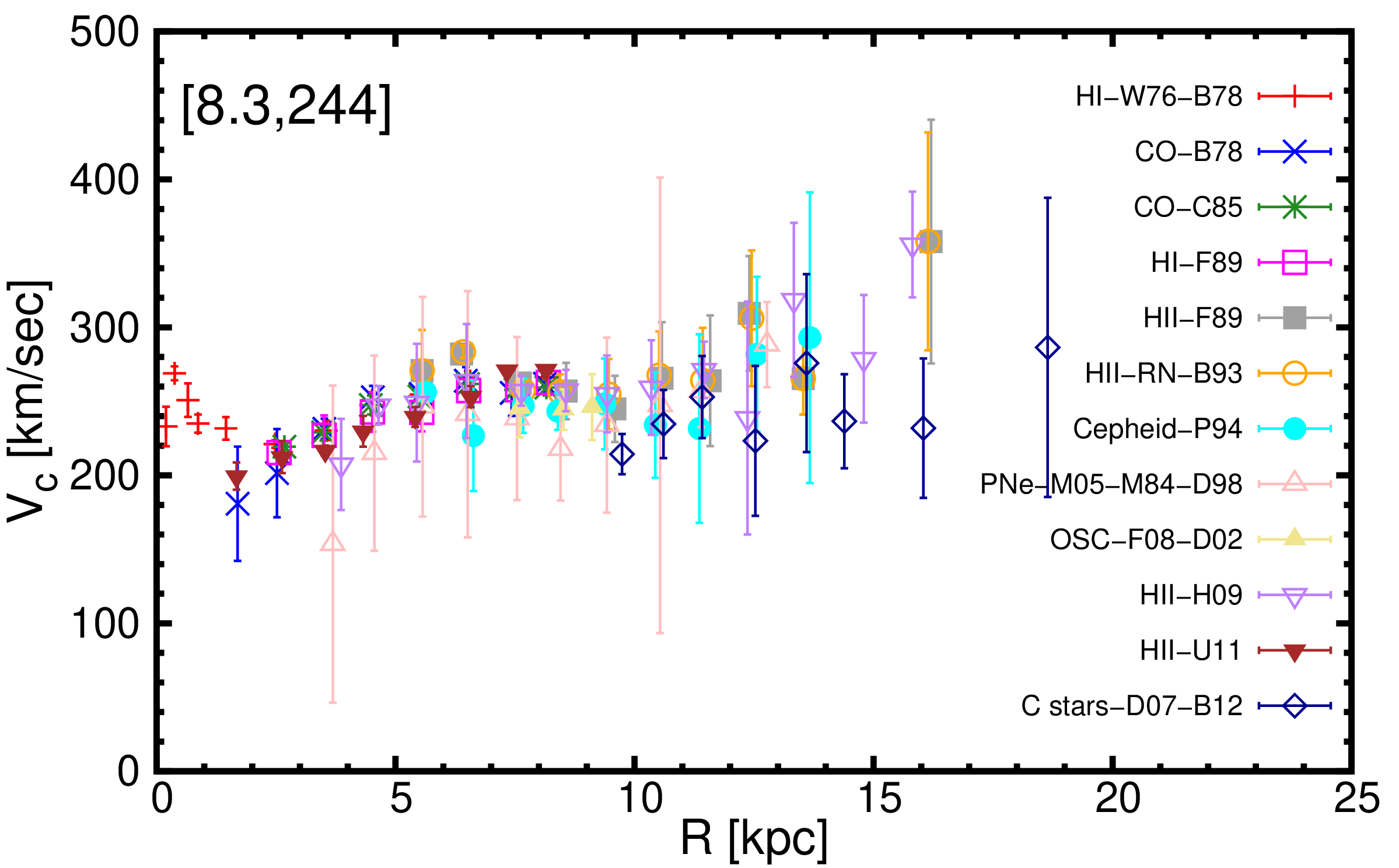,angle=270,width=\columnwidth} & 
\epsfig{file=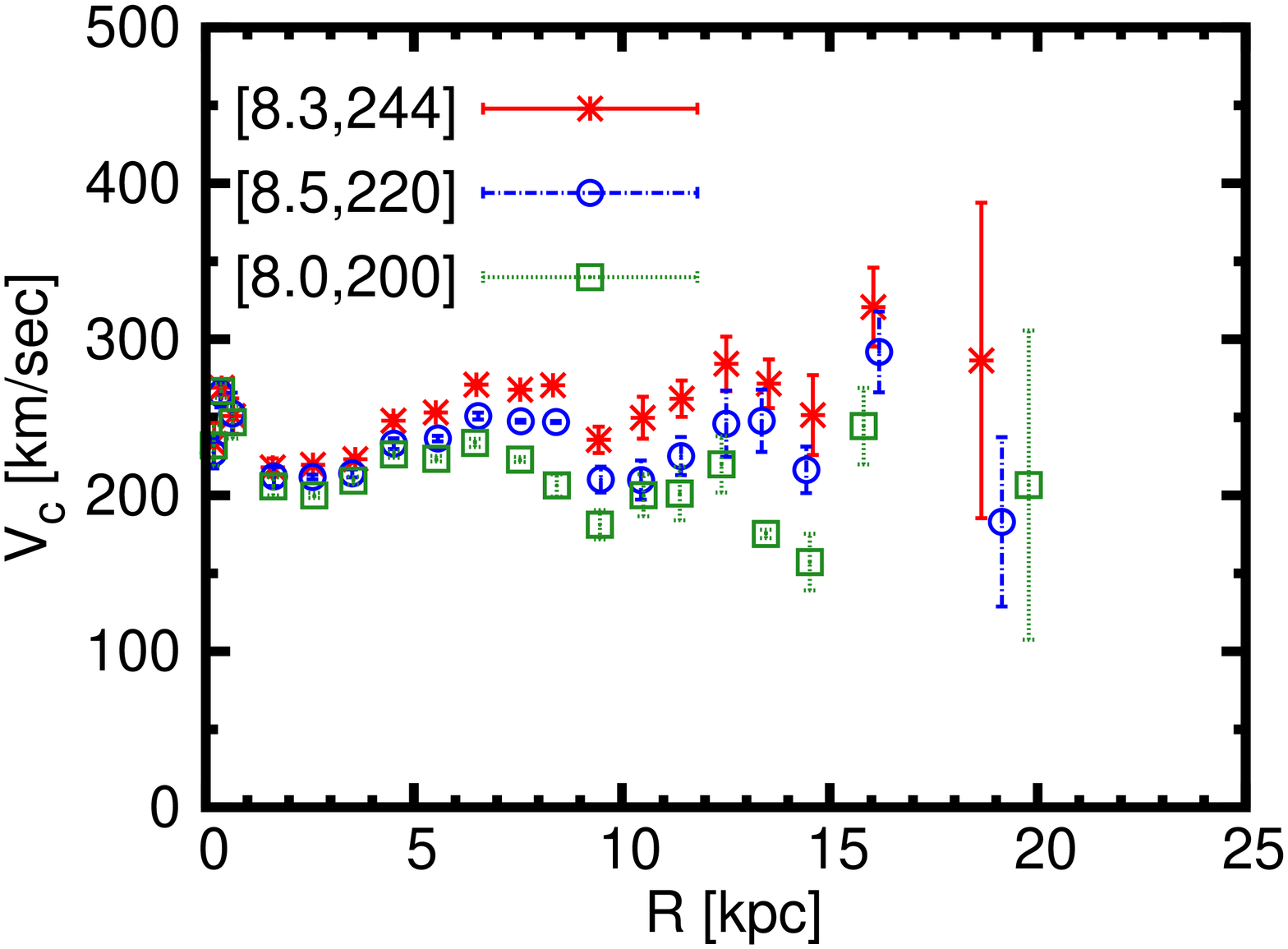,angle=270,width=\columnwidth}
\end{tabular}
\caption{Left: Rotation curves of the Galaxy obtained using the various
different disk tracer samples listed in Table
\ref{Table:Disk_samples} for the Galactic Constants
$\GCset = [8.3, 244]$. See Table \ref{Table:Disk_samples} for
keys to the data points. Right: Averaged rotation curves obtained by 
weighted averaging over the combined $\vc$ data from all the disk tracer 
samples listed in Table \ref{Table:Disk_samples} and shown in the left  
panel, for three different sets of values of $\GCset$ as 
indicated.
}
\label{Fig:Fig_vc_disk_a_samples_8.3_b_coll_dgc_clr}
\end{figure*}

Figure \ref{Fig:Fig_vc_disk_a_samples_8.3_b_coll_dgc_clr} shows our 
calculated RCs for the disk region of the Galaxy. The left panel of 
Figure \ref{Fig:Fig_vc_disk_a_samples_8.3_b_coll_dgc_clr} shows the 
RCs for each of the different tracer samples listed in 
Table \ref{Table:Disk_samples} for the GCs set $\GCset = [8.3, 244]$, and 
the right panel shows the RCs obtained by taking the weighted averages  
of the combined $\vc$ data from all the samples shown in the left panel, 
for three different sets of values of the GCs as indicated. 

The circular velocities and their errors for 
individual disk tracer samples displayed in the 
left panel of Figure \ref{Fig:Fig_vc_disk_a_samples_8.3_b_coll_dgc_clr} are 
obtained in the following way: For each tracer object in a given sample 
we calculate $\vc$ and $R$ for the object from the known position 
coordinates of the object and its measured \los\ velocity as  
described above. We then bin the resulting data ($\vc$ vs.~$R$) in $R$, 
and in each $R$ bin calculate the mean of all the $\vc$ values of all 
the objects contained within that bin and assign it to the mean $R$ 
value of the objects in that bin. The error bars on $\vc$ 
correspond simply to the standard deviation (s.d.) of the $\vc$ values 
in that bin~\footnote{Note that the \los\ velocities $\vh$ of 
individual tracer objects are measured fairly accurately and 
their measurement errors contribute negligibly little to the final 
errors on the $\vc$ values.}. We have taken a bin size of 0.25 kpc for 
$0<R\leq 1\kpc$, 1.0 kpc for $1<R\leq 15\kpc$, and 
2.5 kpc for $15<R\leq 17.5\kpc$. The objects with $R> 17.5\kpc$ are 
few in number and are placed in one single bin. The above choices 
of the bin widths in $R$ for various ranges of $R$, arrived 
at by trial and error, are ``optimal" in the sense that the bin widths 
are large enough so that there are sufficient number of objects 
in each bin (to allow the mean value of $\vc$ in the bin to be a 
reasonably good representative of the true value of $\vc$ at the value 
of $R$ under consideration), while at the same time being not too large 
as to miss the fine features of the RC. The RCs 
in the right panel of Figure 
\ref{Fig:Fig_vc_disk_a_samples_8.3_b_coll_dgc_clr} are obtained by 
combining the $\vc$ data from all the samples shown in the left 
panel in the same $R$ bins as above and then calculating the mean 
circular speed ($\vc$) and its $1 \sigma$ uncertainty ($\Delta \vc$) 
within each bin by the standard weighted average method 
\citep{Bevington_3rdEd_2003}: 
\begin{equation}
\vc =  \frac{\sum_i w_i V_{c,i}}{\sum_i w_i}\,, {\rm \hskip 0.5cm 
and 
\hskip 0.5cm} 
\, \Delta \vc = \sqrt{\frac{1}{\sum_i w_i}}\,,
\label{eq:weighted_av}
\end{equation}
with $w_i = 1/(\Delta V_{c,i})^2$, where $V_{c,i}$ and $\Delta V_{c,i}$ 
are the $\vc$ value and its $1\sigma$ error, respectively, of the 
$i$-th data point within the bin.

As seen from Figure \ref{Fig:Fig_vc_disk_a_samples_8.3_b_coll_dgc_clr}, the 
RC in the disk region depends significantly on the choice of GCs. As 
expected, at any given $R$ the circular velocity is higher for higher 
value of $V_0$. 

\section{Rotation curve from non-disk tracers} 
\label{sec:nondisk}
In order to extend the RC beyond the Galactic disk we next consider 
tracer objects populating the stellar halo of the Galaxy. Unlike the 
nearly circularly rotating disk tracers the non-disk tracers do not  
exhibit any systematic circular motion. Hence the formalism described 
in the previous section cannot be used to derive the RC at large 
galactocentric distances beyond the Galactic disk. Instead, 
we use the Jeans equation~\citep[see, 
e.g.,][p.349]{Binney_Tremaine_2008} for spherical systems relating the 
number density and radial as well as transverse velocity 
dispersions of the tracers at radius $r$ to the circular velocity $\vc$ 
at that radius: 
\begin{equation}
\vc^2(r) = \frac{GM(r)}{r} = - \sigma^2_r 
\left(\frac{d\ln\ntracer}{d\ln r} 
+ \frac{d\ln\sigma^2_r}{d\ln r} + 2\beta\right)\,.
\label{eq:Jeans_eqn}
\end{equation}
Here $r=\left(\rsun^2 + \rh^2 - 2 \rsun ~\rh \cos b \cos 
l\right)^{1/2}$ is the galactocentric radial distance of a tracer (see 
Figure \ref{Fig:Fig_coordinates_a_axes_b_TPM_clr}), and $\ntracer$, $\sigmar$ 
and $\beta$ are, respectively, the number density of the tracer 
population, their galactocentric radial velocity dispersion, and the 
velocity anisotropy parameter, at $r$. The velocity anisotropy $\beta$ 
is defined as  
\begin{equation}
\beta=1-\frac{\sigmat^2}{2\sigmar^2}\,, 
\label{eq:beta_anisotropy_parameter}
\end{equation}
where $\sigmat$ is the galactocentric transverse velocity 
dispersion of the tracers. 

In this work we have chosen two independent classes of  
non-disk stellar tracers, namely, a sample of 4985 Blue Horizontal 
Branch (BHB) stars from SDSS-DR8 compiled 
by \citet{Xue_etal_BHB_SDSS_DR8_2011} and a set 
of 4781 K Giant (KG) stars from SDSS-DR9 
\citep{Xue_etal_KGiants_SDSS_DR9_2012}. These two samples allow us to 
probe the Galactic halo up to a galactocentric distance of $\sim 100 
\kpc$. In order to reach out further we 
consider an additional heterogeneous (Hg) sample  of 430 
objects comprising of 143 Globular Clusters (GCl)  
\citep{Harris_GCl_2010_1996}, 118 red halo giants (RHG) 
\citep{Carney_etal_2003_2008}, 108 field blue horizontal branch  (FHB) stars 
\citep{Clewley_etal_2004}, 38 RR-Lyrae stars (RRL)  
\citep{Kinman_etal_2012}, and 23 dwarf spheroidals (dSph)  
\citep{McConnachie_DG_2012}. To  ensure that the sample comprises of 
only halo objects, we apply a cut on the $z$ and $R$ coordinates of the 
tracers, leaving out objects with $r<25\kpc$ in all the non-disk tracer 
samples mentioned above. After these cuts, we are left with a ``BHB" 
sample of 1457 blue horizontal branch stars, a ``KG" sample of 2227 
K-giant stars and a ``Hg" sample of 65 objects comprising of 16 GCls, 
28 FHB stars and 21 dSphs, with which we shall 
construct our RC for the non-disk region. The last sample allows us to 
extend the RC to a galactocentric distance of 190 kpc, the mean $r$ of 
the objects in the furthest radial bin in the Hg sample. The spatial 
distributions of the three final non-disk tracer samples (after position 
cuts mentioned above) in terms of 
$x$-$z$, $y$-$z$ and $x$-$y$ scatter plots are shown in the Appendix 
(Figure \ref{Fig:Fig_nondisk_scatter_BHB_KG_HS_clr}). 

The number density of the tracers, $\ntracer$, appearing in the Jeans 
equation (\ref{eq:Jeans_eqn}) is estimated in the following way. We 
radially bin the objects in a given sample and estimate the 
tracer density from the star counts in the annular volume of each bin  
and assign it at the mean radius of the objects contained within 
that bin. In order to ensure a reasonably good number of objects per bin 
we adopt a variable bin size increasing with distance. For the BHB 
sample, a uniform bin size of 2 kpc is used over its entire range of $r$ 
from 25 to 55 kpc. For the KG samples, the bin widths are 2 kpc for 
$25\kpc < r \leq 55\kpc$ and 4 kpc for $55\kpc < r \leq 103\kpc$; 
objects with $r > 103\kpc$ (up to 110 kpc) are all placed in one single 
bin. For the Hg sample, because of the relatively small total number 
(65) of objects, we adopt the following optimal, ``object wise" binning 
in increasing order of the galactocentric distance $r$ of the objects:  
the first 6 radial bins contain 8 objects in each bin; the next 2 bins 
contain 6 objects in each bin; and, finally, the remaining 5 objects are 
placed in one single bin. Uncertainties in the number density estimates 
are obtained from Poissonian errors on the tracer counts in each bin. 

\begin{figure*}[!htb]
\begin{tabular}{cc}
\hskip 1cm
\epsfig{file=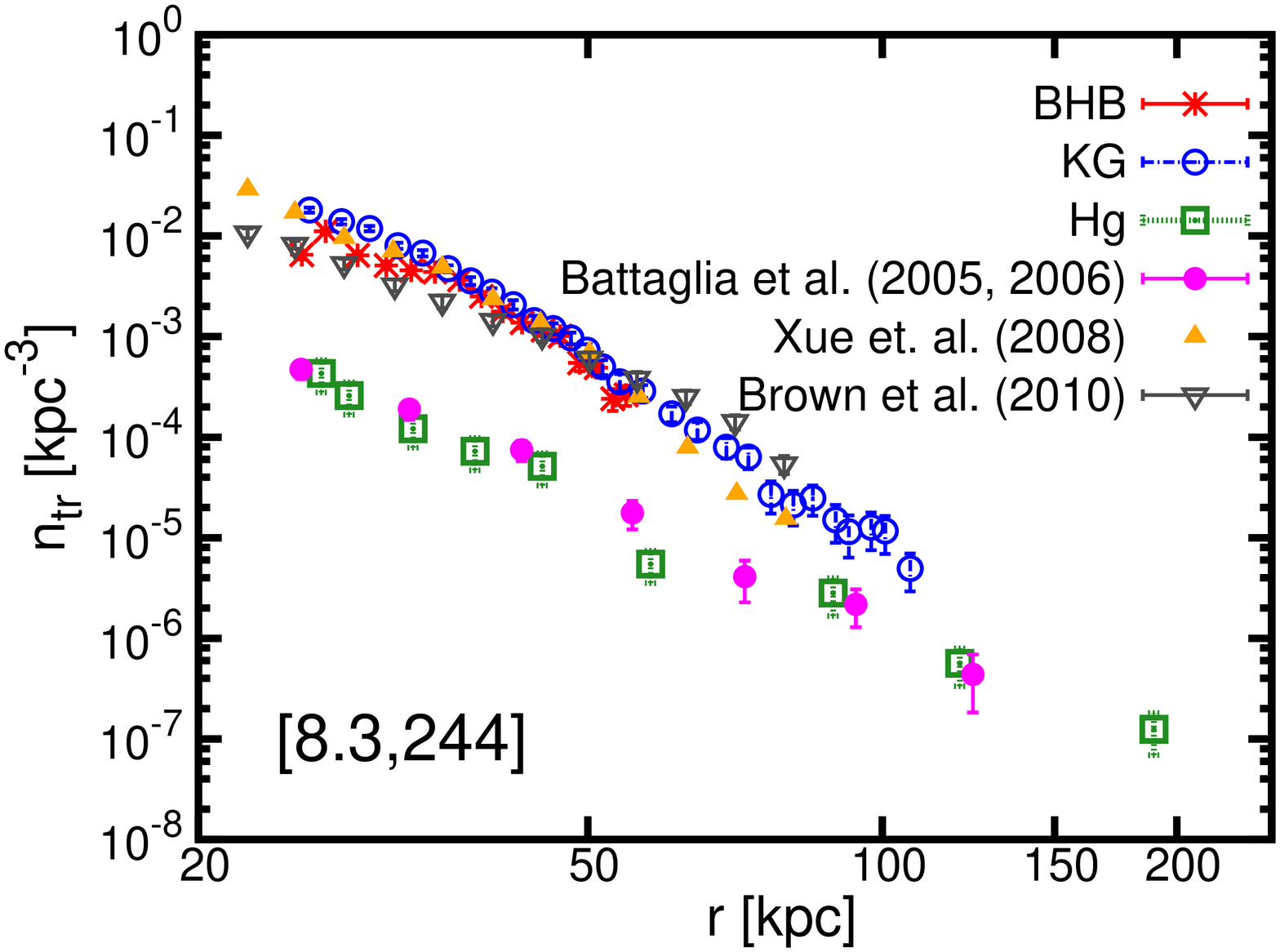,angle=270,width=0.9\columnwidth}&
\epsfig{file=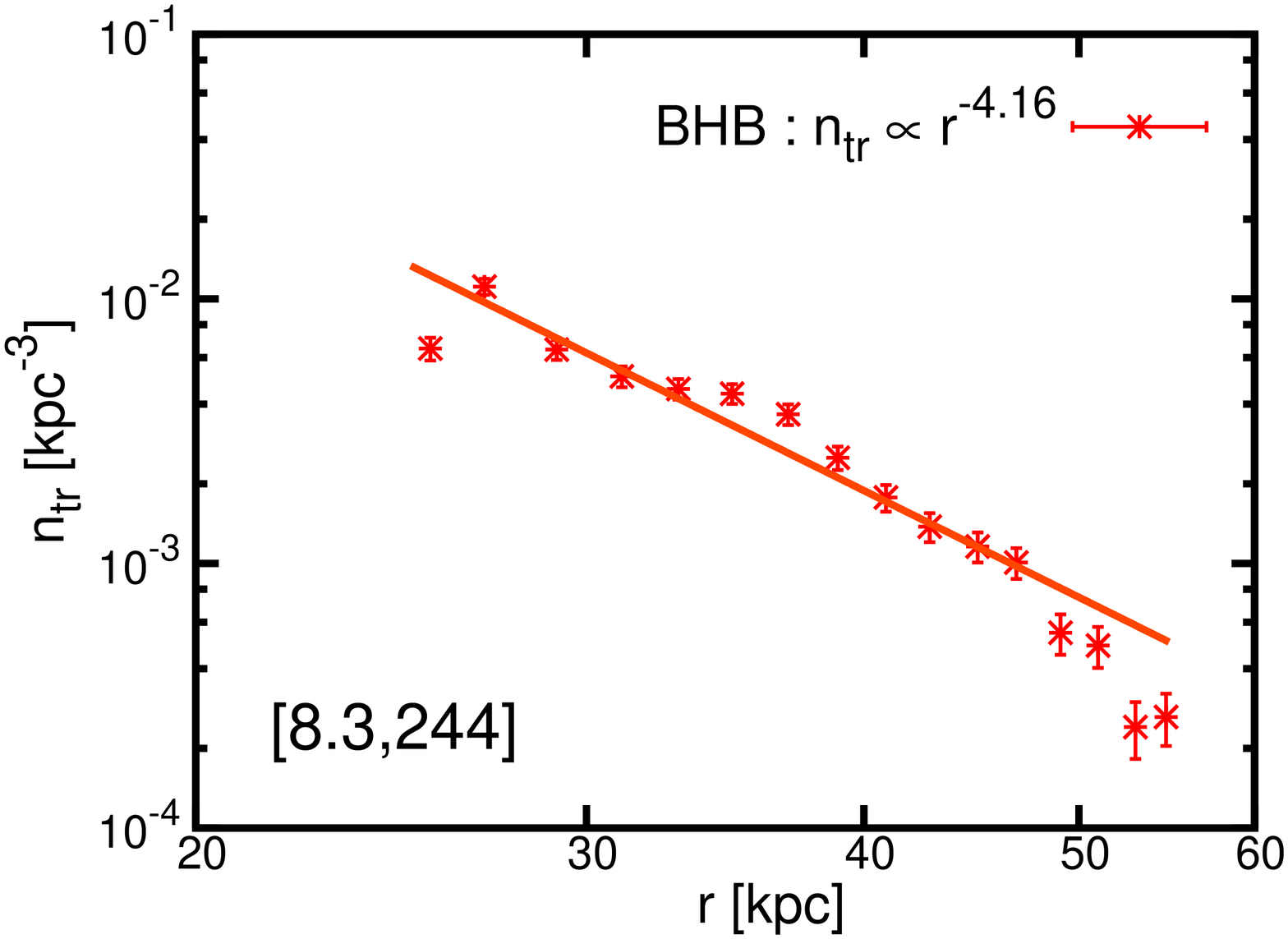,angle=270,width=0.9\columnwidth}\\
\hskip 1cm
\epsfig{file=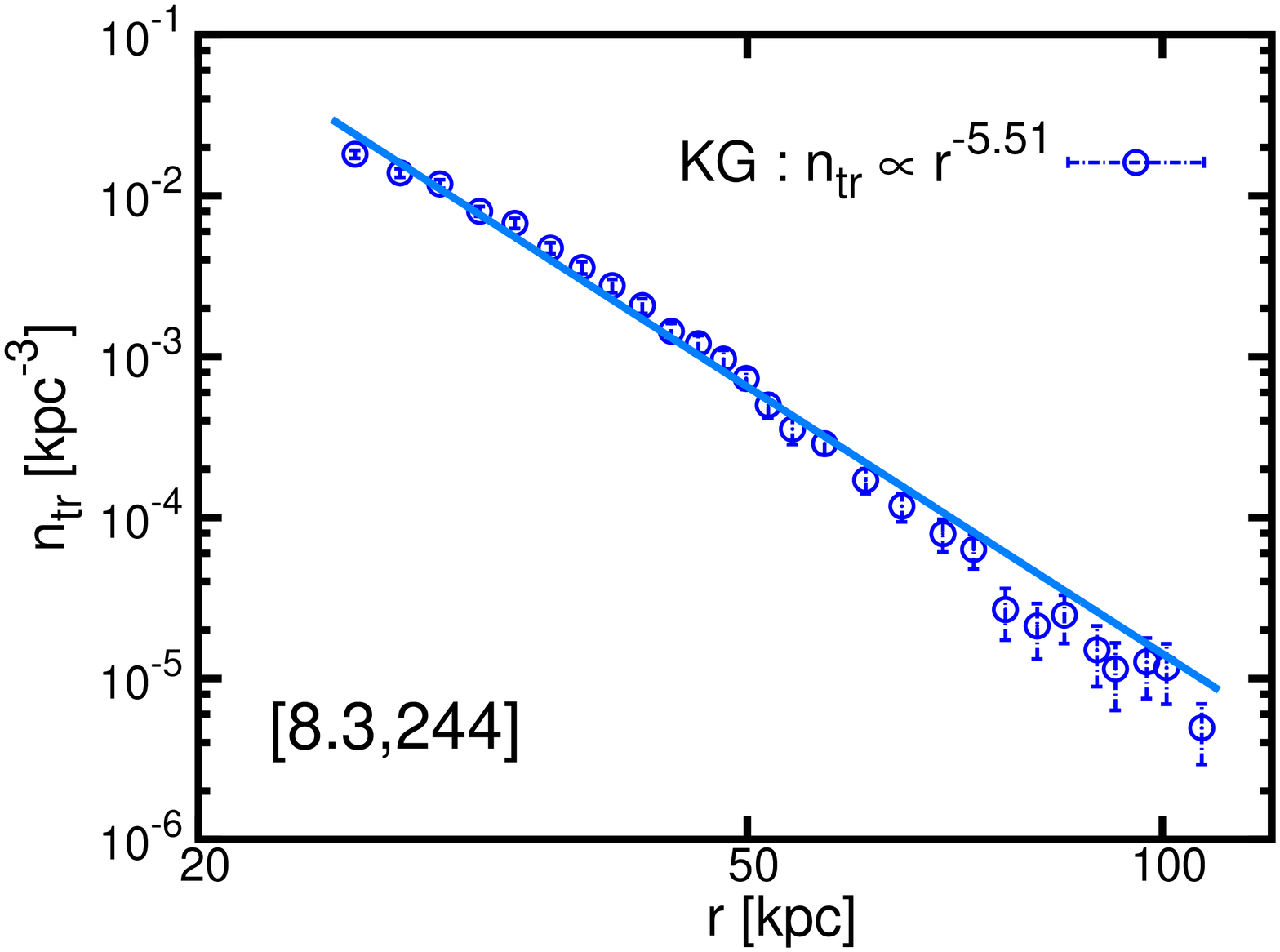,angle=270,width=0.9\columnwidth}&
\epsfig{file=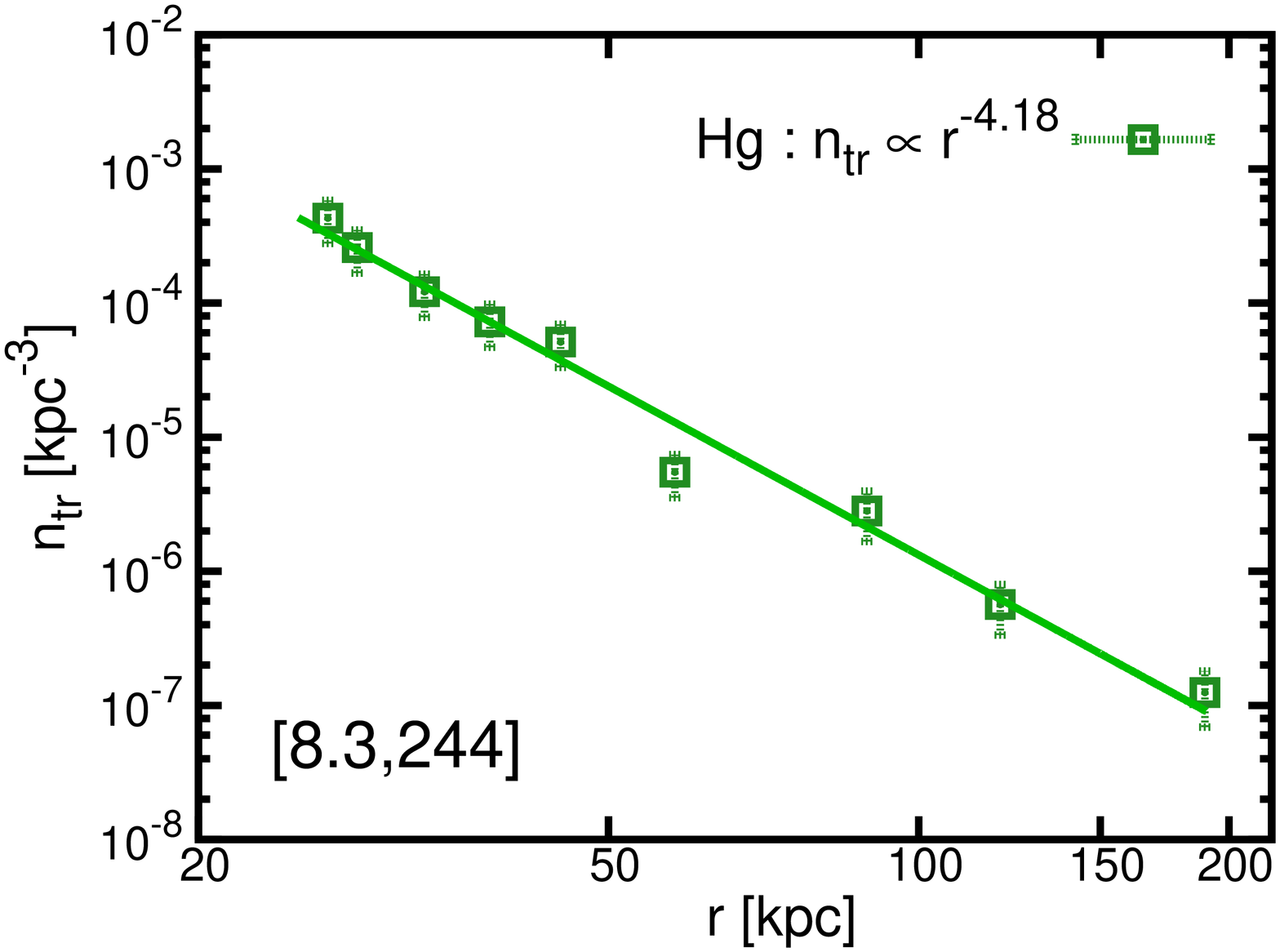,angle=270,width=0.9\columnwidth}\\
\end{tabular}
\caption{The tracer number density, $\ntracer$, for the three
non-disk tracer samples considered in this paper (see
text and Figure
\ref{Fig:Fig_nondisk_scatter_BHB_KG_HS_clr} for details and source references
for the samples). The top left panel shows, for comparison, the tracer
densities obtained in some earlier studies
\citep{Battaglia_etal_2005_2006,Xue_etal_2008,Brown_etal_2010} which
used different tracer samples. The other three panels show the best
power law fits to the radial profiles of $\ntracer$ for the three
non-disk samples. The GC set used is  $\GCset=[8.3,244]$.
}
\label{Fig:Fig_ntr_8.3_a_compare_bcd_powfits_clr}
\end{figure*}

The resulting density estimates for the three 
samples mentioned above with the GCs set $\GCset=[8.3,244]$ are shown in 
Figure \ref{Fig:Fig_ntr_8.3_a_compare_bcd_powfits_clr}, where we also show 
for comparison (see the top left panel of Figure 
\ref{Fig:Fig_ntr_8.3_a_compare_bcd_powfits_clr}) the tracer densities from 
some earlier studies that used different tracer samples. Our results are 
seen to be in reasonably good agreement with those obtained in the 
previous studies. 

We then perform power-law fits ($\ntracer (r) \propto r^{-\gamma}$) to 
the radial profile of the tracer number density for each of the three 
samples separately. The resulting best power-law fits are also shown in 
Figure \ref{Fig:Fig_ntr_8.3_a_compare_bcd_powfits_clr}. The values of the 
parameters of the best power-law fit for each tracer sample are given 
in Table \ref{Table:nd_n_sigma_fit_params_bf}. Within 
each sample, there is no significant difference in the values of 
$\ntracer$ for the three different sets of GCs, as also seen from the 
values of the power-law fit parameters given in Table 
\ref{Table:nd_n_sigma_fit_params_bf}.

Next, we have to calculate the galactocentric radial velocity 
dispersion, $\sigmar$, that appears in the Jeans equation 
(\ref{eq:Jeans_eqn}), for our non-disk samples. To do this we first 
transform the observed heliocentric \los\ velocity, $\vh$, of each 
individual tracer object to $\vgsr$, the velocity that would be measured 
in the Galactic Standard of Rest (GSR) frame. This is 
easily done by correcting for the circular motion of the LSR 
$(\vsun)$ and 
solar peculiar motion with respect to LSR, $(U_\odot, V_\odot, W_\odot)$ 
(see Figure \ref{Fig:Fig_coordinates_a_axes_b_TPM_clr}): 
%
\beqa
\vgsr & = & \vh+ U_\odot\cos b \cos l + V_\odot \cos b \sin l 
\nonumber \\
 & \, \, & + W_\odot 
\sin b +\vsun~\cos b ~\sin l\,.
\label{eq:vh2vgsr_nd}
\eeqa
%

For large samples like the BHB and KG stars described above, we 
calculate the $\vgsr$ for all the individual tracers in the same radial 
bins as used in the estimation of the tracers' number density described 
above, calculate their dispersion, $\sigmagsr$, and assign it to the 
mean radius of all the tracers contained within that bin. The 
corresponding uncertainty, $\Delta\sigmagsr$, in our estimate of  
$\sigmagsr$ in each bin is calculated by using the standard formula  
$\Delta\sigmagsr = \sqrt{1/[2(N-1)]}\sigmagsr$ 
\citep{LC_sigma_error_1998,EHP_sigma_error_1993,GKP_sigma_error_1994}, 
where $N$ is the number of objects in the bin. 

For the Hg sample, however, owing to its small size, we follow a 
different method, similar to that used in 
\citet{Battaglia_etal_2005_2006}, for calculating the $\sigmagsr$ and 
its uncertainty in each radial bin: we 
randomly generate a sample of 10,000 mock values of $\vh$ for each 
tracer object in a radial bin using a Gaussian centered at the observed 
value of $\vh$ and a width of typically $\sim (10 - 20)\%$ of this $\vh$ 
value. We then transform these 10,000 $\vh$ values for 
each tracer in the bin to get the corresponding 10,000 values of $\vgsr$ 
using equation (\ref{eq:vh2vgsr_nd}), and calculate the associated 
dispersions $\sigmagsr$ in that bin. We assign the mean 
value of the $\sigmagsr$ values for all the 
objects in a given bin to the mean radius of all the objects 
in the bin. The corresponding uncertainty in $\sigmagsr$ is taken 
to be the r.m.s. deviation of the $\sigmagsr$ values in that bin.  

\begin{figure*}[!htb]
\begin{tabular}{cc}
\hskip 1cm
\epsfig{file=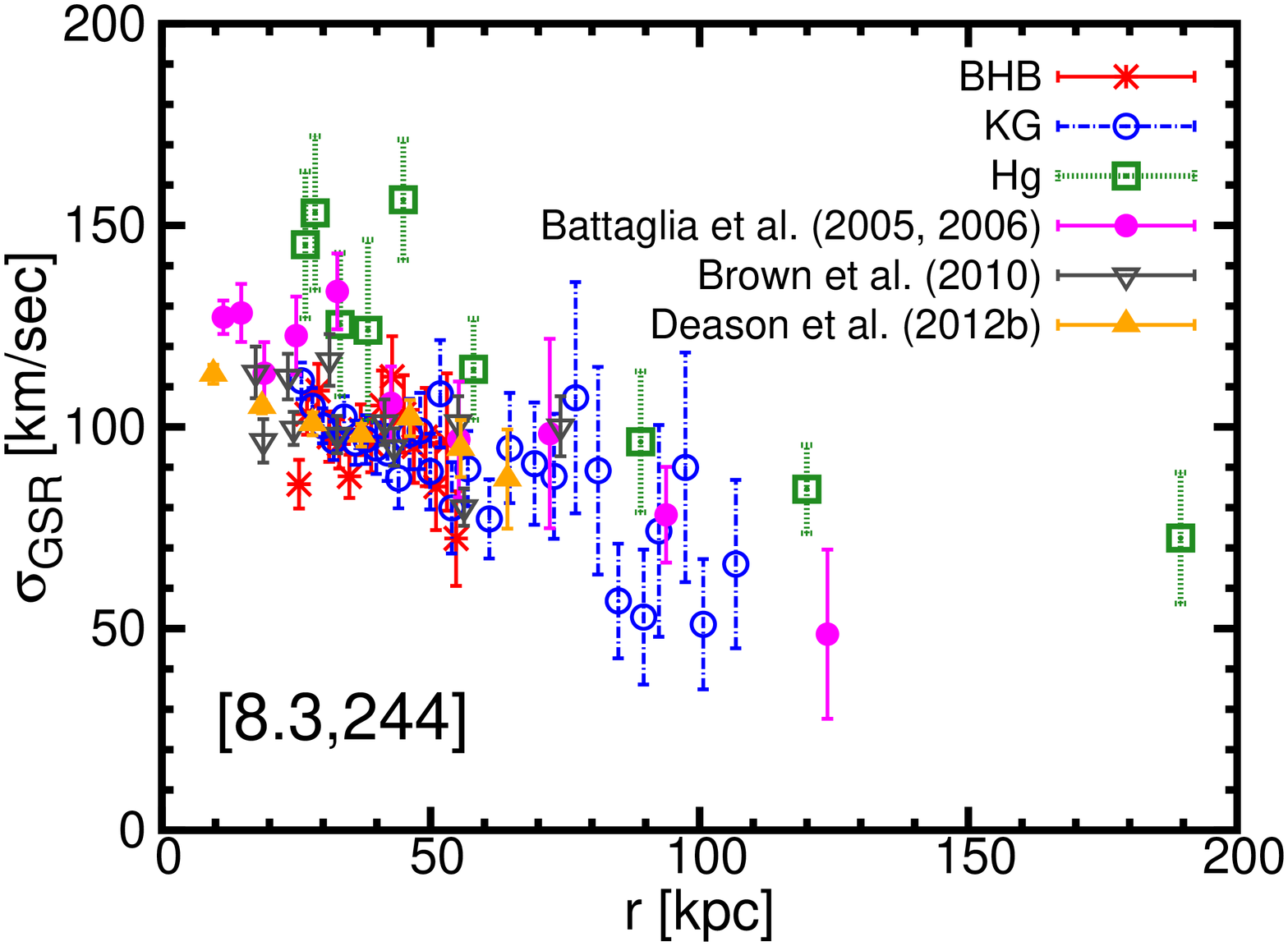,angle=270,width=0.9\columnwidth}&
\epsfig{file=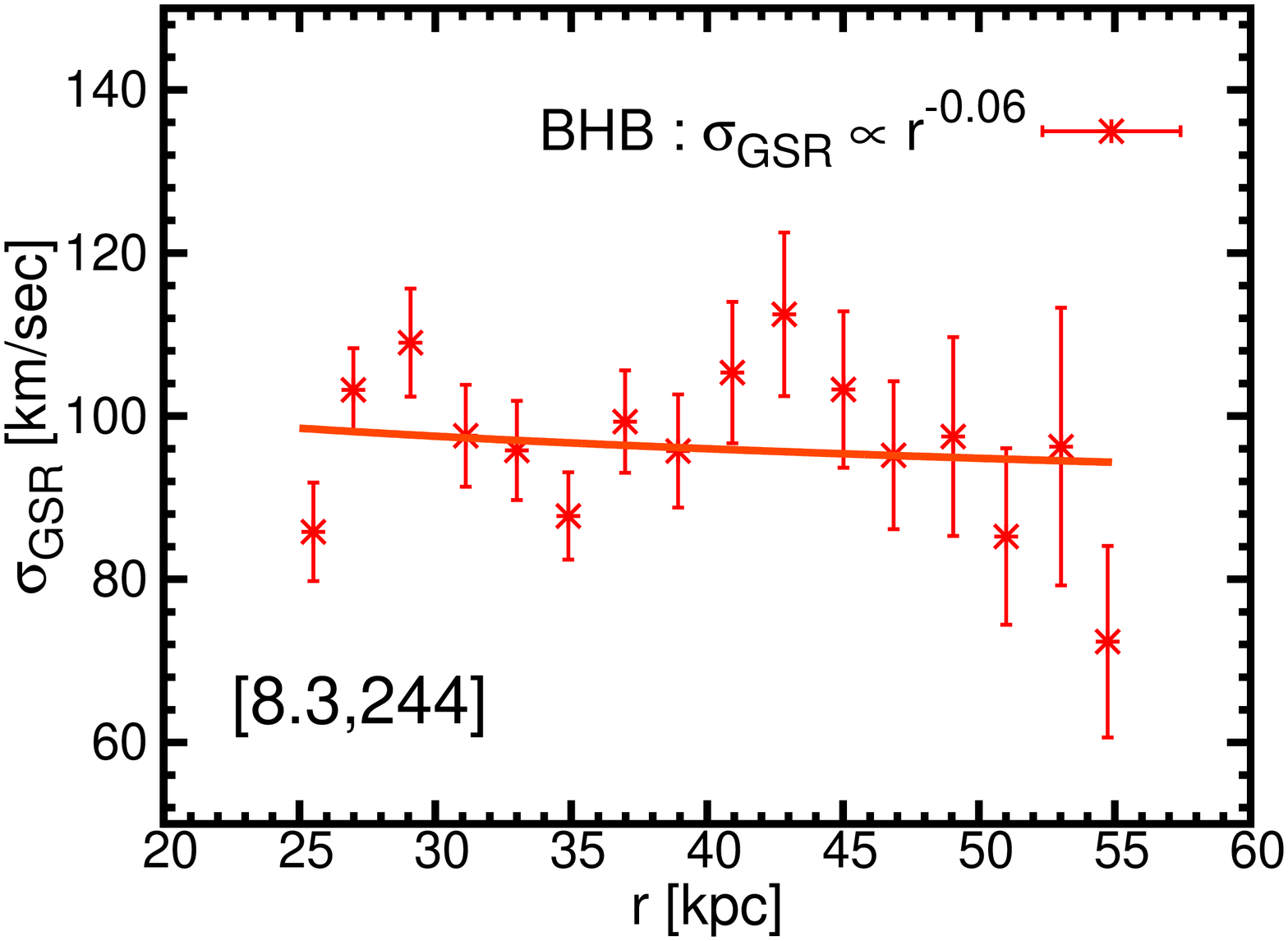,angle=270,width=0.9\columnwidth}\\
\hskip 1cm 
\epsfig{file=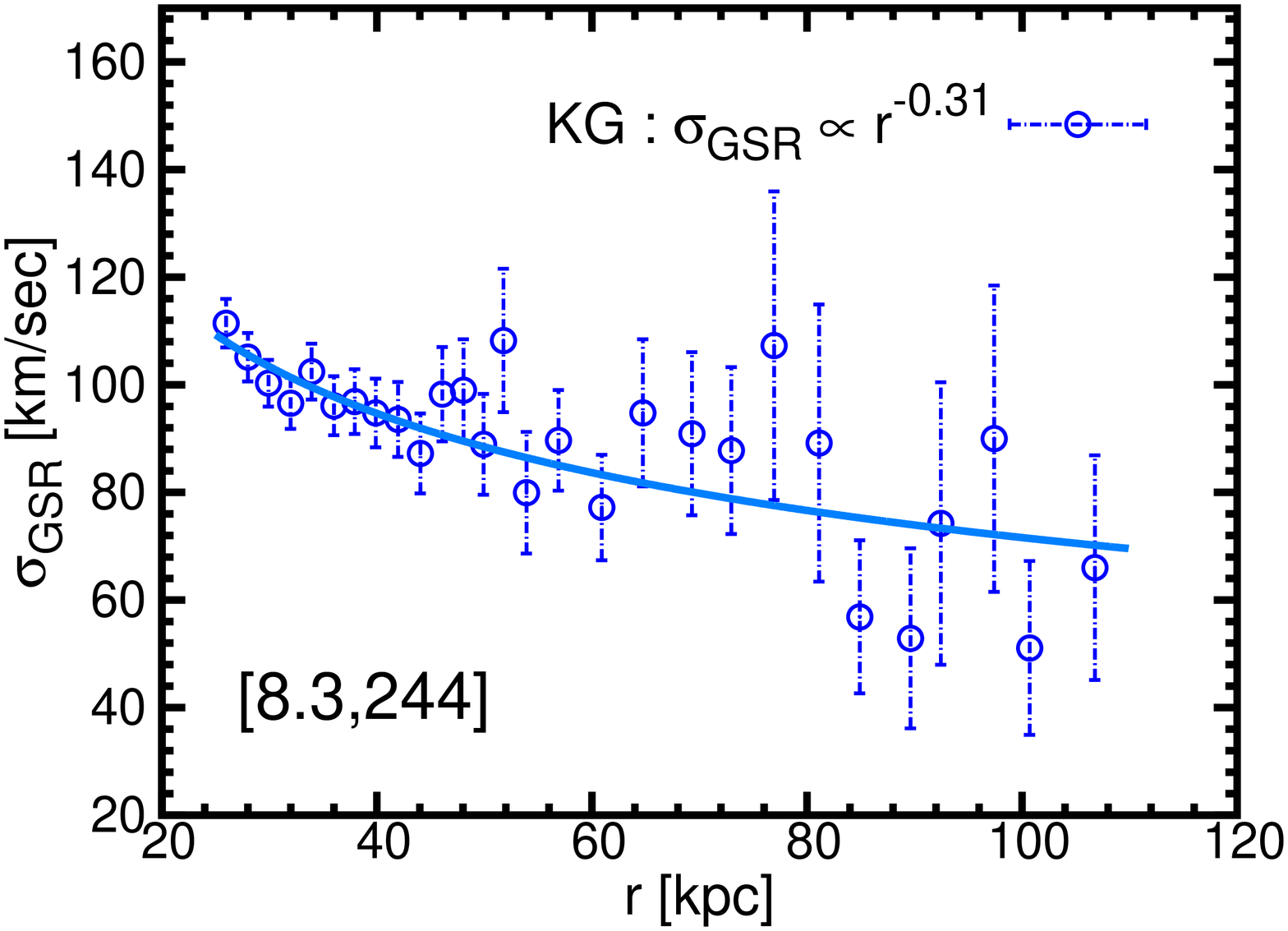,angle=270,width=0.9\columnwidth}&
\epsfig{file=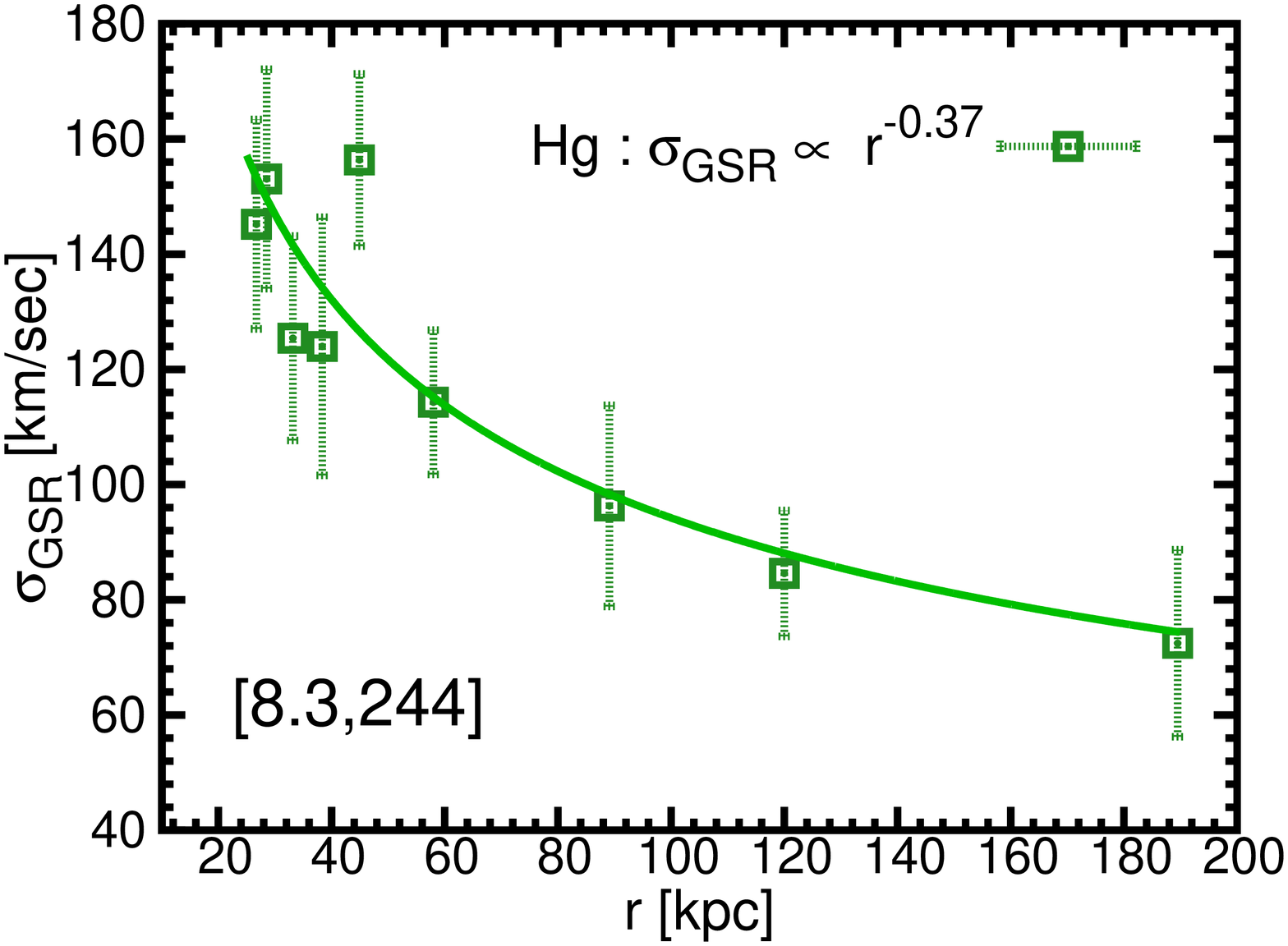,angle=270,width=0.9\columnwidth}
\end{tabular}
\caption{The GSR frame \los\ velocity dispersion of the tracers,
$\sigmagsr$, for the three non-disk tracer samples considered in
this paper (see text and Figure \ref{Fig:Fig_nondisk_scatter_BHB_KG_HS_clr} for
details and source references for the samples). The top left panel also
shows, for comparison, the $\sigmagsr$
obtained in some earlier studies
\citep{Battaglia_etal_2005_2006,Brown_etal_2010,Deason_etal_coldveil_2012b} 
which used different tracer samples. The other three panels show the 
best power-law fits to the radial profiles of $\sigmagsr$ for the three
non-disk samples. The GC set used is $\GCset=[8.3,244]$.
}
\label{Fig:Fig_sigmatr_8.3_a_compare_bcd_powfits_clr}
\end{figure*}

Our results for $\sigmagsr$ for the three tracer samples are shown in 
Figure~\ref{Fig:Fig_sigmatr_8.3_a_compare_bcd_powfits_clr} in which 
we also show for comparison (see the top left panel of Figure 
\ref{Fig:Fig_sigmatr_8.3_a_compare_bcd_powfits_clr}) the $\sigmagsr$ values 
obtained in some earlier studies using different samples, which, again, 
are seen to be in reasonably good agreement with our results. 
%

%

The other three panels of 
Figure~\ref{Fig:Fig_sigmatr_8.3_a_compare_bcd_powfits_clr} 
show the best power-law fits ($\sigmagsr (r) \propto r^{-\alpha}$) 
to the radial profiles of $\sigmagsr$ for each of the three non-disk 
samples. The values of the parameters of the best power-law fits for the 
three tracer samples are given in Table 
\ref{Table:nd_n_sigma_fit_params_bf}. 
Again, as in the case of $\ntracer$, the effect of variation of the 
Galactic Constants on $\sigmagsr$ is negligible.

Finally, the galactocentric radial velocity dispersion, $\sigmar$, can 
be obtained from $\sigmagsr$ by using the relation 
\citep{Battaglia_etal_2005_2006}
\begin{equation}
\sigmar=\frac{\sigmagsr}{\sqrt{1-\beta H(r)}}\,,
\label{eq:sigmagsr2sigmar}
\end{equation}
where 
\begin{equation}
H(r) = \frac{r^2+\rsun^2}{4r^2} 
-\frac{\left(r^2-\rsun^2\right)^2}{8r^3\rsun} 
\ln\frac{r+\rsun}{r-\rsun}\,,  
\quad \quad (r>\rsun)
\label{eq:H_r}
\end{equation}
and $\beta$ is the velocity anisotropy of the tracers defined in 
equation (\ref{eq:beta_anisotropy_parameter}). Equation 
(\ref{eq:sigmagsr2sigmar}) 
is derived by decomposing the $\vgsr$'s into their galactocentric 
radial and transverse components and taking the averages of the squares 
of the velocity components.\footnote{Note that equation (3) given in 
the 2005 paper of \citet{Battaglia_etal_2005_2006} is incorrect. The 
correct equation, same as equation (\ref{eq:sigmagsr2sigmar}) above, is 
given in the 2006 (Erratum) paper of \citet{Battaglia_etal_2005_2006} 
and also in \citet{Dehnen_etal_2006}.}

The last quantity that remains to be specified before we can solve the 
Jeans equation (\ref{eq:Jeans_eqn}) is the velocity anisotropy 
parameter, $\beta$, of the tracers. There is not much definite 
observational information available on the value of $\beta$ of the 
tracers because of the lack of availability of proper motion 
measurements on sufficiently large number of tracer objects. In 
general $\beta$ can be a function of $r$. A recent maximum 
likelihood analysis \citep{Deason_etal_2012a} of radial velocity data of 
a large sample of halo stars, performed within the context of a model 
for the (in general anisotropic) velocity distribution function of the 
halo stars, indicates the stellar velocity anisotropy being radially biased 
with a value of $\beta\sim 0.5$ for $r$ from $\sim$ 16 kpc up to $r\sim 
48 \kpc$. This is also indicated by the recent results 
from the large numerical simulation study of 
\citet{Rashkov_etal_eris_2013}, 
which finds the velocity distribution of the Galaxy's stellar 
population at large $r$ to be radially biased ($\beta > 0$) with 
stellar orbits tending to purely radial ($\beta \to 1$) at $r\gsim 
100\kpc$. Based on these considerations, to explore various 
possibilities for $\beta$, in this paper we shall 
calculate our RCs for (a) three representative constant  
values of $\beta$, namely, $\beta=0$ (isotropic), 0.5 (mildly radially 
biased anisotropy), and 1 (fully radially anisotropic), (b) a radially 
varying $\beta$ of the OM form
\citep[see][p.297-298]{Binney_Tremaine_2008} given by
$\beta(r)=(1+r_a^2/r^2)^{-1}$, $r_a$ being the ``anisotropy radius",  
and (c) a radial profile of $\beta$
obtained from the recent large high resolution
hydrodynamical simulations done by \citet{Rashkov_etal_eris_2013}. In 
principle, $\beta$ and its radial profile may be different for 
different tracer samples. But since 
currently no reliable measurements of 
$\beta$ for the different samples extending to large galactocentric 
distances are available, any choice of different $\beta$ for different 
samples would be necessarily arbitrary. For simplicity, therefore, we 
assume the same values of $\beta$ and its radial profile for our 
three tracer samples. 

\begin{table*}[!hbt]
\begin{center}
\setlength\extrarowheight{3pt}
\caption{Best-fit parameter values for power-law fits to the  
radial profiles of the number density, $\ntracer$, and the  
Galactic Standard of Rest (GSR) frame \los\ velocity 
dispersion, $\sigmagsr$, of the tracers for the three non-disk tracer 
samples considered in this paper (see text and Figure 
\ref{Fig:Fig_nondisk_scatter_BHB_KG_HS_clr} for details and source 
references for the samples). The parameter values are given for  
three different sets of values of the GCs, $\GCset$. 
}
\label{Table:nd_n_sigma_fit_params_bf}
\vspace{0.5cm}
\begin{tabular}{||c|c|c|c|c||}
\hline\hline
 & \multicolumn{4}{c| |}{Number densities and radial velocity 
dispersions}\\
& \multicolumn{4}{c| |}{of non-disk tracers}\\ 
\multicolumn{1}{||c|}{$\GCset$} & \multicolumn{4}{c| |}{$\ntracer=n_0\,  
(\frac{r}{50\kpc})^{-\gamma}\,, \,\, \sigmagsr=\sigma_0\, 
(\frac{r}{50\kpc})^{-\alpha}$}\\
\cline{2-5}
\cline{2-5}
 & $\frac{n_0}{\kpc^3}$ & $\gamma$ & 
$\sigmazerokmps$ & $\alpha$ \\
 \cline{2-5}
& \multicolumn{4}{c||}{BHB} \\
\hline\hline
$[8.3,244]$ & $7.51\times 10^{-4}$ & $4.16$ & $93.0$ & $0.06$ \\
$[8.5,220]$ & $7.66\times 10^{-4}$ & $4.15$ & $94.45$ & $0.07$ \\
$[8.0,200]$ & $7.45\times 10^{-4}$ & $4.17$ & $93.58$ & $0.05$ \\
\hline\hline
\cline{2-5}
 & \multicolumn{4}{c|}{KG} \\
\cline{2-5}
\hline\hline
$[8.3,244]$ & $6.57\times 10^{-4}$ & $5.51$ & $86.75$ & $0.31$ \\
$[8.5,220]$ & $6.53\times 10^{-4}$ & $5.51$ & $88.23$ & $0.30$ \\
$[8.0,200]$ & $6.40\times 10^{-4}$ & $5.51$ & $87.89$ & $0.29$ \\
\hline\hline
\cline{2-5}
 & \multicolumn{4}{c|}{Hg} \\
\cline{2-5}
\hline\hline
$[8.3,244]$ & $2.37\times 10^{-5}$ & $4.18$ & $121.21$ & $0.37$ \\
$[8.5,220]$ & $2.39\times 
10^{-5}$ & $4.18$ & 
$117.51$ & $0.40$ \\
$[8.0,200]$ & $2.38\times 10^{-5}$ & $4.17$ & $115.34$ & $0.42$ \\
\hline\hline
\end{tabular}
\end{center}
\end{table*}

With $\ntracer$, $\sigmar$ and $\beta$ thus specified, we can now 
proceed to solve the Jeans equation (\ref{eq:Jeans_eqn}) to obtain 
the $\vc$ 
profiles for the three different tracer samples described above. For 
each tracer sample we calculate the $\vc$'s in the same radial bins as 
used in calculating the $\ntracer$'s and $\sigmagsr$'s, and the best-fit 
power-law forms of $\ntracer$ and $\sigmagsr$ described above are used 
for calculating the radial derivatives appearing in the Jeans equation 
(\ref{eq:Jeans_eqn}). The corresponding $1\sigma$ error, $\Delta \vc$, 
on $\vc$ within each radial bin is calculated from those of $\ntracer$ 
and $\sigmagsr$ in the bin by standard quadrature.  
\newpage
\begin{figure*}[!htb]
\begin{tabular}{cc}
\epsfig{file=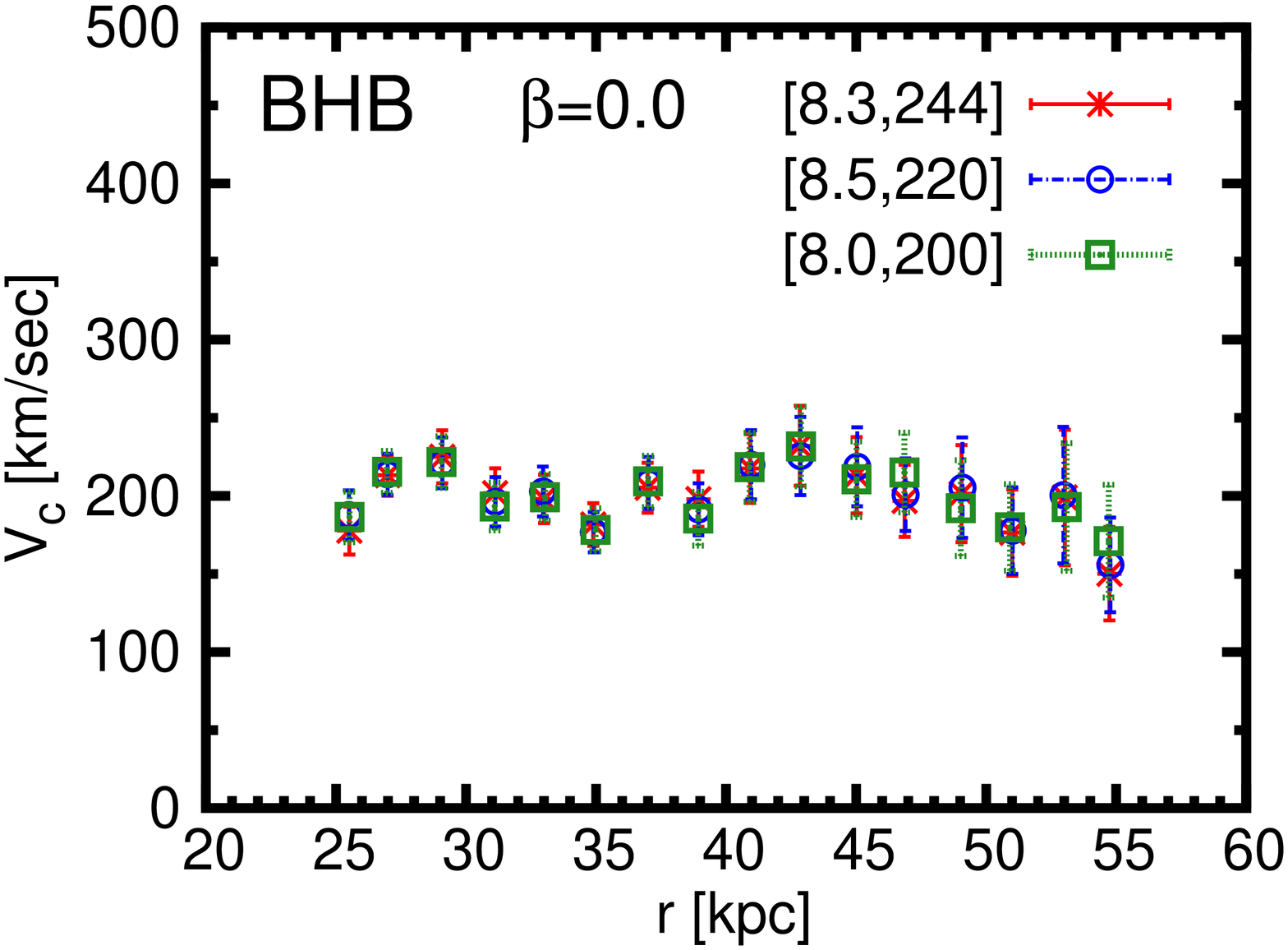,angle=270,width=\columnwidth}&
\epsfig{file=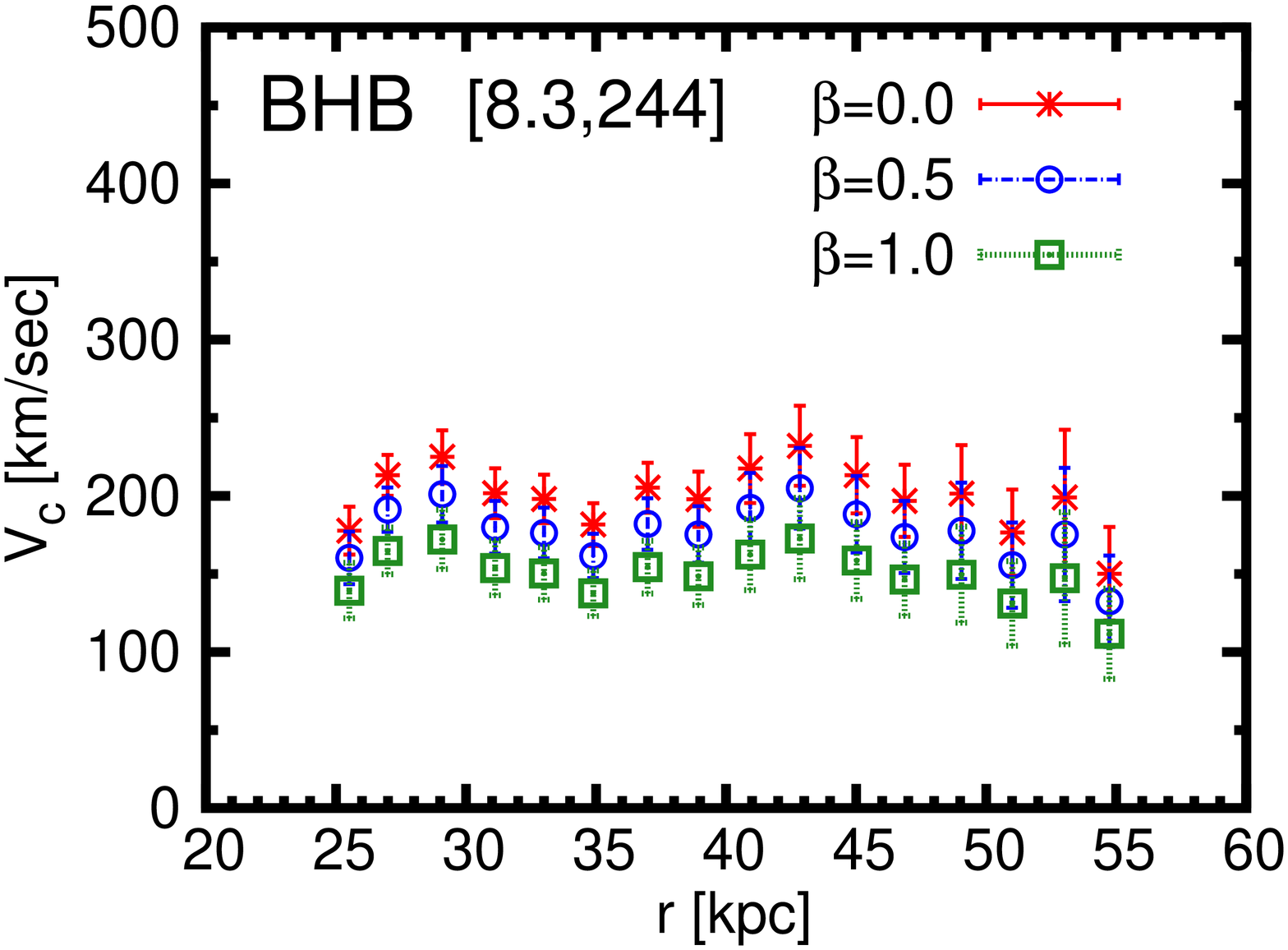,angle=270,width=\columnwidth}\\
\epsfig{file=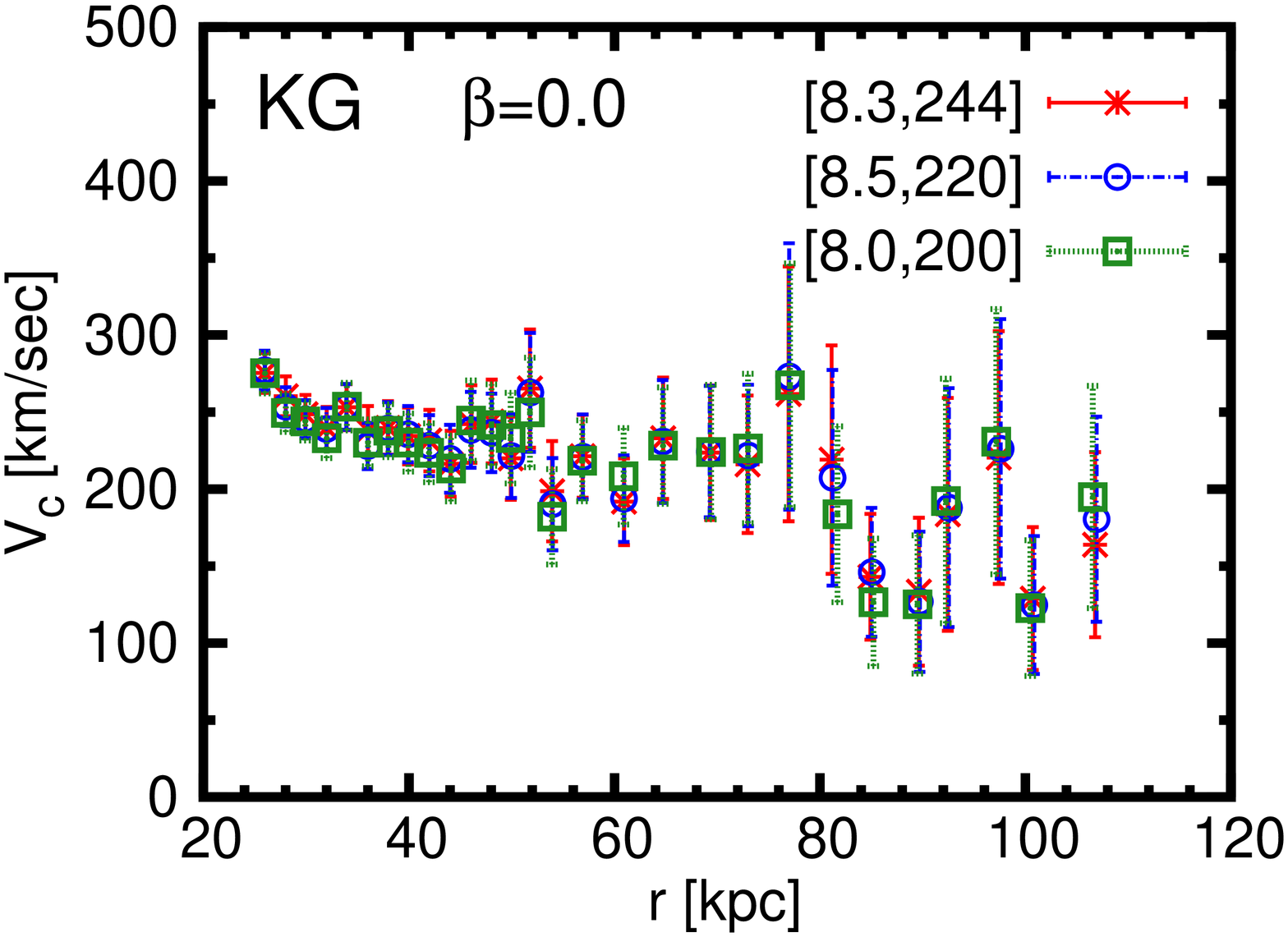,angle=270,width=\columnwidth}&
\epsfig{file=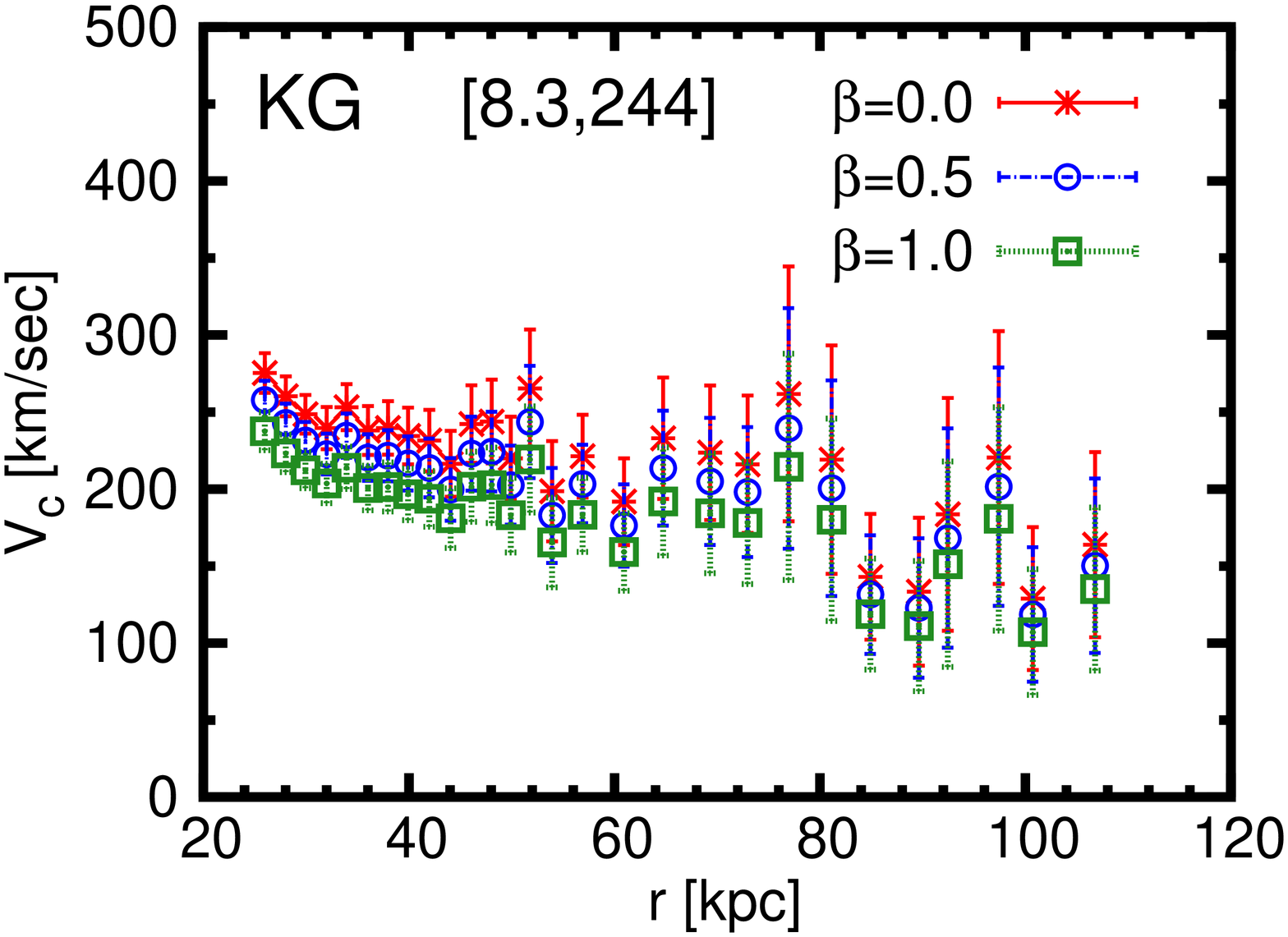,angle=270,width=\columnwidth}\\
\epsfig{file=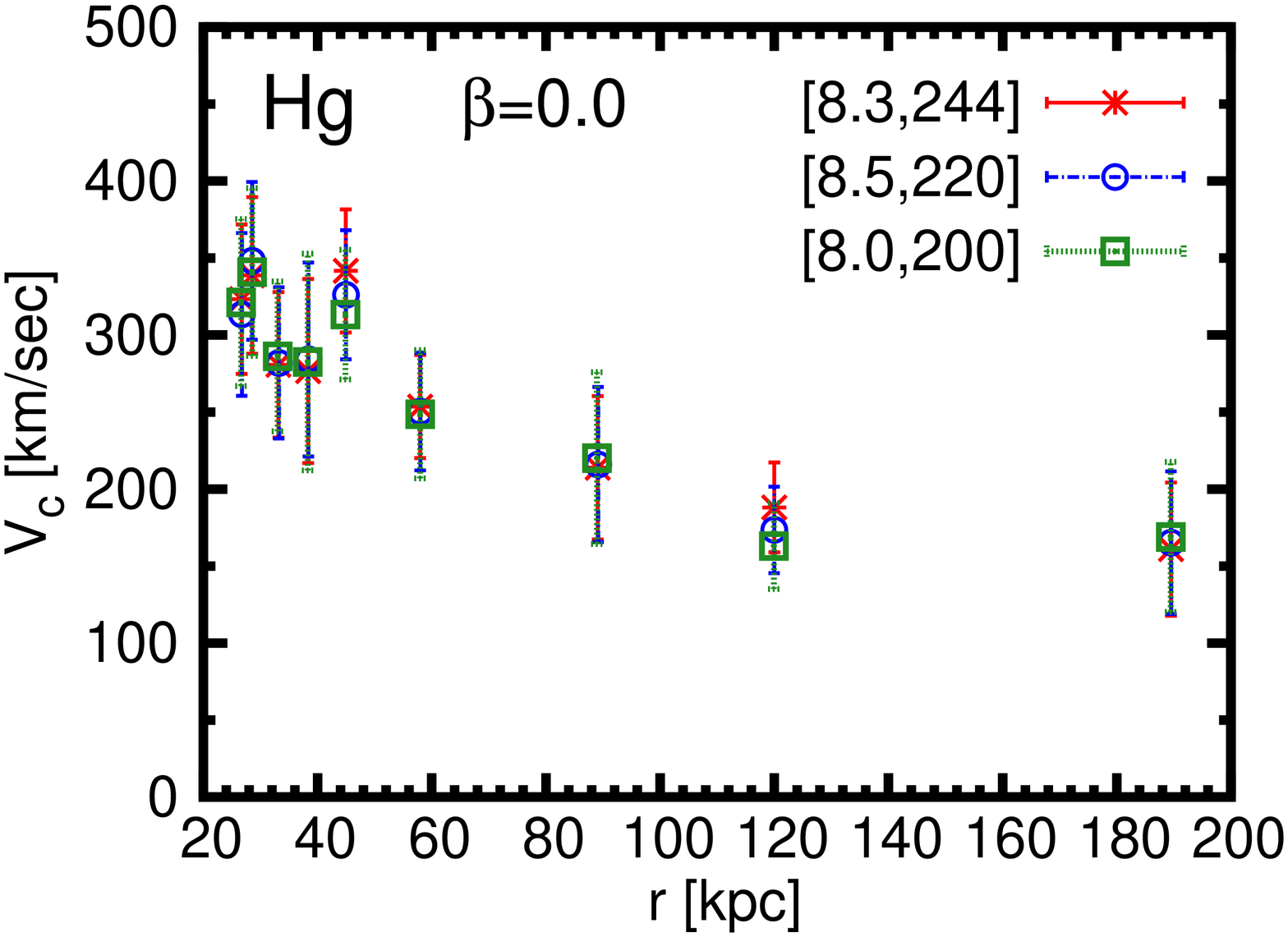,angle=270,width=\columnwidth}&
\epsfig{file=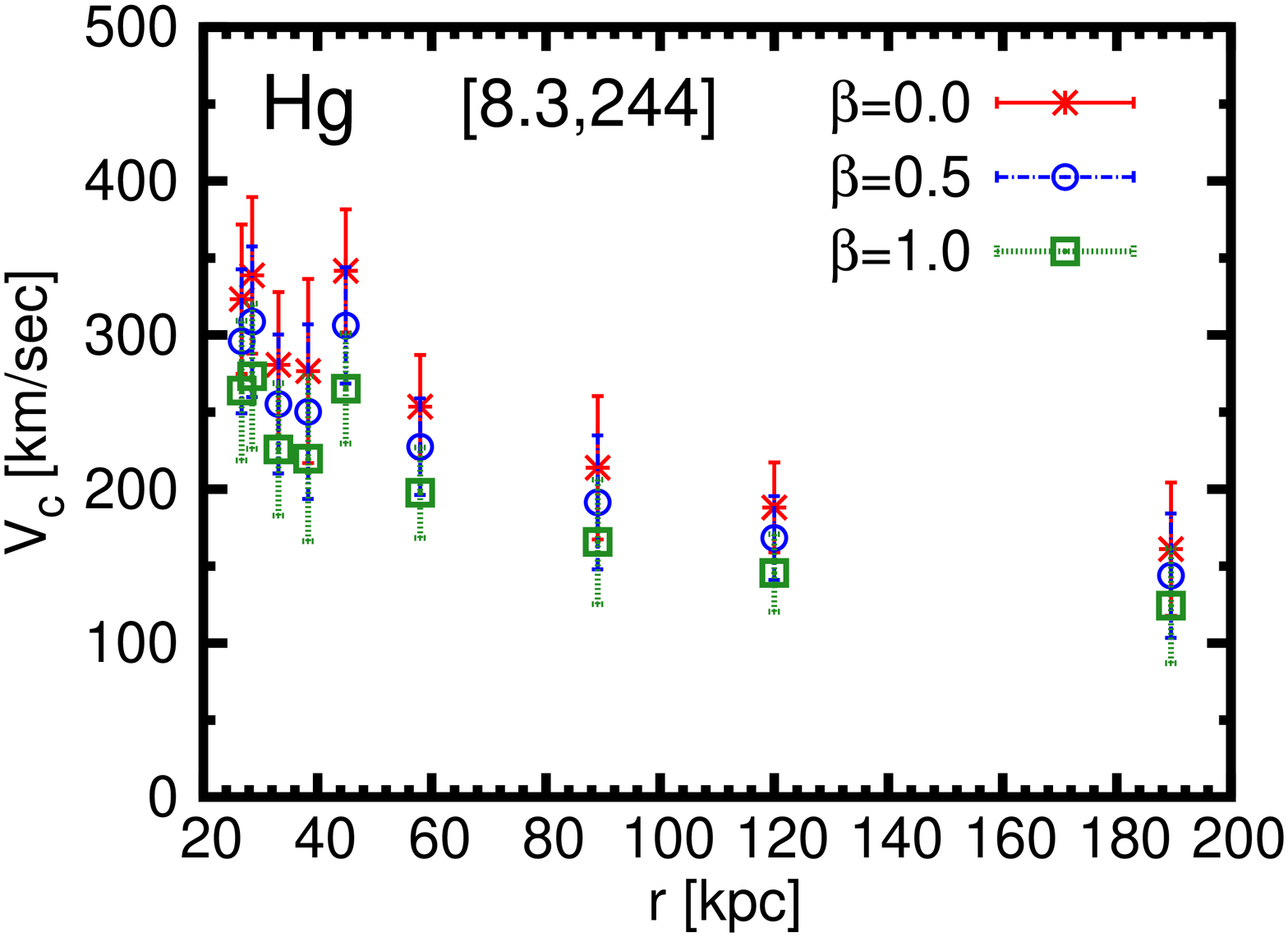,angle=270,width=\columnwidth}
\end{tabular}
\caption{Circular velocities with their $1\sigma$ error bars for the 
three different non-disk tracer samples used in this paper (see text and 
Figure \ref{Fig:Fig_nondisk_scatter_BHB_KG_HS_clr} for details and source 
references for the samples). The left panels are for tracer velocity 
anisotropy $\beta=0$ and three different sets of values of the Galactic 
constants, $\GCset$, as indicated, whereas the right panels show the 
results for three different constant ($r$-independent) values of $\beta$ 
= 0, 0.5 and 1, with $\GCset = [8.3,244]$.  
}
\label{Fig:Fig_vc_nd_diff_samples_leftcol_dgc_beta0_rightcol_8.3_dbeta_clr}
\end{figure*}

\newpage
The resulting RCs for the three tracer samples are shown in Figure 
\ref{Fig:Fig_vc_nd_diff_samples_leftcol_dgc_beta0_rightcol_8.3_dbeta_clr}. 
As clear from the left panels of Figure 
\ref{Fig:Fig_vc_nd_diff_samples_leftcol_dgc_beta0_rightcol_8.3_dbeta_clr} 
the RCs for  
different choices of GCs almost overlap, thus indicating that the  
RC at large galactocentric distances beyond a few tens of 
kpc is fairly insensitive to the precise values of the GCs. 
Instead, the main uncertainty in the RC comes from the unknown value of 
the tracers' velocity anisotropy parameter $\beta$, as evident from the 
right panels of Figure
\ref{Fig:Fig_vc_nd_diff_samples_leftcol_dgc_beta0_rightcol_8.3_dbeta_clr}. 
As expected, 
the lowest rotation speeds obtain for the most radially biased velocity 
anisotropy ($\beta=1$).
%
\section{Combined rotation curves to $r\sim200\kpc$}
\label{sec:combined_vc}
%
\begin{figure*}[!htb]
\begin{tabular}{cc}
\epsfig{file=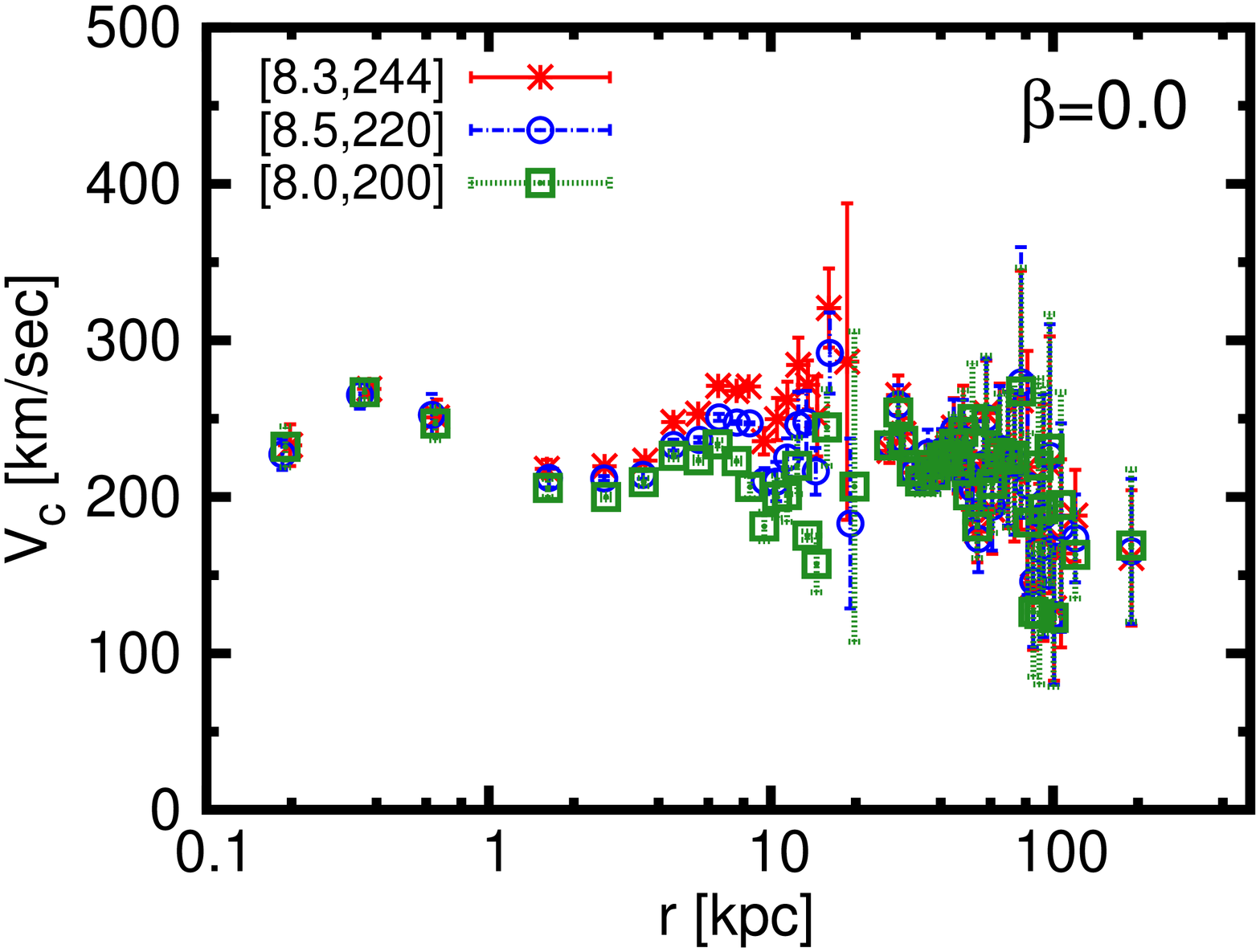,angle=270,width=\columnwidth} 
&
\epsfig{file=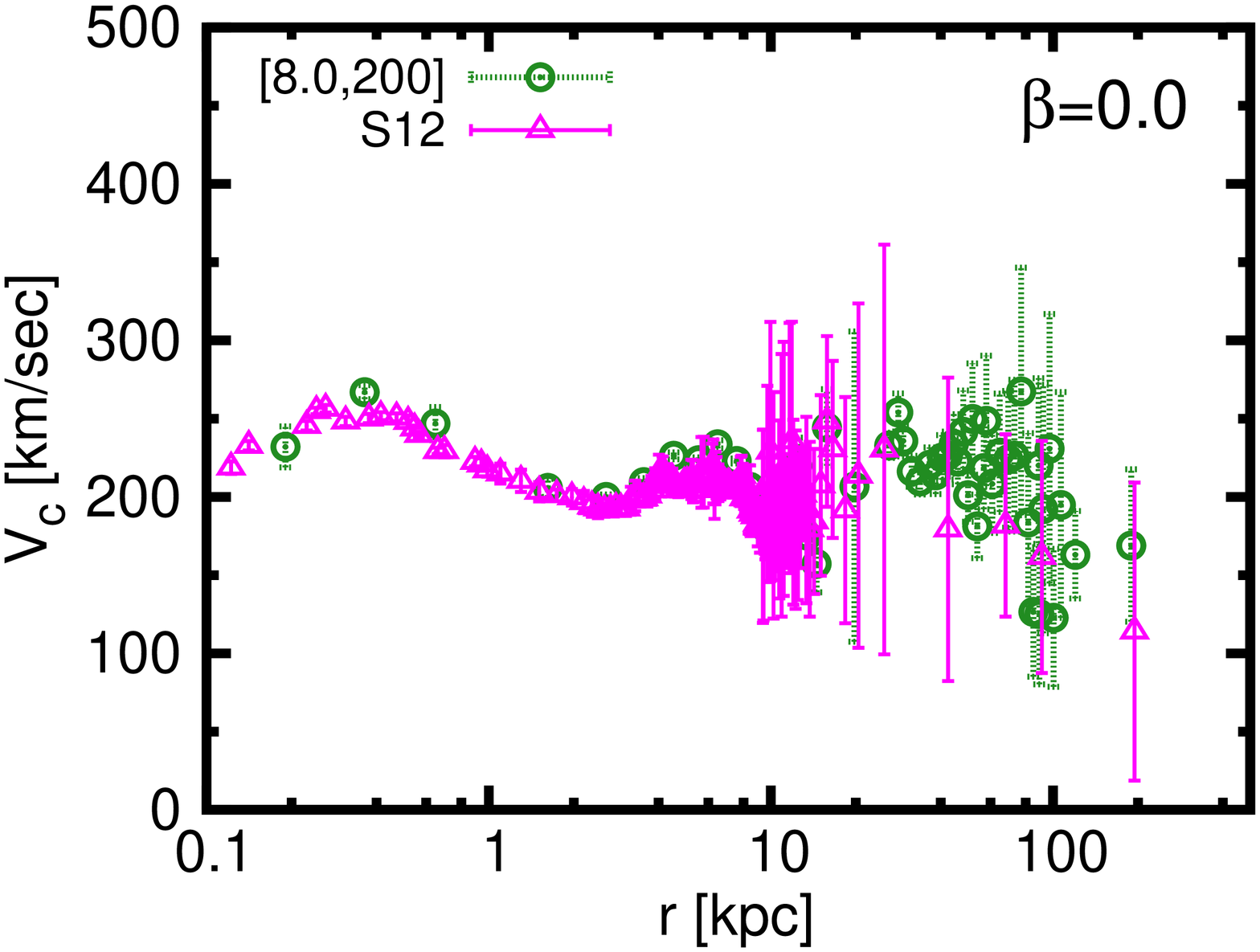,angle=270,width=\columnwidth}
\end{tabular}
\caption{Left: Rotation curve of the Galaxy for three different sets of 
values of the Galactic constants $\GCset$ as indicated and non-disk 
tracers' velocity anisotropy parameter $\beta=0$. The data 
points and their $1\sigma$ error bars shown here are obtained by 
weighted averaging over the combined $\vc$ data obtained from different 
disk and non-disk tracer samples (see Figures 
\ref{Fig:Fig_vc_disk_a_samples_8.3_b_coll_dgc_clr} and 
\ref{Fig:Fig_vc_nd_diff_samples_leftcol_dgc_beta0_rightcol_8.3_dbeta_clr}).  Right: Rotation curve of the Galaxy for $\GCset = [8.0,200]$ 
and non-disk 
tracers' velocity anisotropy parameter $\beta=0$ compared with that obtained by \citet{Sofue_GrandRC_2012} (S12). 
}
\label{Fig:Fig_vcfull_samplecollapse_a_dgc_0.0_log_b_8.0_0.0_log_sofue12_clr}
\end{figure*}

\noindent We now combine the rotation curves obtained from disk and 
non-disk tracers (Figures 
\ref{Fig:Fig_vc_disk_a_samples_8.3_b_coll_dgc_clr} 
and \ref{Fig:Fig_vc_nd_diff_samples_leftcol_dgc_beta0_rightcol_8.3_dbeta_clr})  to 
construct the rotation curve of the Galaxy up to $\sim 200  
\kpc$.
\begin{figure*}[!htb]
\begin{tabular}{cc}
\epsfig{file=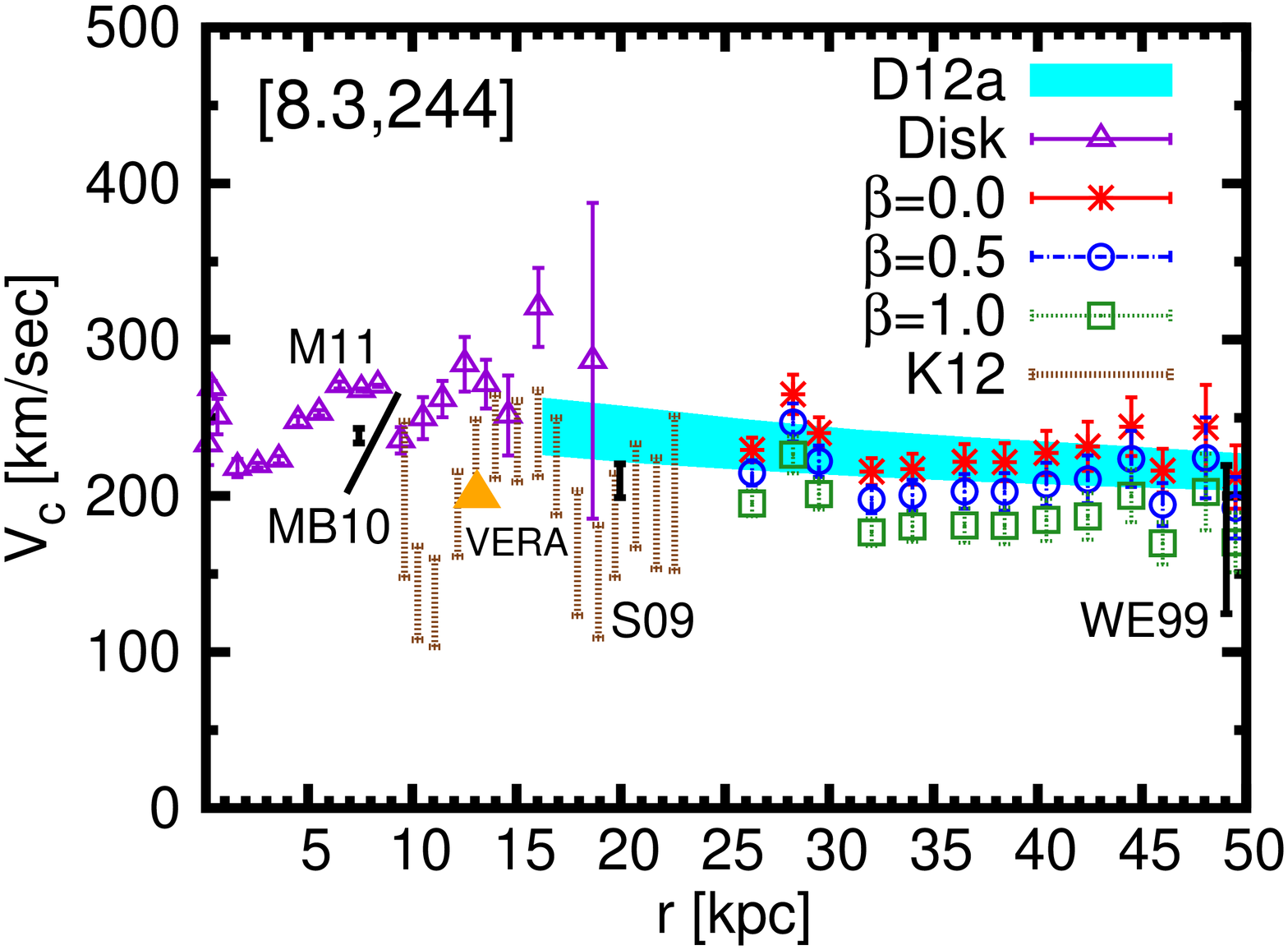,angle=270,width=\columnwidth}
& 
\epsfig{file=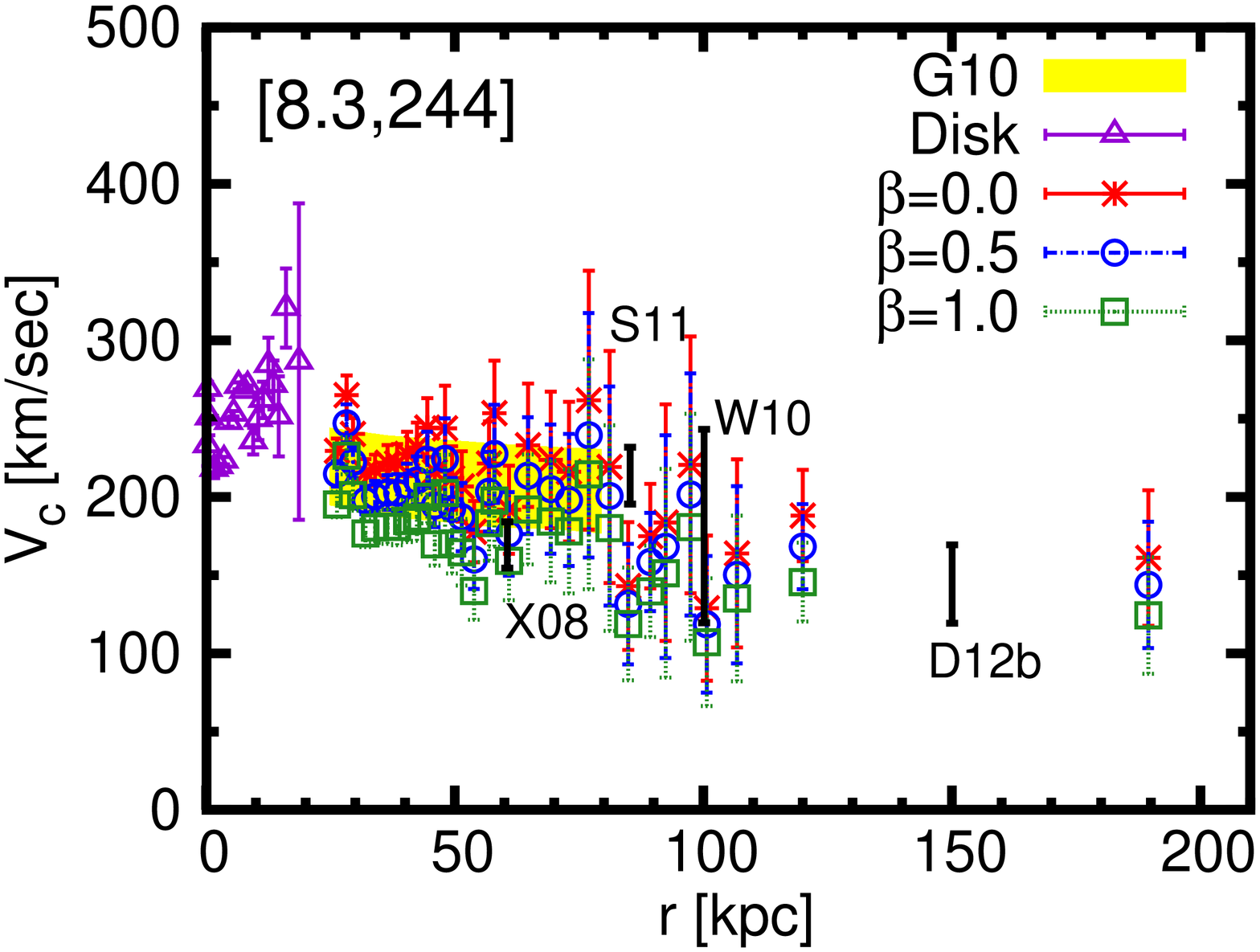,angle=270,width=\columnwidth}
\end{tabular}
\caption{Rotation Curve for $\GCset=[8.3,244]$ and various 
values of $\beta$. The shaded bands marked D12a and G10 in the left and 
right panels, respectively, represent the RCs and their uncertainty 
bands obtained earlier by \citet{Deason_etal_2012a} (D12a) (up to 
$r\sim50\kpc$) and \citet{Gnedin_etal_2010} (G10) (up to 
$r\sim 80\kpc$), respectively. In addition, estimates of  
circular velocities at certain specific 
values of $r$ obtained from various independent considerations by 
\citet{Kafle_etal_2012} (K12),
\citet{McMillan_2011} (M11), \citet{MB_2010} (MB10), \citet{Honma_etal_VERA_2007} (VERA), 
\citet{Sofue_etal_2009} (S09), \citet{WE_1999} (WE99), 
\citet{Xue_etal_2008} (X08), \citet{Samurovich_etal_2011} (S11), 
\citet{Watkins_etal_2010} (W10), and \citet{Deason_etal_coldveil_2012b} 
(D12b) are shown for comparison. 
}
\label{Fig:Fig_vcfull_8.3_dbeta_linear_bm_a_below50kpc_b_above50kpc_clr}
\end{figure*}

For the disk region ($r<25 \kpc$) we take the averaged $\vc$ data 
for a chosen set of GCs from the 
right panel of Figure \ref{Fig:Fig_vc_disk_a_samples_8.3_b_coll_dgc_clr}. 
For the non-disk region ($r\geq 25\kpc$), we combine the $\vc$ data from 
Figure 
\ref{Fig:Fig_vc_nd_diff_samples_leftcol_dgc_beta0_rightcol_8.3_dbeta_clr} 
for the 
three tracer samples in every 2 kpc radial bins and calculate the 
resulting mean circular speed ($\vc$) and its $1 \sigma$ uncertainty 
($\Delta \vc$) within a bin by weighted averaging  
as described in section \ref{sec:disk} [see equation 
(\ref{eq:weighted_av})].  

The resulting rotation curves for $\beta=0$ and three sets of values 
of the GCs are shown in Figure 
\ref{Fig:Fig_vcfull_samplecollapse_a_dgc_0.0_log_b_8.0_0.0_log_sofue12_clr}, 
and 
those for different values of $\beta$, for one particular set of GCs, 
$\GCset=[8.3,244]$, are shown in Figure 
\ref{Fig:Fig_vcfull_8.3_dbeta_linear_bm_a_below50kpc_b_above50kpc_clr}. 
For comparison, 
we also show in Figure
\ref{Fig:Fig_vcfull_8.3_dbeta_linear_bm_a_below50kpc_b_above50kpc_clr}
estimates of circular velocities at specific values of $r$
obtained from a variety of independent considerations in some earlier
studies by various authors. 

\begin{figure*}[!htb]
\begin{center}
\epsfig{file=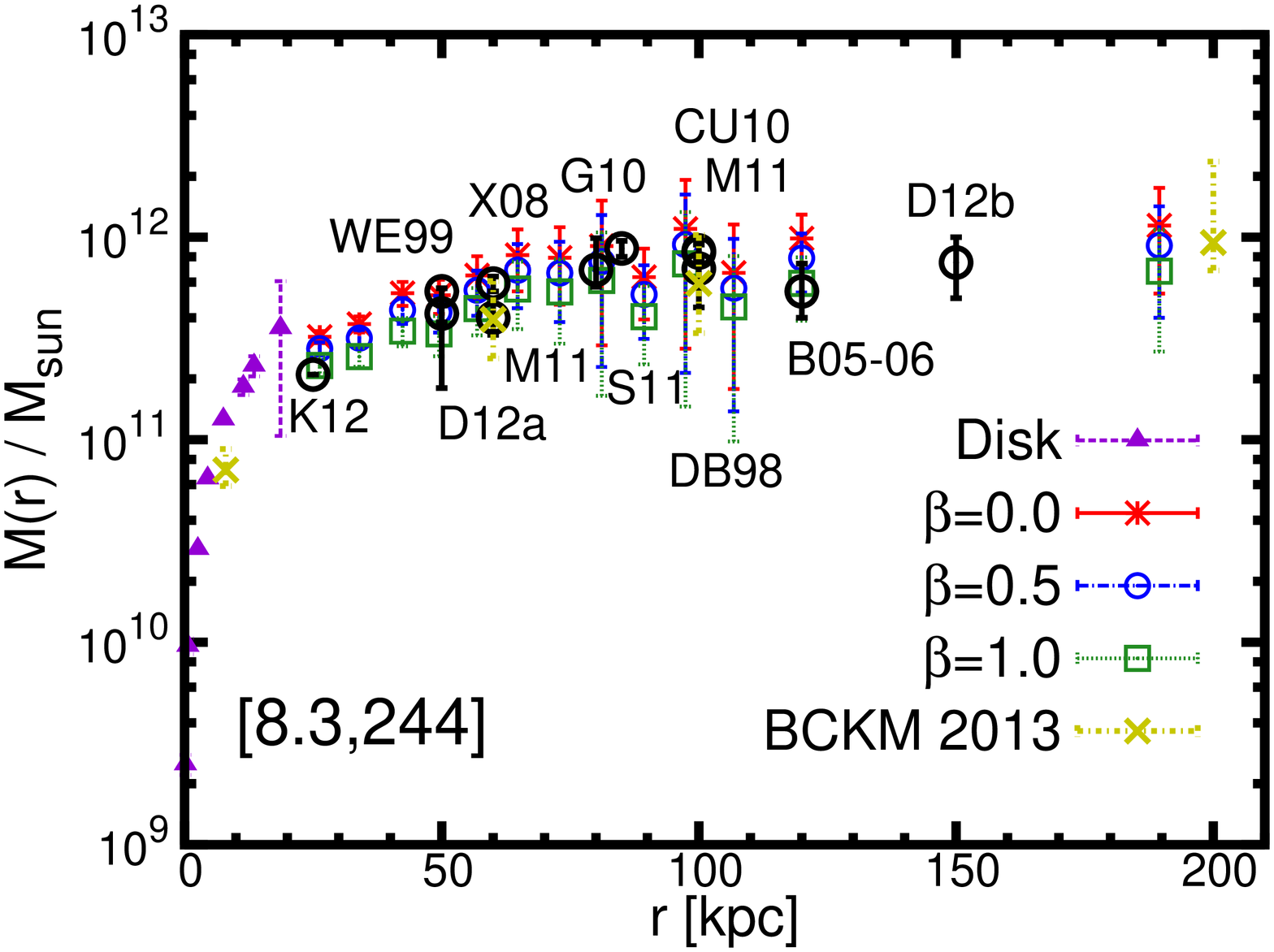,angle=270,width=\columnwidth}
\end{center}
\caption{The mass, $M(r)=r \vc^2(r)/G$, within $r$, as a function of 
$r$, obtained from the RCs shown in Figure 
\ref{Fig:Fig_vcfull_8.3_dbeta_linear_bm_a_below50kpc_b_above50kpc_clr} for 
$\GCset=[8.3,244]$ and various values of the tracers' velocity 
anisotropy parameter $\beta$. Estimates of $M(r)$ at certain 
specific values of $r$ obtained from 
various independent considerations in some earlier works, namely, 
\citet{Kafle_etal_2012} (K12), \citet{WE_1999} (WE99), 
\citet{Deason_etal_2012a} (D12a), \citet{Xue_etal_2008} (X08), 
\citet{McMillan_2011} (M11), \citet{Gnedin_etal_2010} (G10), 
\citet{Samurovich_etal_2011} (S11), \citet{Catena_Ullio_2010} (CU10), \citet{Dehnen_Binney_1998} (DB98), 
\citet{Battaglia_etal_2005_2006} (B05-06), 
\citet{Deason_etal_coldveil_2012b} (D12b), and 
\citet{BCKM_PRD_2013} (BCKM 2013), are shown for comparison. 
}
\label{Fig:Fig_mr_8.3_dbeta_bm_clr}
\end{figure*}

The $\beta$ dependence of the radial profile of the cumulative mass, 
$M(r)=r \vc^2(r)/G$, is shown in Figure 
\ref{Fig:Fig_mr_8.3_dbeta_bm_clr}. Again, estimates of $M(r)$ from 
various independent considerations and given at 
certain specific values of $r$ in some earlier works, are also shown in 
Figure \ref{Fig:Fig_mr_8.3_dbeta_bm_clr} for comparison.  

Note that the lowest mass of the 
Galaxy corresponds to $\beta=1$, which 
allows us to set a lower limit on the mass of the Galaxy, 
$M(\sim200\kpc)\geq (6.8\pm4.1)\times 10^{11}\Msun$. 

\begin{figure*}[!htb]
\begin{tabular}{cc}
\epsfig{file=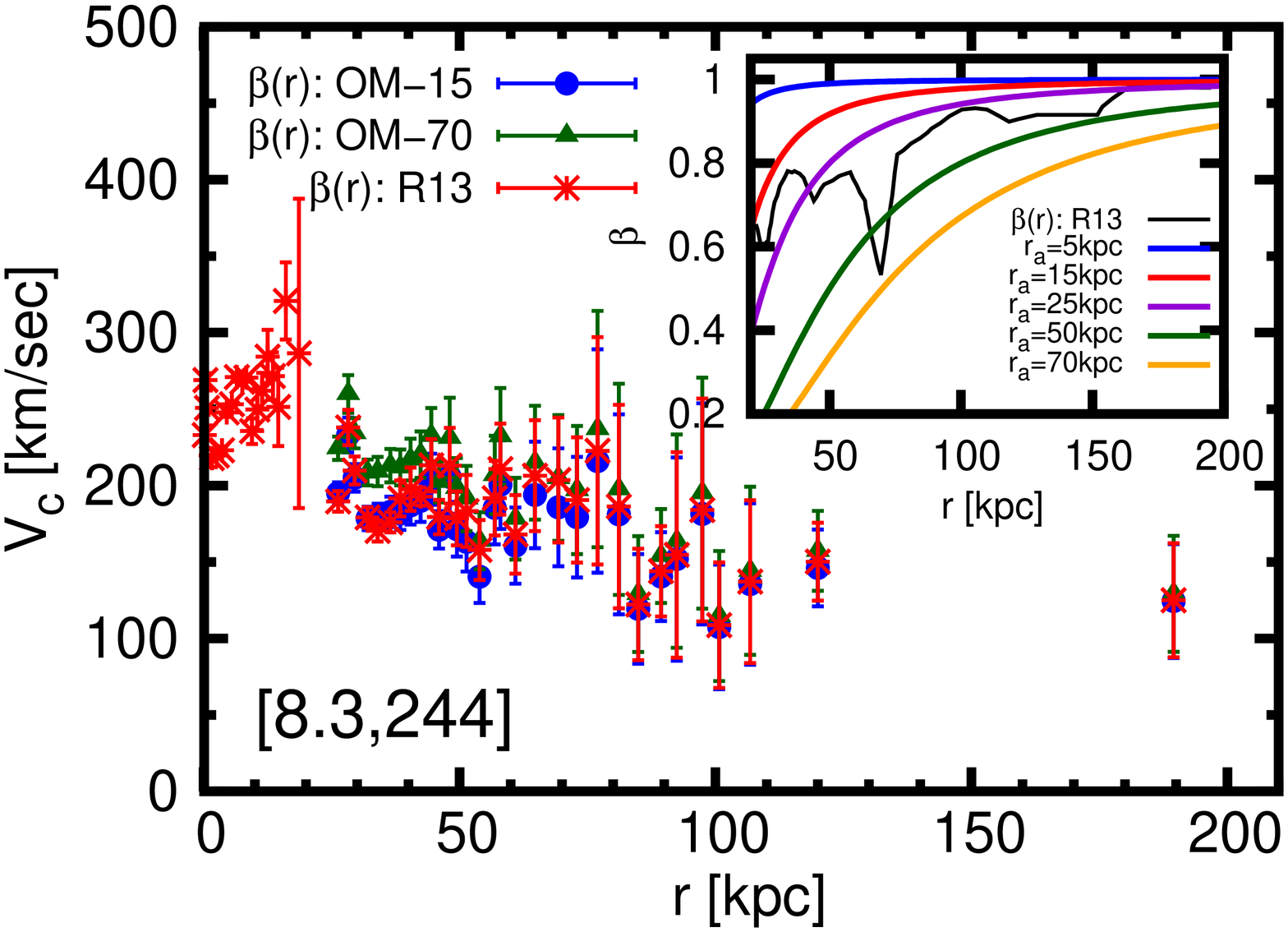,angle=270,width=\columnwidth}
&
\epsfig{file=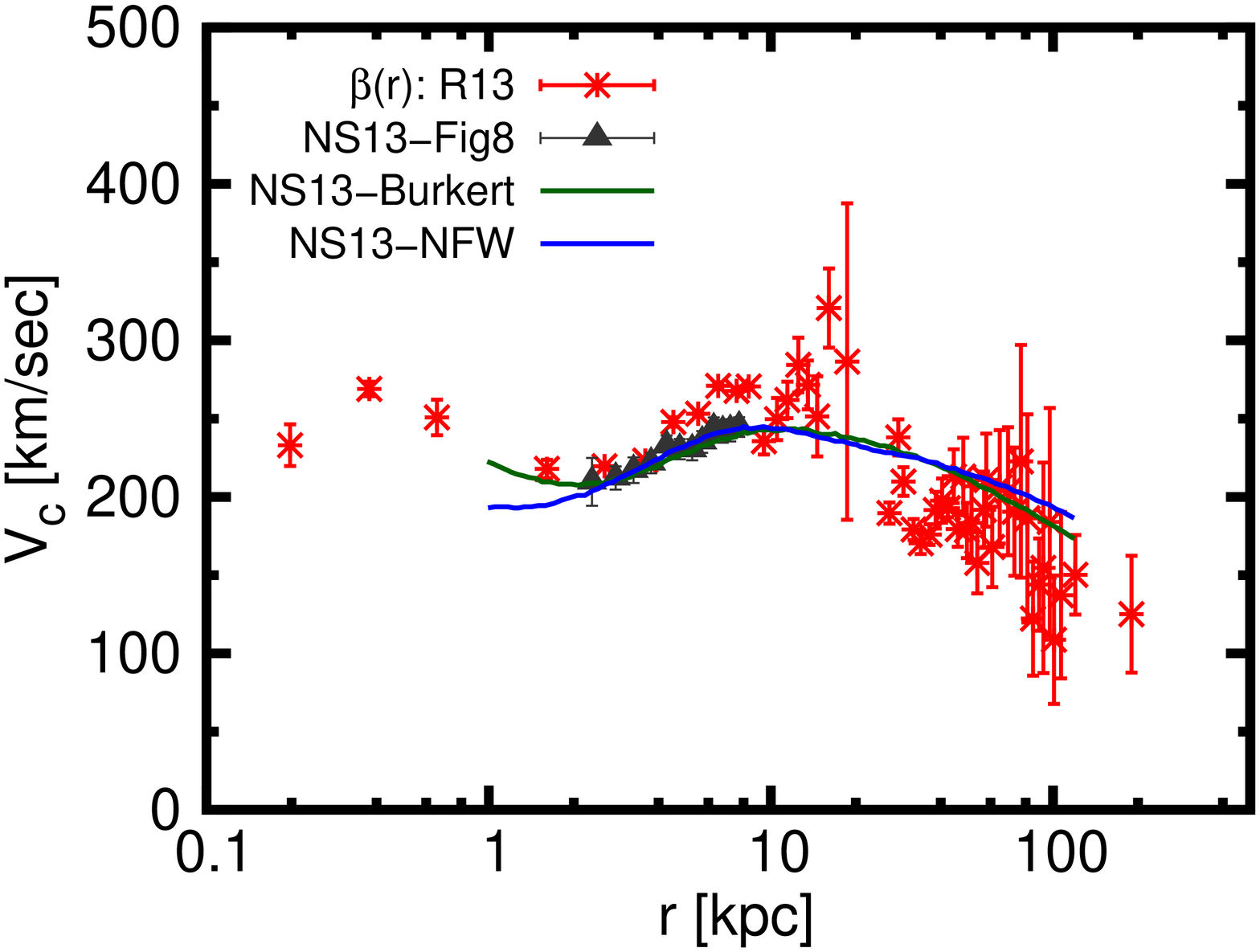,angle=270,width=\columnwidth}
\end{tabular}
\caption{Left: Rotation curve of the Milky Way to $\sim 200\kpc$ for 
$\GCset=[8.3,244]$ and for a radial profile of the non-disk tracers' 
velocity anisotropy parameter $\beta$ of the 
Osipkov-Merritt (OM) form, $\beta(r)=(1+r_a^2/r^2)^{-1}$, with two 
values of the ``anisotropy radius" $r_a=$15 kpc (OM-15) and 70 kpc 
(OM-70). The RC data generated with a radial profile of $\beta$ derived 
from Figure 2 of \citet{Rashkov_etal_eris_2013} (R13) (data points 
marked $\beta(r)$:R13) are also shown for comparison. The inset shows 
the OM $\beta$ profile for various values of $r_a$ together with 
the $\beta$ profile from Figure 2 of R13. Right: $\beta(r)$:R13 from the 
left panel, but in log scale and compared with the circular velocity 
data from terminal velocities (NS13-Fig8) and rotation curve fits for 
Burkert (NS13-Burkert) and NFW (NS13-NFW) models from 
\citet{Nesti_Salucci_2013}. The numerical data for $\beta(r)$:R13 are 
given in Table \ref{Table:RC_data_Rashkov_beta_8.3_244}. The full 
rotation curve data in machine readable form for the $\beta$ profile of 
R13 and three sets of values of $\GCset=$ 
[8.3,244], [8.5,220] and [8.0,200]    
are available in the online version of the paper.} 
\label{Fig:Fig_vcfull_grand_8.3_a_linear_R13_OM_inset_b_log_R13_NScomp_clr}
\end{figure*}

In Figure
\ref{Fig:Fig_vcfull_grand_8.3_a_linear_R13_OM_inset_b_log_R13_NScomp_clr} 
we show the full rotation curve of the Galaxy out to $\sim 200\kpc$ for  
$\GCset=[8.3,244]$ and for a radial profile of the non-disk tracers' 
velocity anisotropy parameter $\beta$ of the OM form, 
$\beta(r)=(1+r_a^2/r^2)^{-1}$, for two different values of 
$r_a=$ 15 kpc and 70 kpc. In addition, we show the RC generated with a 
$\beta$ profile extracted from Figure 2 of 
\citet{Rashkov_etal_eris_2013} with the corresponding numerical data in 
tabular form given in Table 
\ref{Table:RC_data_Rashkov_beta_8.3_244}. The inset in the left panel of 
Figure 
\ref{Fig:Fig_vcfull_grand_8.3_a_linear_R13_OM_inset_b_log_R13_NScomp_clr} 
shows the OM $\beta$ profile for various values of $r_a$ as well as the 
$\beta$ profile obtained in \citet{Rashkov_etal_eris_2013}. The latter 
is seen to roughly follow the OM form and is 
reasonably well bracketed within OM $\beta$ profiles with $r_a=15\kpc$ 
and $r_a=70\kpc$. In Figure 
\ref{Fig:Fig_vcfull_grand_8.3_a_linear_R13_OM_inset_b_log_R13_NScomp_clr} 
we also show the circular 
velocity data from terminal velocities and rotation curve 
fits for the Burkert and NFW models of the DM halo of the Galaxy given 
in \citet{Nesti_Salucci_2013} (up to $\sim 100\kpc$) in comparison with 
our RC generated with the $\beta$ profile 
of \citet{Rashkov_etal_eris_2013}.  

\begin{table*}[!htb]
\begin{center}
\setlength\extrarowheight{3pt}
\caption{The circular velocity, $V_c$,   
and its 1-$\sigma$ error, $\Delta\vc$, for various 
values of the Galactocentric distance, $r$, for a radial profile of the 
non-disk tracers' velocity anisotropy parameter $\beta$ derived from 
Figure 2 of \citet{Rashkov_etal_eris_2013}, with $\GCset = [8.3,244]$.}
\label{Table:RC_data_Rashkov_beta_8.3_244}
\vspace{0.5cm}
\begin{tabular}{||c|c|c||c|c|c||}
\hline\hline
$r$ & $\vc$ & $\Delta\vc$ & $r$ & $\vc$ & $\Delta\vc$ \\
($\kpc$) & ($\kmps$) & ($\kmps$) & ($\kpc$) & ($\kmps$) & ($\kmps$)\\
\hline\hline
  0.20   &     233.0      &   13.32    &   38.41   &     191.57     &   11.73 \\
  0.38   &     268.92     &    4.67    &   40.42   &     197.59     &   14.12 \\
  0.66   &     250.75     &   11.35    &   42.40   &     192.79     &    5.92 \\
  1.61   &     217.83     &    5.81    &   44.49   &     213.22     &   17.17 \\
  2.57   &     219.58     &    1.48    &   45.99   &     179.39     &   11.23 \\
  3.59   &     223.11     &    2.43    &   48.06   &     213.03     &   24.72 \\
  4.51   &     247.88     &    2.99    &   49.49   &     178.57     &   17.63 \\
  5.53   &     253.14     &    1.69    &   51.39   &     183.31     &   23.58 \\
  6.50   &     270.95     &    2.19    &   53.89   &     157.89     &   19.57 \\
  7.56   &     267.80     &    0.96    &   56.89   &     191.76     &   24.35 \\
  8.34   &     270.52     &    0.66    &   57.98   &     210.72     &   29.81 \\
  9.45   &     235.58     &    8.44    &   60.92   &     168.02     &   25.67 \\
 10.50   &     249.72     &   13.44    &   64.73   &     206.47     &   36.27 \\
 11.44   &     261.96     &   11.71    &   69.31   &     203.62     &   40.89 \\
 12.51   &     284.30     &   17.50    &   72.96   &     190.53     &   40.98 \\
 13.53   &     271.54     &   15.57    &   76.95   &     222.72     &   74.37 \\
 14.59   &     251.43     &   25.60    &   81.13   &     186.29     &   66.53 \\
 16.05   &     320.70     &   25.27    &   84.90   &     122.25     &   36.46 \\
 18.64   &     286.46     &  101.18    &   89.35   &     143.95     &   29.49 \\
 26.30   &     189.64     &    6.74    &   92.44   &     154.66     &   67.23 \\
 28.26   &     237.99     &   11.54    &   97.41   &     184.0      &   72.86 \\
 29.51   &     209.82     &    9.16    &  100.72   &     108.68     &   40.99 \\
 32.04   &     179.14     &    6.65    &  106.77   &     137.15     &   53.17 \\
 33.99   &     170.37     &    6.93    &  119.98   &     150.18     &   25.46 \\
 36.49   &     175.92     &    6.62    &  189.49   &     125.01     &   37.32 \\
\hline\hline
 \end{tabular}
\end{center}
\end{table*}

As already mentioned, a noticeable feature of the 
rotation curve, irrespective of the velocity anisotropy of the 
tracer objects, is its clearly declining nature beyond about $\sim$60 
kpc, as would be expected of an effectively finite size of the dark 
matter halo of the Galaxy.     

We emphasize that, for any given $\beta$, the rotation curve and 
mass profile of the Galaxy shown in 
Figures 
\ref{Fig:Fig_vcfull_8.3_dbeta_linear_bm_a_below50kpc_b_above50kpc_clr} 
and 
\ref{Fig:Fig_mr_8.3_dbeta_bm_clr}, respectively, are based entirely 
on 
observational data, and are obtained without making any models of the 
mass distributions of the various components (the bulge, disk and dark 
matter halo) of the Galaxy. 
%
%
\section{Summary}
\label{sec:summary}
In this paper, we have constructed the rotation curve (RC) of the Galaxy 
from a galactocentric distance of $\sim 0.2\kpc$ out to $\sim200\kpc$ by 
using kinematical data on a variety of both disk and non-disk objects 
that trace the gravitational potential of the Galaxy, 
without assuming any theoretical models of the visible and dark
matter components of the Galaxy. We have studied 
the dependence of the RC on the choice of the Galactic 
constants (GCs) and also studied the dependence on the velocity 
anisotropy parameter $\beta$ of the non-disk tracers. The RC in the 
disk region is found to depend significantly on the choice of values 
of the GCs. The rotation curve at large distances beyond 
the stellar disk, however, depends more significantly on the parameter 
$\beta$ than on the values of the GCs. In general, the 
mean RC is found to steadily decline beyond $r\sim 60\kpc$, irrespective 
of the value of $\beta$. At any given galactocentric distance $r$, the 
circular speed is lower for larger values of $\beta$. Considering that 
the largest allowed value of $\beta$ is unity (complete radial 
anisotropy), this allows us to set a model-independent lower limit on 
the total mass 
of the Galaxy, giving $M(\lsim 200\kpc)\geq 
(6.8\pm4.1)\times10^{11}\Msun$. We have also noted 
that recent results from high resolution hydrodynamical simulations of 
formation of galaxies like Milky Way \citep{Rashkov_etal_eris_2013} 
indicate an increasingly radially biased velocity ellipsoid of the 
Galaxy's stellar population at large distances, with stellar orbits 
tending to be almost purely radial ($\beta\to 1$) beyond $\sim 
100\kpc$. This implies that the above lower 
limit on the Galaxy's mass (obtained from our results with $\beta=1$) may 
in fact be a good estimate of the actual mass of the Galaxy out to 
$\sim 200\kpc$. 

\acknowledgements We thank G.~Battaglia, W.~Brown, A.~Deason, 
O.~Gnedin, P.~Kafle, S.~Sharma, Y.~Sofue, M.~Weber, and X.~Xue for 
useful communications. PB thanks R.~Cowsik for discussions 
and for support under a Clark Way Harrison Visiting Professorship at the 
McDonnell Center for the Space Sciences and Physics Department at 
Washington University in St. Louis. We thank the anonymous referee for 
useful comments and suggestions.  
\vfill\eject

\newpage


\vfill\eject

\newpage
\vskip2cm
\appendix
\noindent
In this Appendix, we collect together some details of the disk- and 
non-disk tracers used in this paper for calculating the rotation
curves. 

\begin{table*}[!htb]
\begin{center}
\setlength\extrarowheight{3pt}
\caption{Disk tracer types, their source references and $(l,~b)$ ranges 
of the data sets used in this paper. Superscript `$a$' denotes the 
tracers limited within the solar circle ($R<\rsun$) where tangent point 
method has been used to derive the rotation speeds. The identifier for 
each tracer data set used in the paper is given within parentheses in 
the first column under the respective tracer type for subsequent  
references in the paper. } 
\label{Table:Disk_samples}  
\vskip 1pt
\begin{tabular}{p{4.0cm} p{5.2cm} p{6cm}}
\tableline\tableline
Tracer Type   & Data Source &  $(l, b)$ Ranges \\
\tableline
HI regions$^a$ \newline (HI-W76-B78)   & 
\citet{Westerhout_HI_1976};\newline\citet{Burton_Gordon_CO_plus_HI_1978}	
&  $ 1{\textdegree} < l < 90{\textdegree}$   \\

CO clouds$^a$ \newline (CO-B78)  & 
\citet{Burton_Gordon_CO_plus_HI_1978}	
&   
$ 9{\textdegree} < l < 82{\textdegree}$  \\

CO clouds$^a$ \newline (CO-C85)  & \citet{Clemens_CO_1985} &  
$ 13{\textdegree}  < l < 86{\textdegree}$ \\

HI regions$^a$ \newline (HI-F89)  & \citet{Fich_et_al_1989} &  
$ 15{\textdegree} < l < 89{\textdegree}$ and $271{\textdegree} 
< l < 345{\textdegree}$\\

HII regions \newline (HII-F89)  & \citet{Fich_et_al_1989} &  
$ 10{\textdegree} < l < 170{\textdegree}$ and $ 190{\textdegree} < l < 
350{\textdegree}$ \\

HII regions \& \newline reflection nebulae \newline (HII-RN-B93)  & 
\citet{Brand_Blitz_HII_1993} &   
$ 10{\textdegree} < l < 170{\textdegree}$ and 
$ 190{\textdegree} < l < 350{\textdegree}$   \\

Cepheids \newline (Cepheid-P94)  & \citet{Pont_Cepheids_1994} &  
$ 10{\textdegree} < l < 170{\textdegree}$ 
and $ 190{\textdegree} < l < 350{\textdegree}$; $|b| < 10{\textdegree}$ \\

Planetary nebulae \newline (PNe-M05-M84-D98)  & 
\citet{Maciel_Lago_PN_2005}; 
\newline\citet{Maciel_PN_distance_1984}; 
\newline\citet{Durand_PN_velocity_1998} &  
$ 15{\textdegree} < l < 345{\textdegree}$; $|b| < 10{\textdegree}$ \\

Open star clusters \newline (OSC-F08-D02)  & 
\citet{Frinchaboy_OSC_2008}; \newline\citet{Dias_OSC_distance_2002} &   
$ 10{\textdegree} < l < 170{\textdegree}$ and 
$ 190{\textdegree} < l < 350{\textdegree}$; $|b| < 9{\textdegree}$\\

HII regions \newline (HII-H09)  & \citet{Hou_HII_2009} &  
$ 10{\textdegree} < l < 170{\textdegree}$ and  
$  190{\textdegree} < l < 350{\textdegree}$ \\

HII regions$^a$ \newline (HII-U11)  & \citet{Urquhart_HII_2011} &  
$10{\textdegree} < l < 65{\textdegree}$ and 
$ 280{\textdegree} < l < 350{\textdegree}$  \\
%
C stars \newline (C stars-D07-B12)  & 
\citet{Demers_Battinelli_2007,Battinelli_etal_2012} & 
$ 54{\textdegree} < l < 150{\textdegree}$; 
$3{\textdegree} < |b| < 9{\textdegree}$  \\
\tableline
\vfill
\end{tabular}
\end{center}
\end{table*}
\vfill\eject

\begin{figure*}[!htb]
\begin{tabular}{ccc}
\centering
\epsfig{file=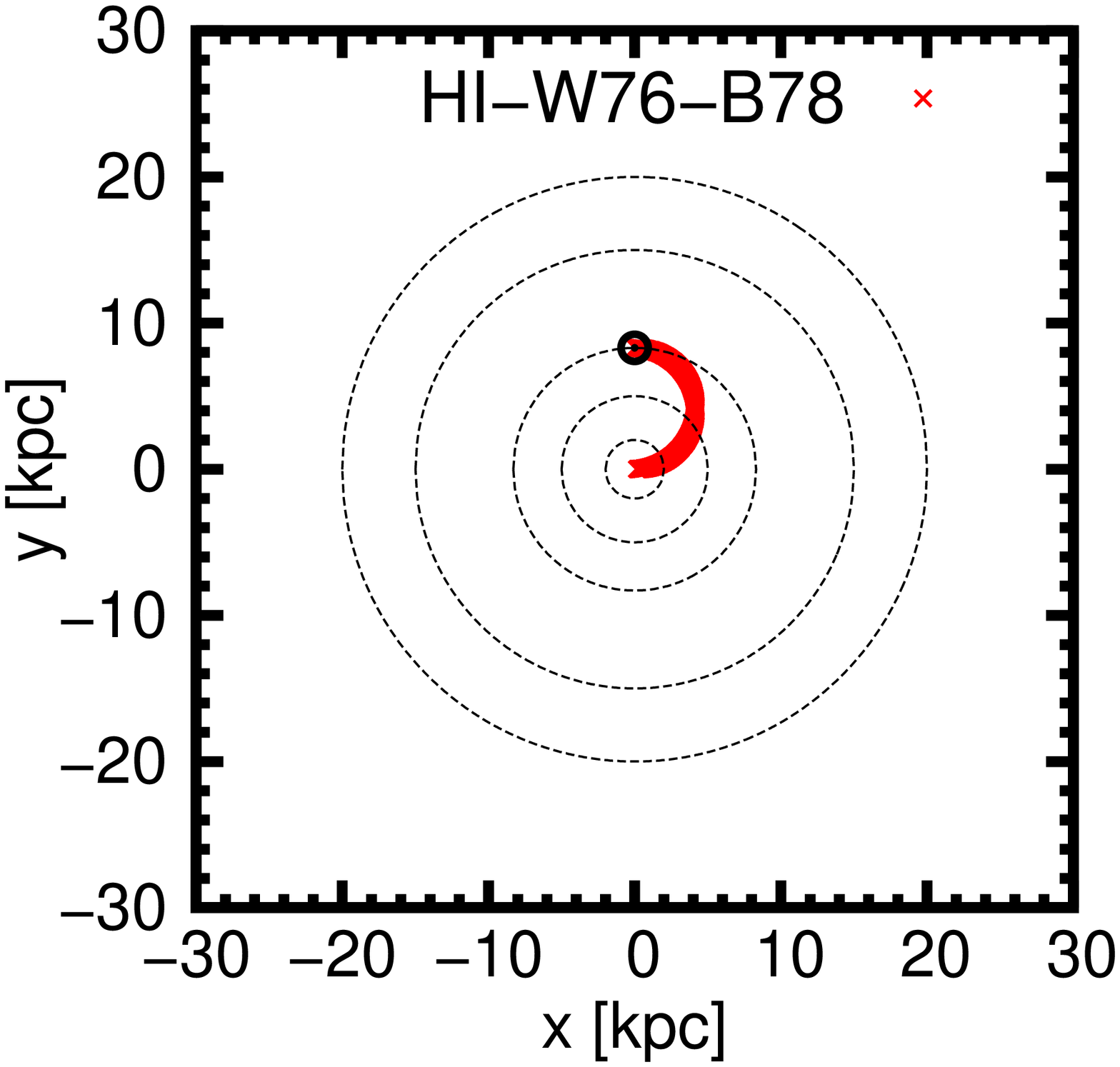,angle=270,width=2.5in}&
\hskip -1.7cm
\epsfig{file=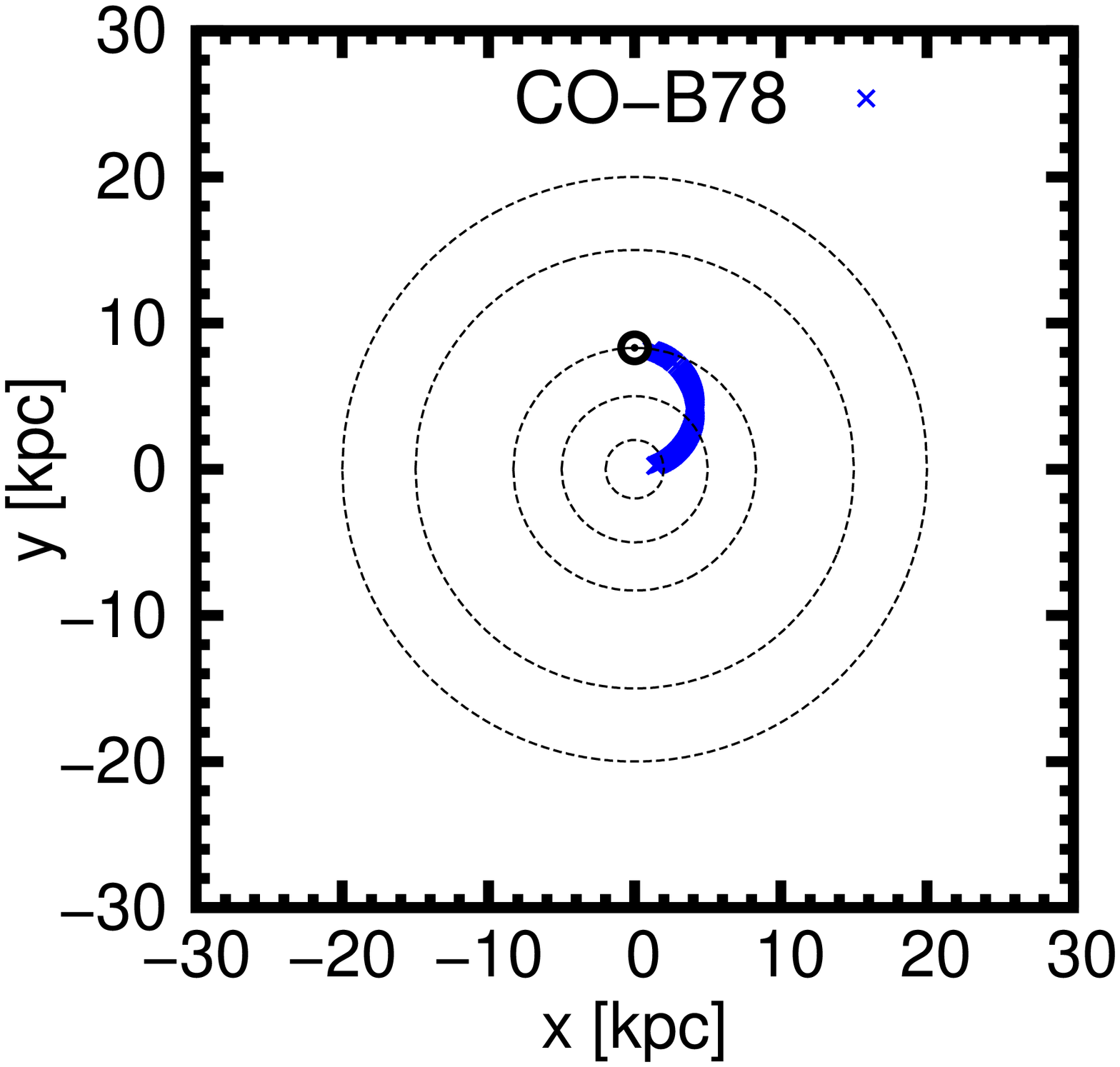,angle=270,width=2.5in}&
\hskip -1.7cm
\epsfig{file=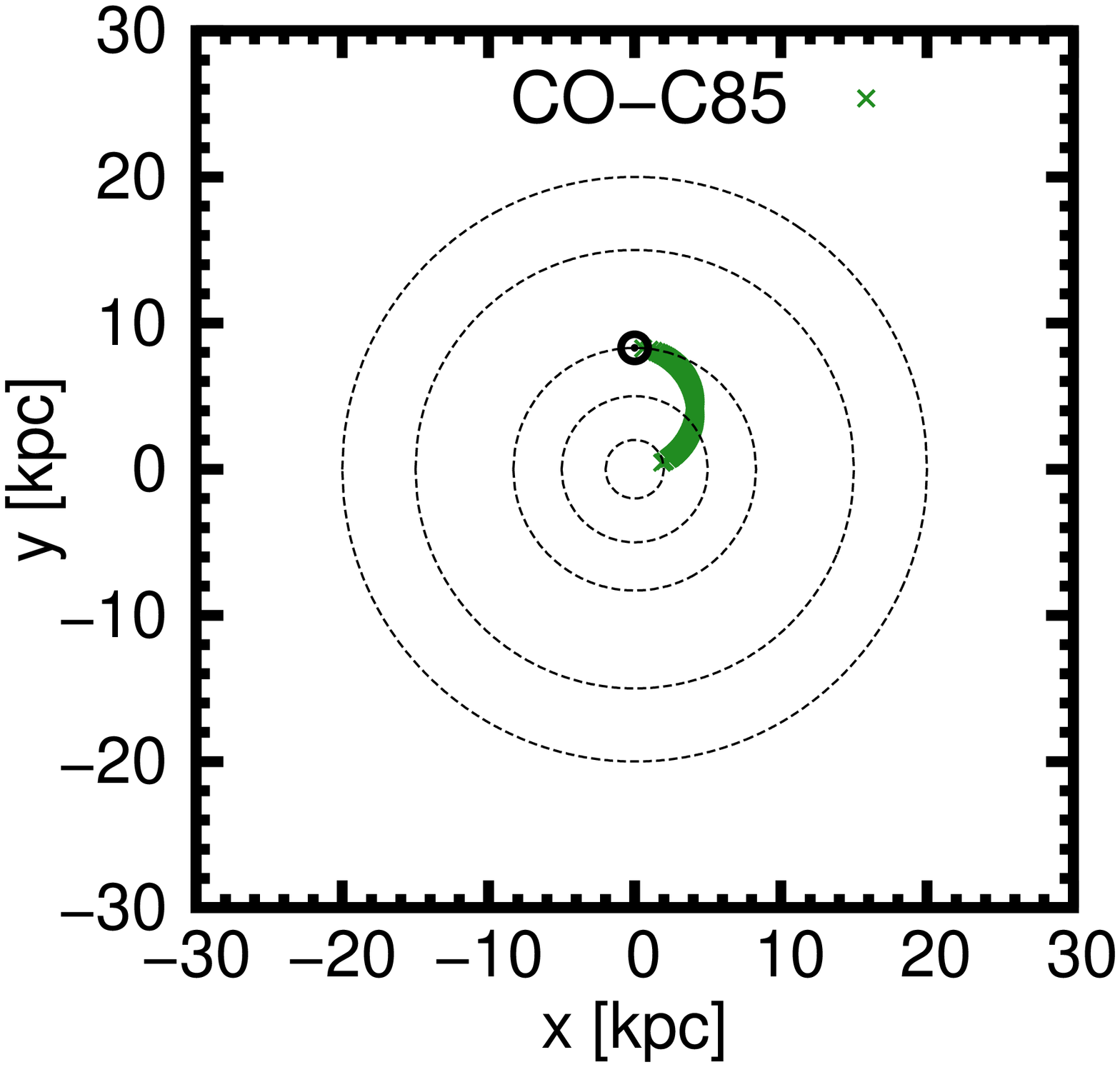,angle=270,width=2.5in}\\
\epsfig{file=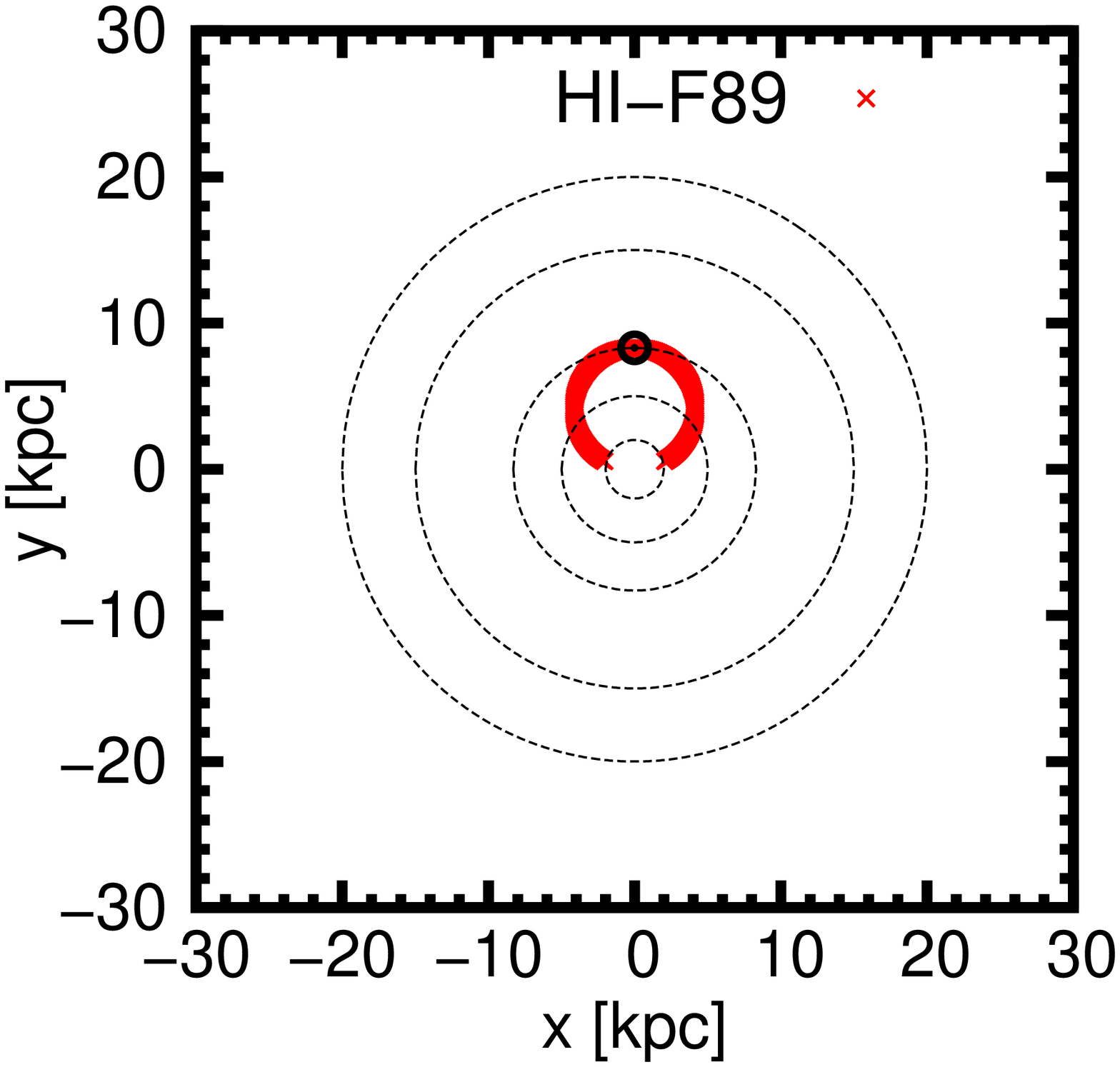,angle=270,width=2.5in}&
\hskip -1.7cm
\epsfig{file=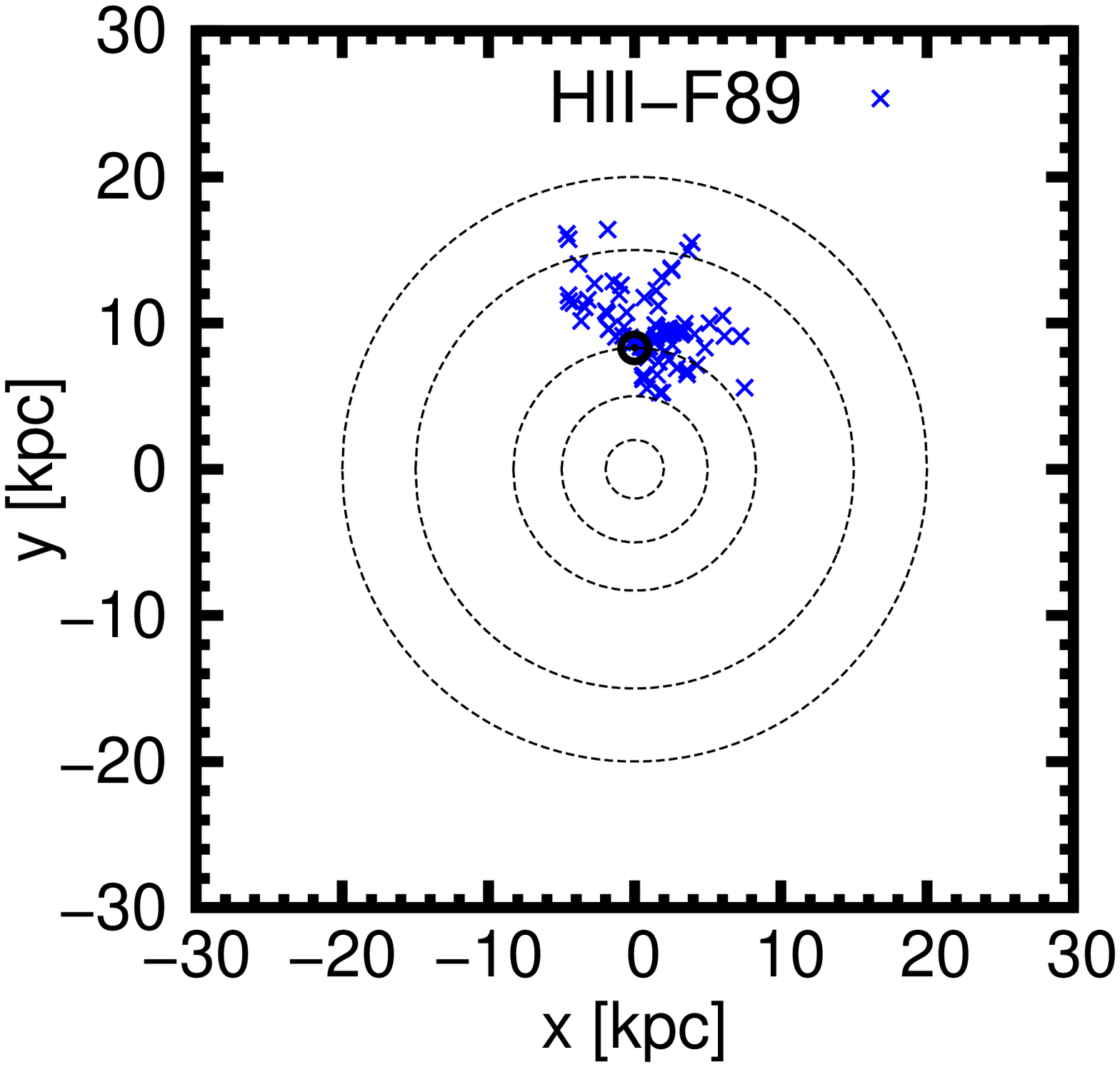,angle=270,width=2.5in}&
\hskip -1.7cm
\epsfig{file=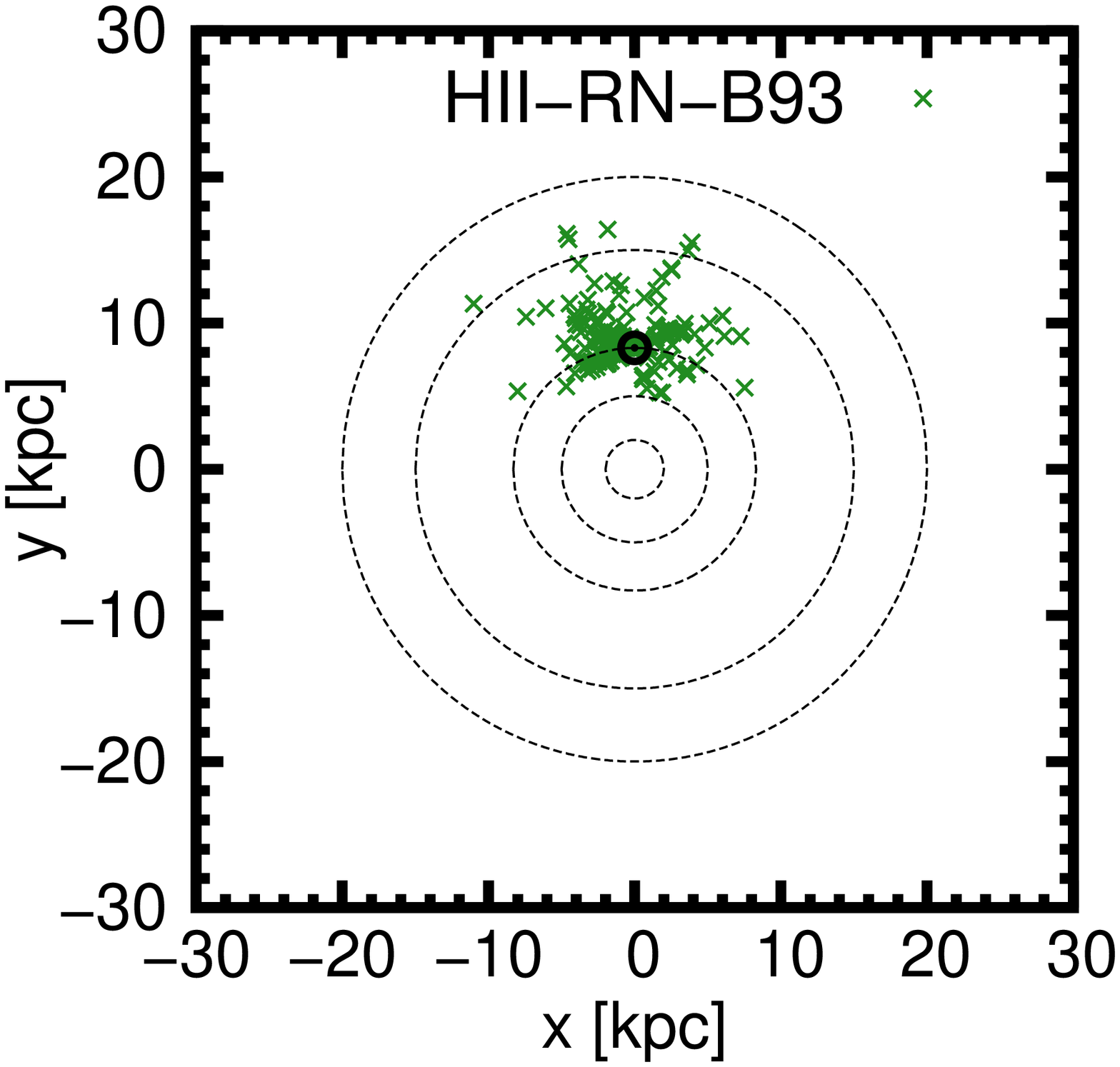,angle=270,width=2.5in}\\
\epsfig{file=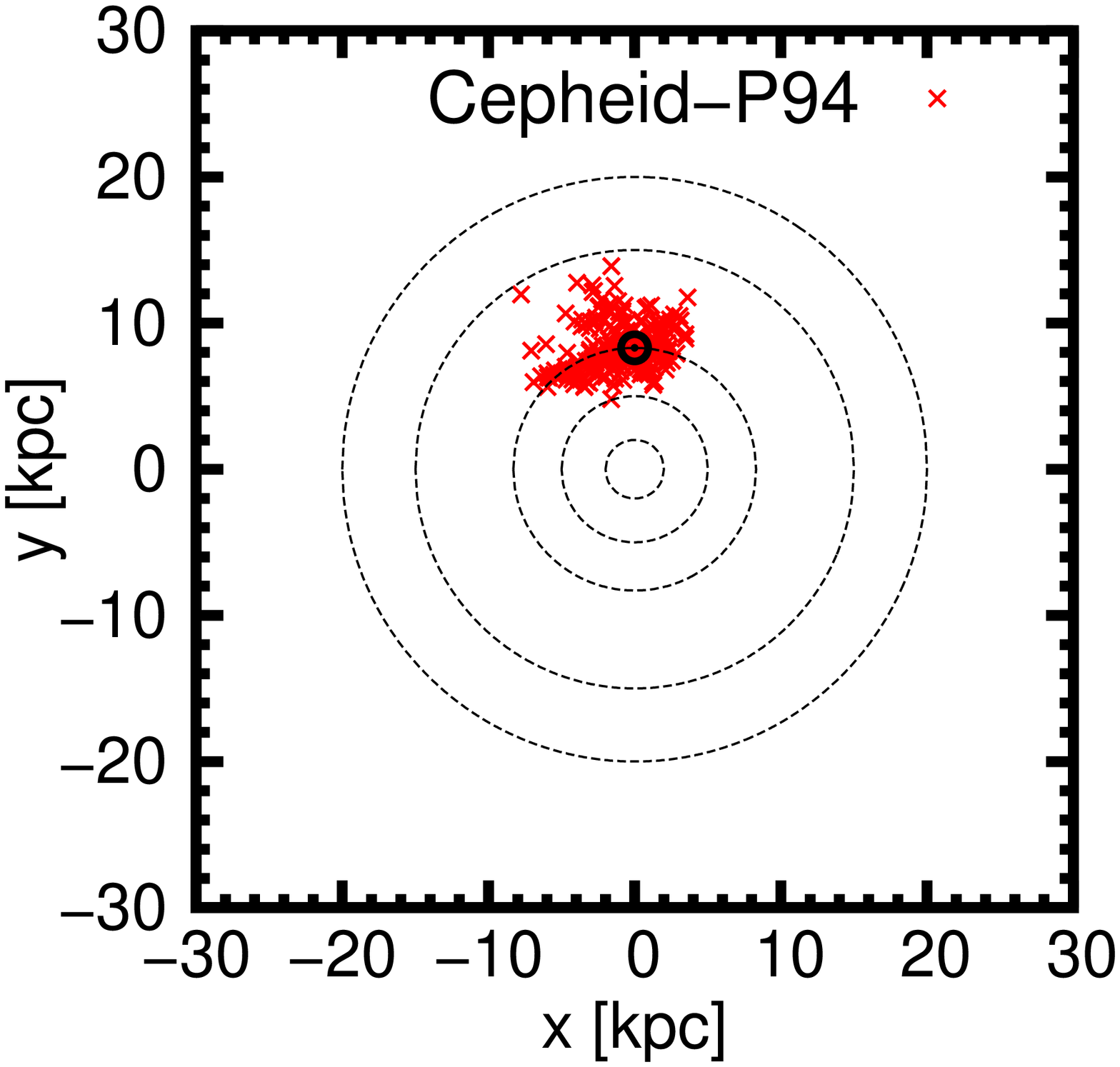,angle=270,width=2.5in}&
\hskip -1.7cm
\epsfig{file=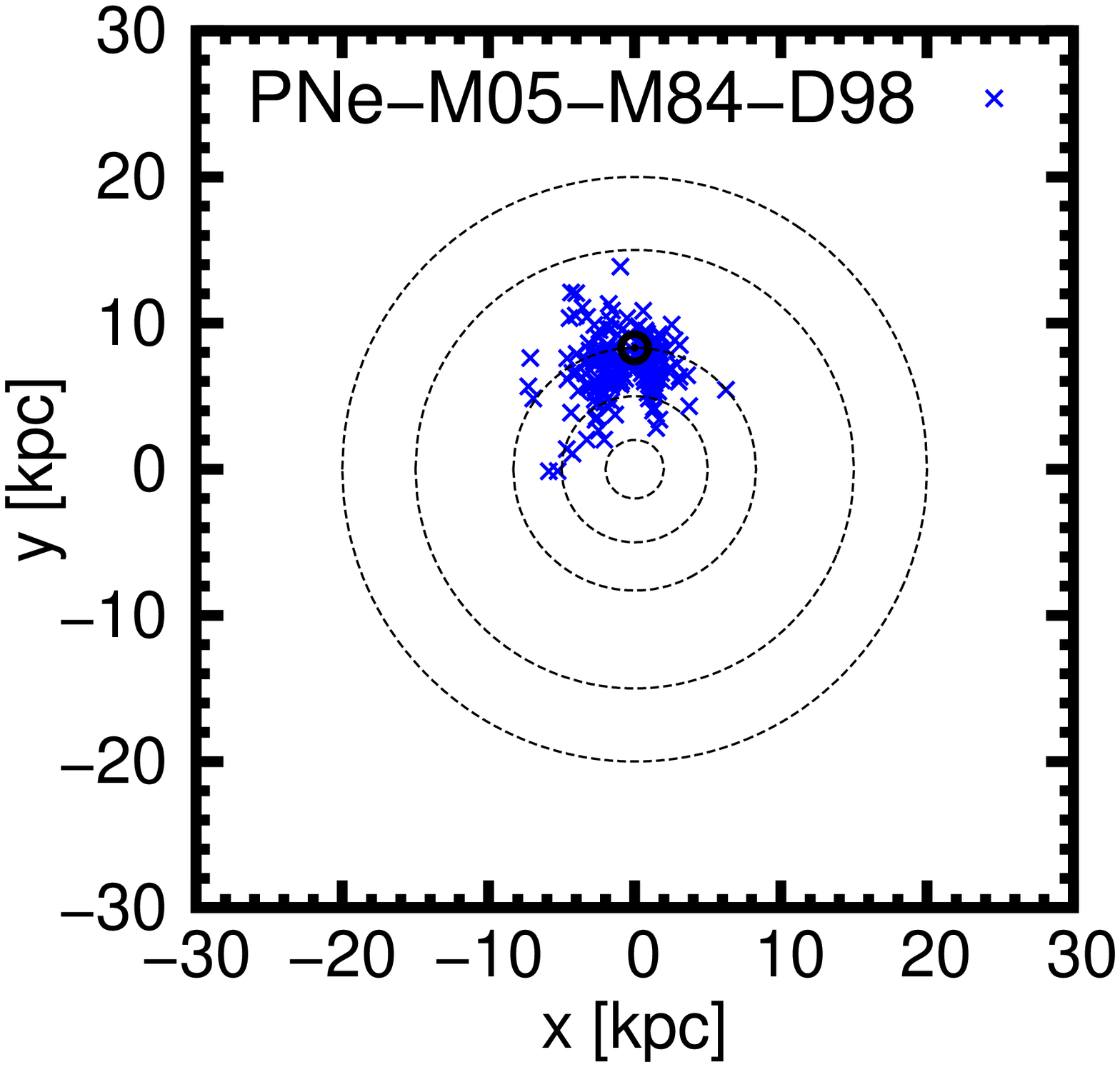,angle=270,width=2.5in}&
\hskip -1.7cm
\epsfig{file=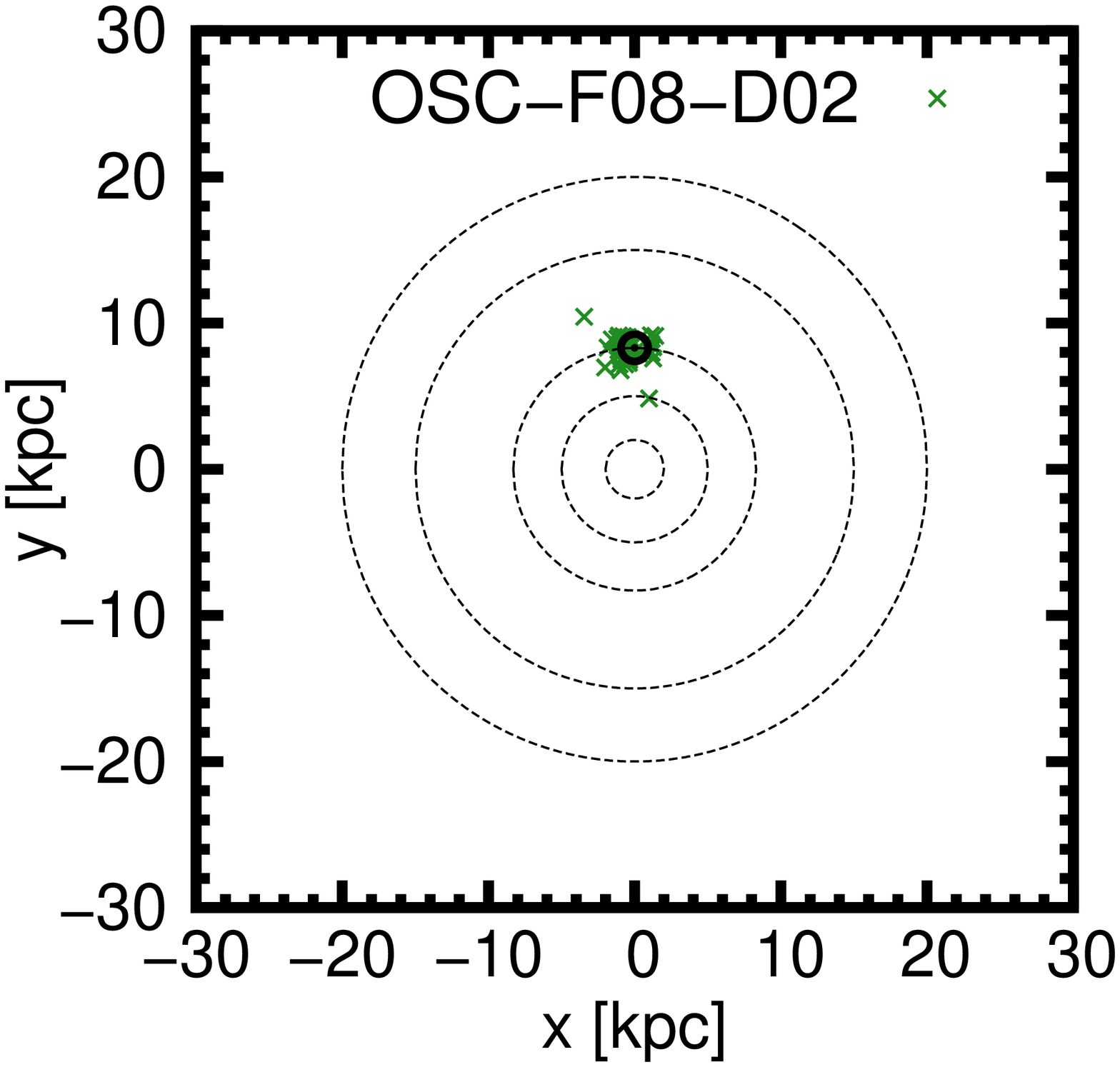,angle=270,width=2.5in}\\
\epsfig{file=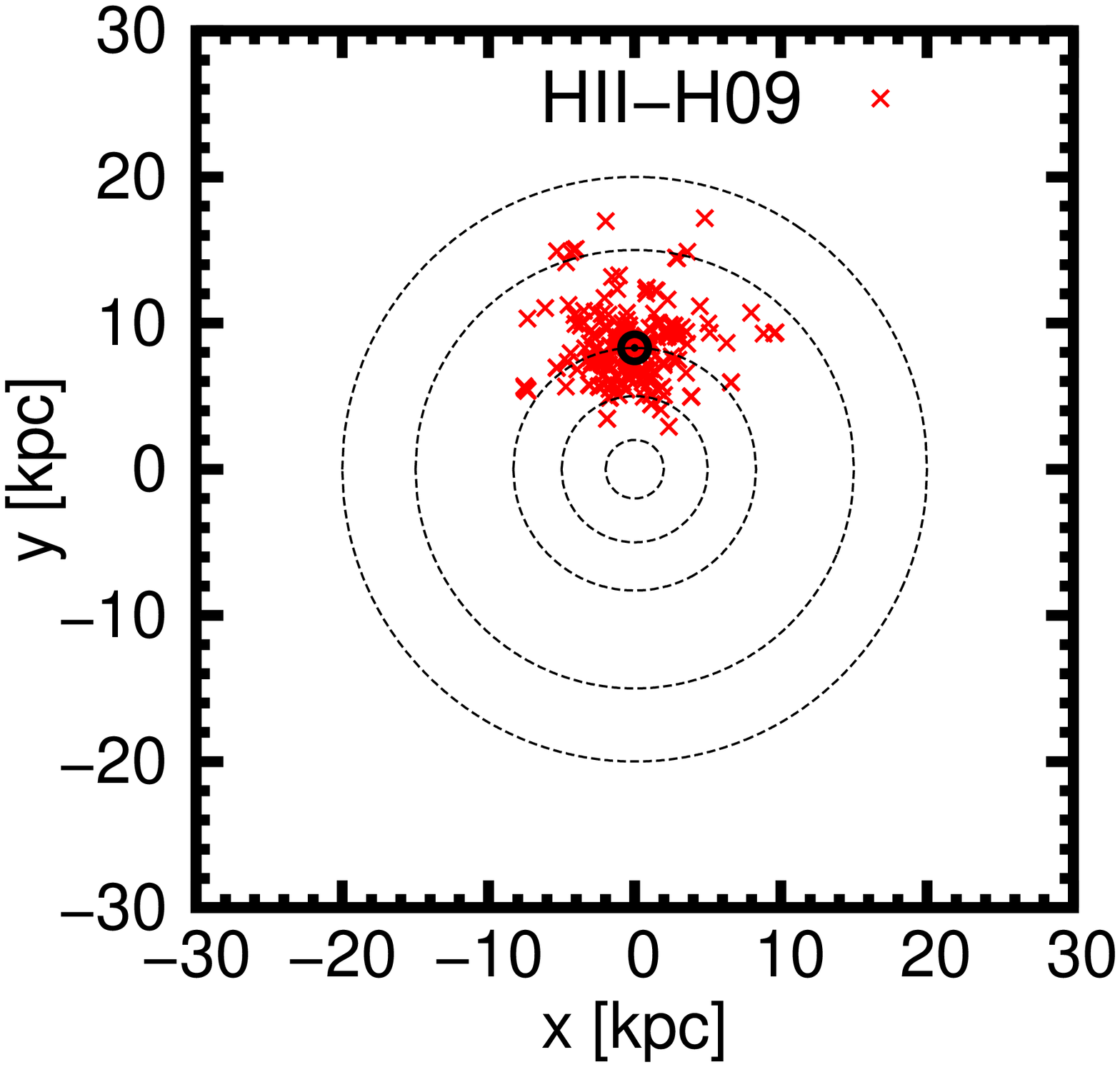,angle=270,width=2.5in}&
\hskip -1.7cm
\epsfig{file=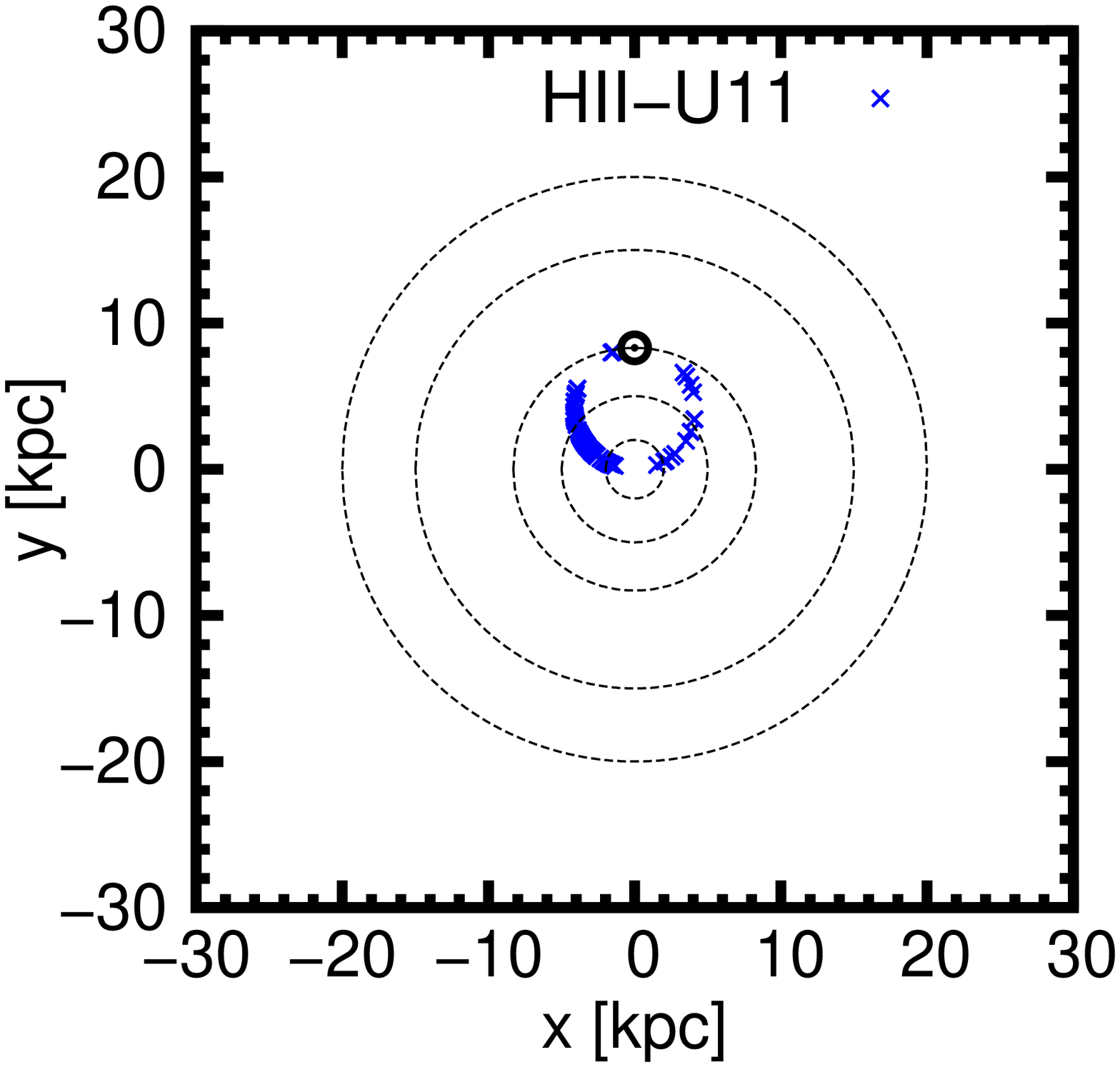,angle=270,width=2.5in}&
\hskip -1.7cm
\epsfig{file=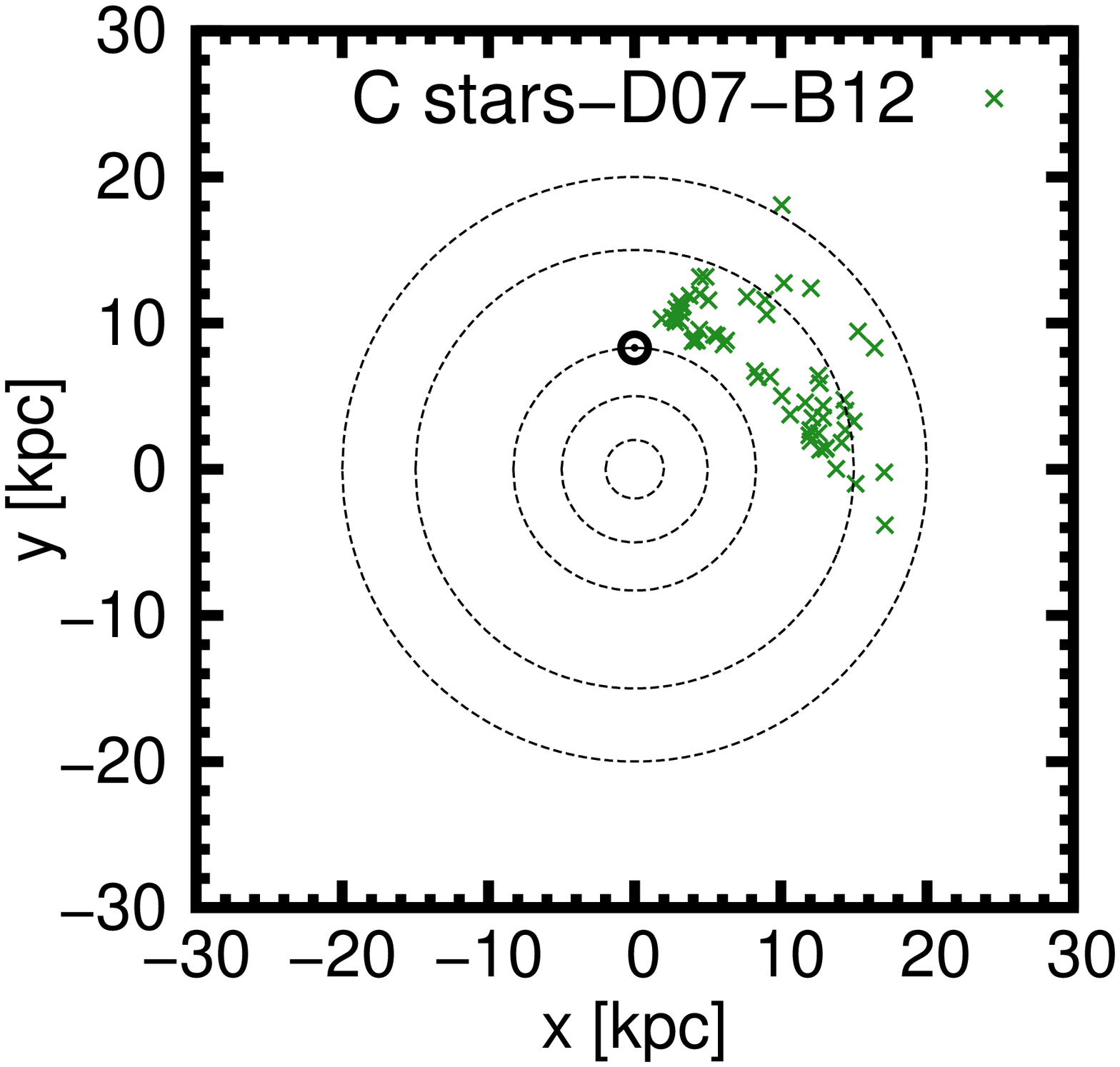,angle=270,width=2.5in}
\end{tabular}
\caption{$x$--$y$ scatter plots for the different 
disk tracer samples
listed in Table \ref{Table:Disk_samples}, for the case 
$\Rsun=8.3\kpc$. The Galactic Center is chosen to be at origin $(0,0)$
with the sun marked by an open circle located at $(0,\Rsun)$.}
\label{Fig:Fig_disk_xy_scatter_clr}
\end{figure*}

\vfill\eject

\begin{figure*}[!htb]
\begin{tabular}{ccc}
\centering
\epsfig{file=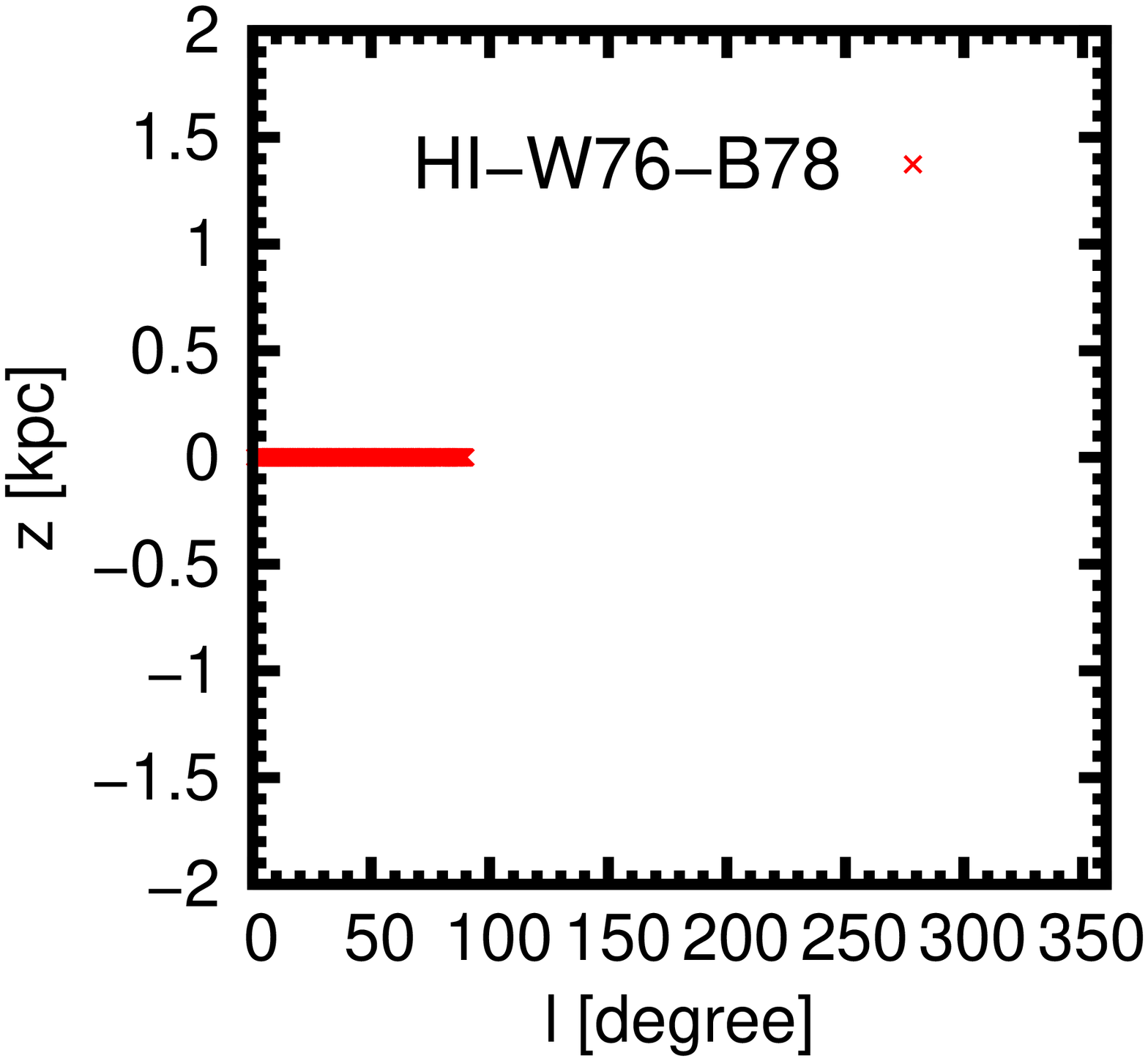,angle=270,width=2.5in}&
\hskip -1.7 cm
\epsfig{file=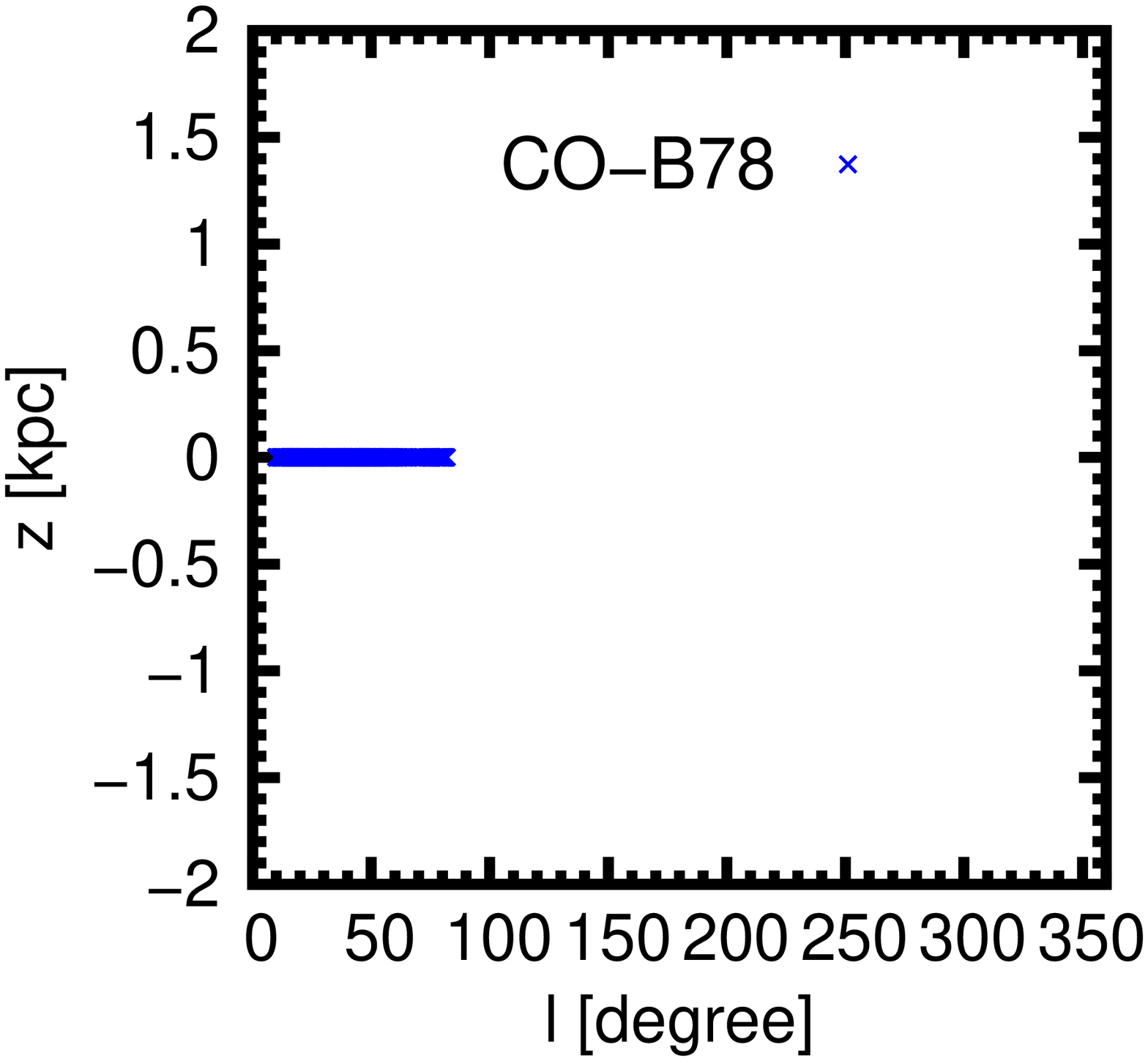,angle=270,width=2.5in}&
\hskip -1.7cm
\epsfig{file=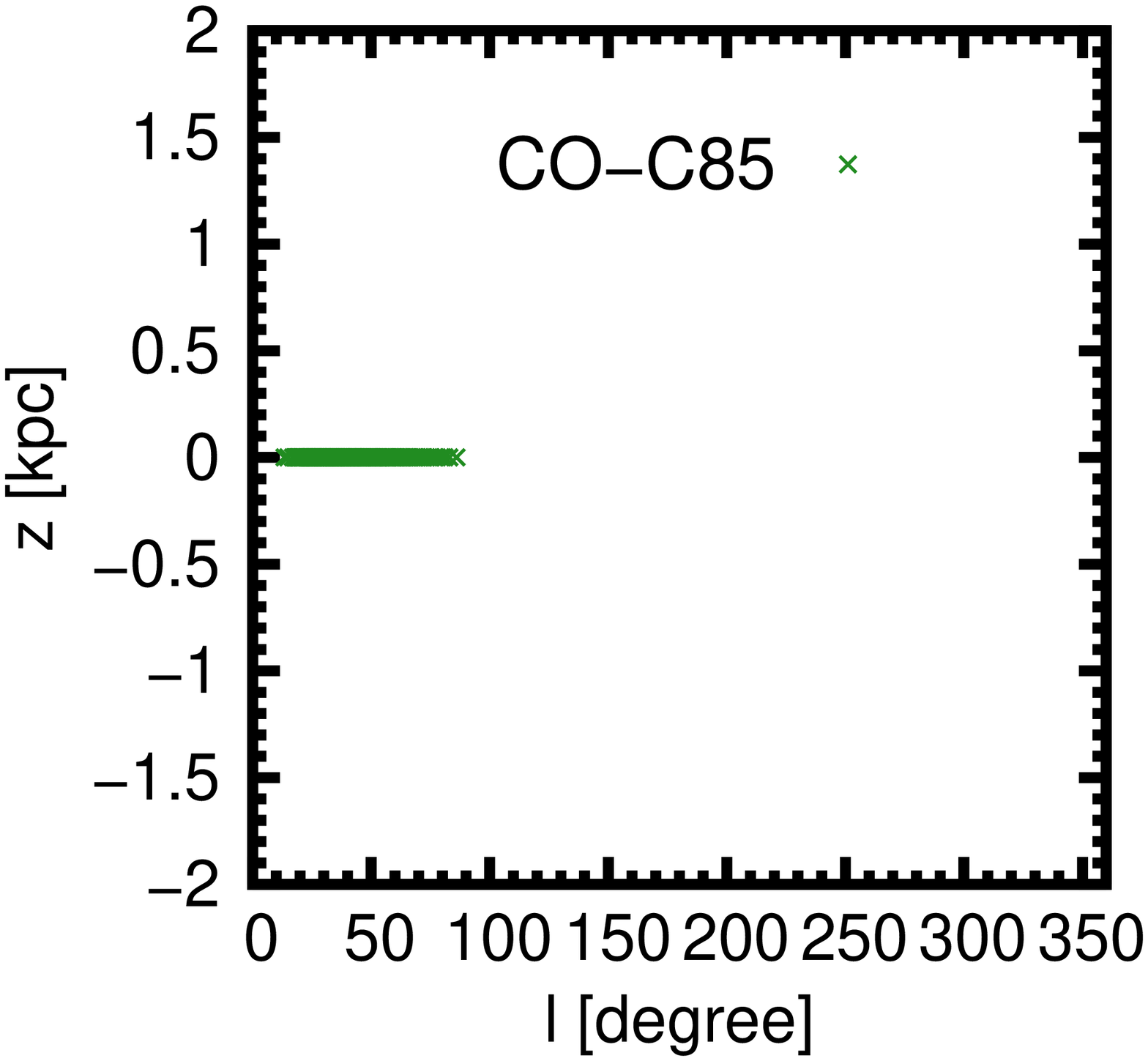,angle=270,width=2.5in}\\
\epsfig{file=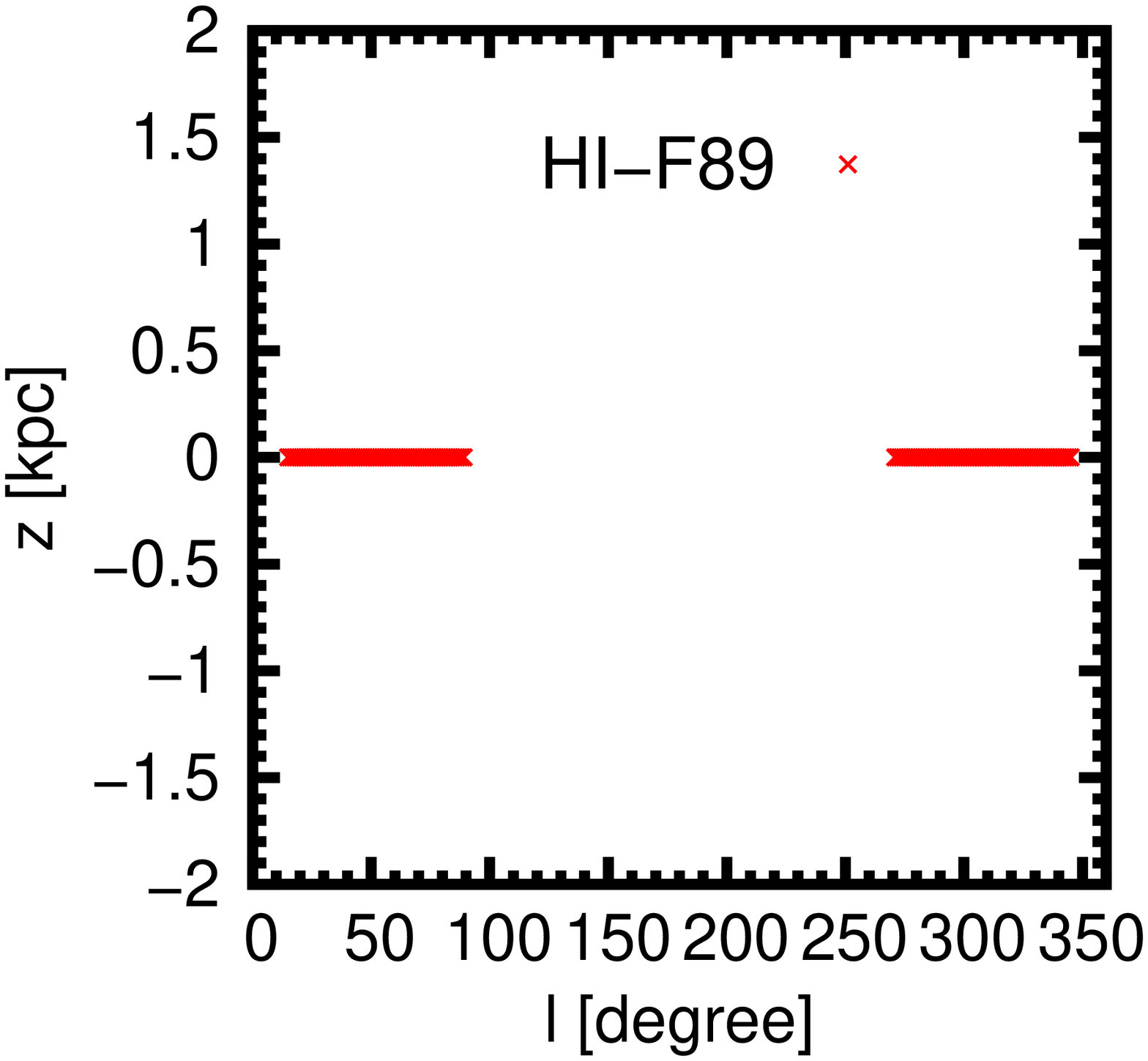,angle=270,width=2.5in}&
\hskip -1.7 cm
\epsfig{file=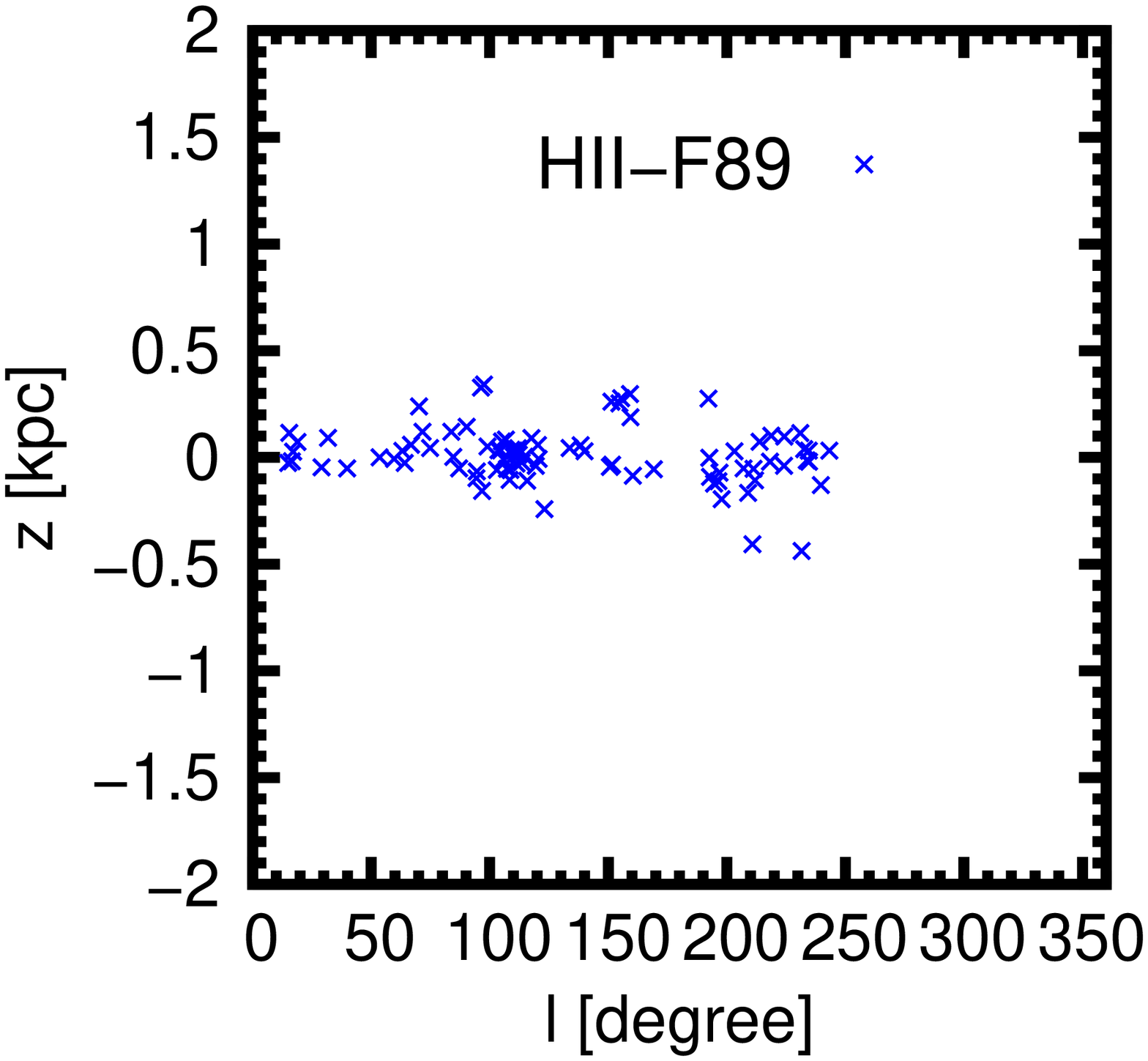,angle=270,width=2.5in}&
\hskip -1.7cm
\epsfig{file=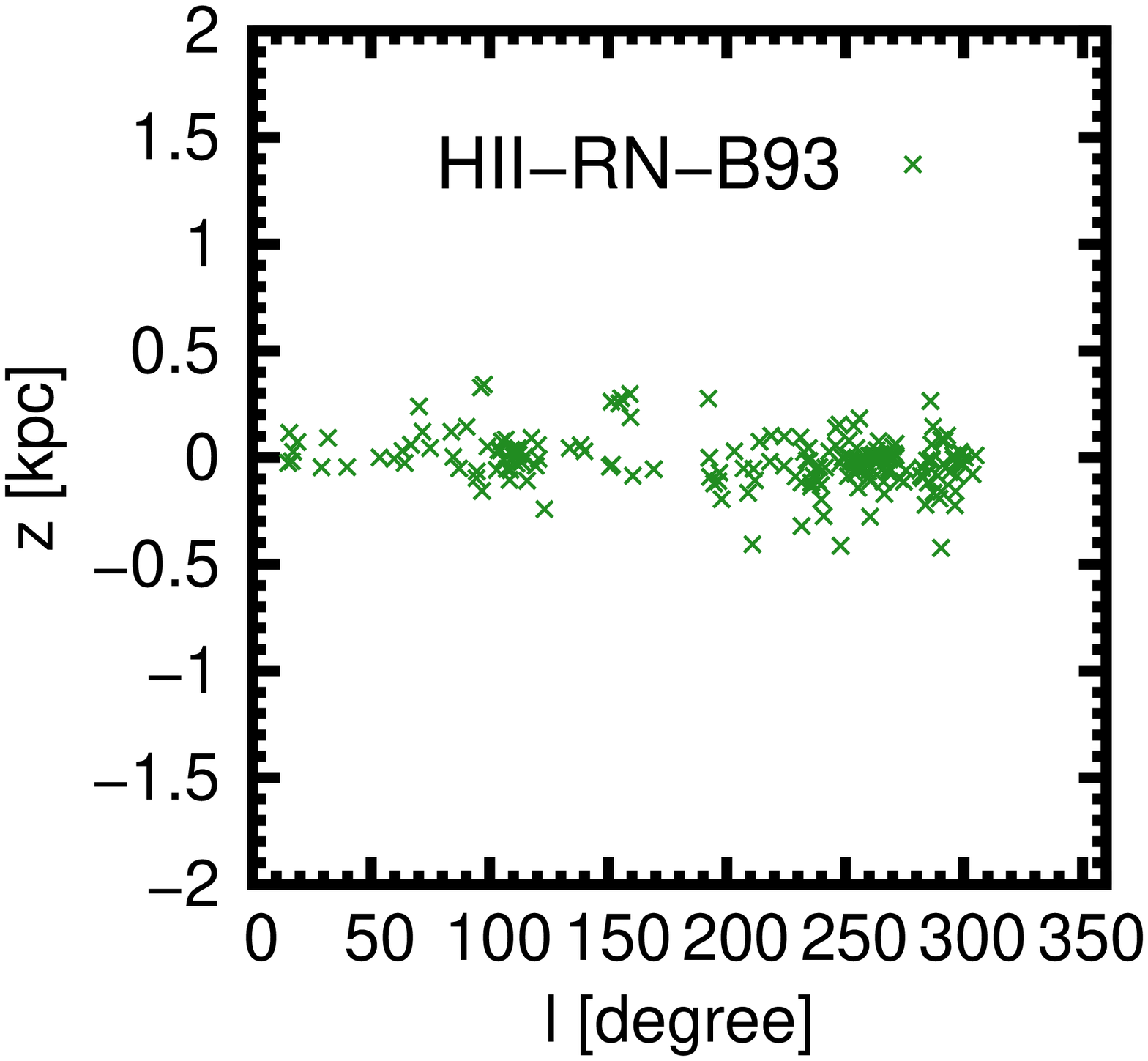,angle=270,width=2.5in}\\
\epsfig{file=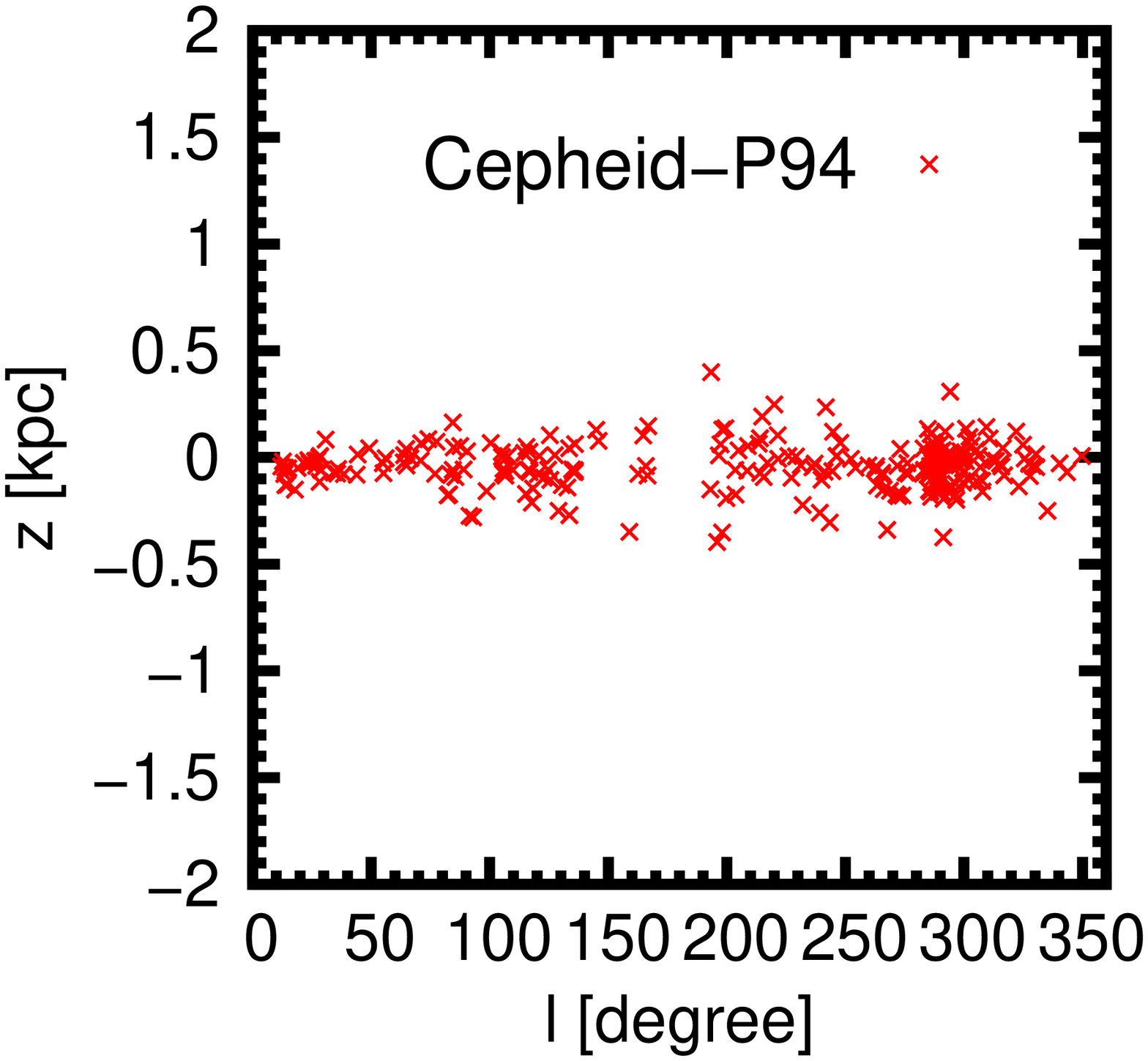,angle=270,width=2.5in}&
\hskip -1.7 cm
\epsfig{file=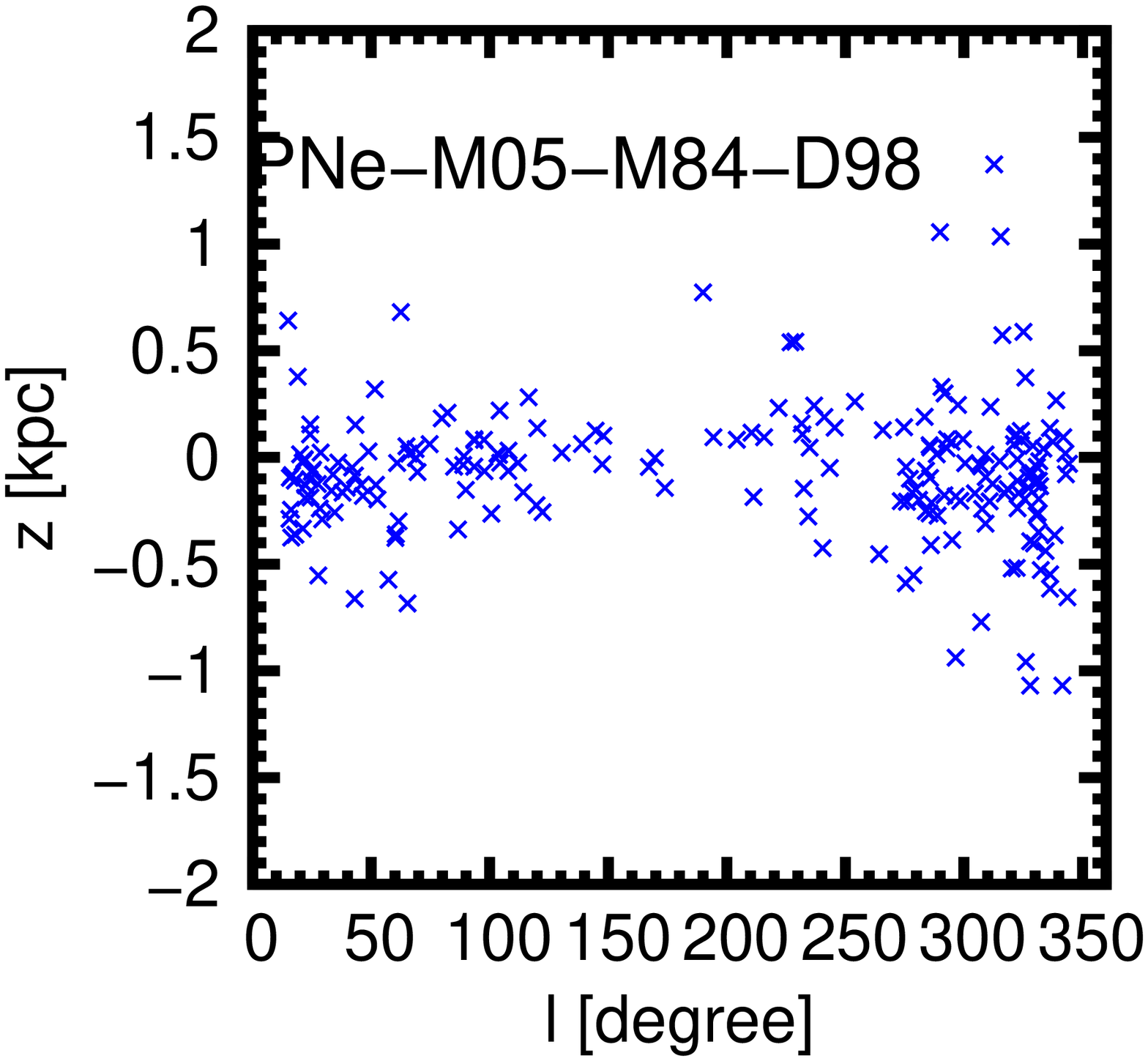,angle=270,width=2.5in}&
\hskip -1.7 cm
\epsfig{file=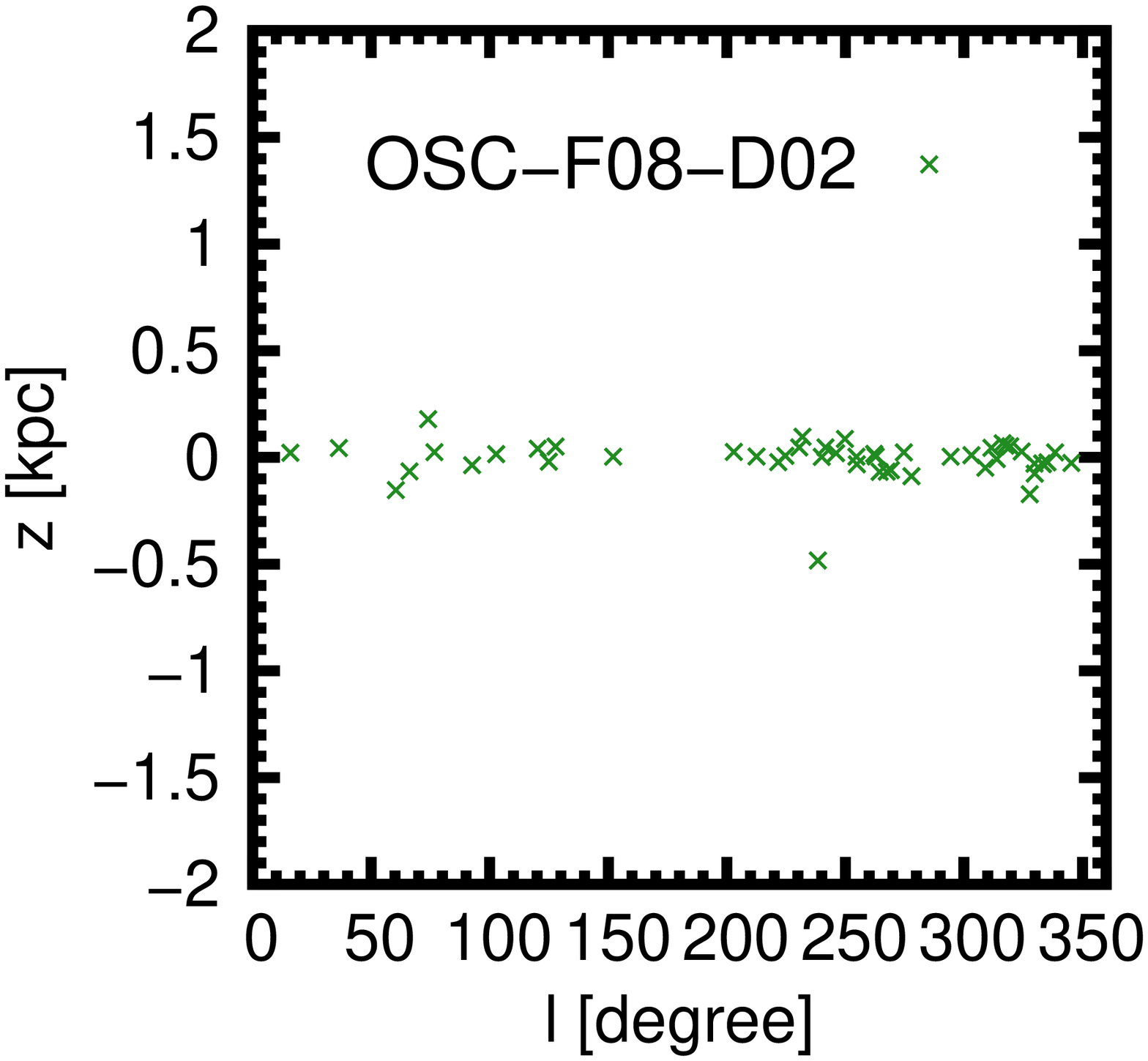,angle=270,width=2.5in}\\
\epsfig{file=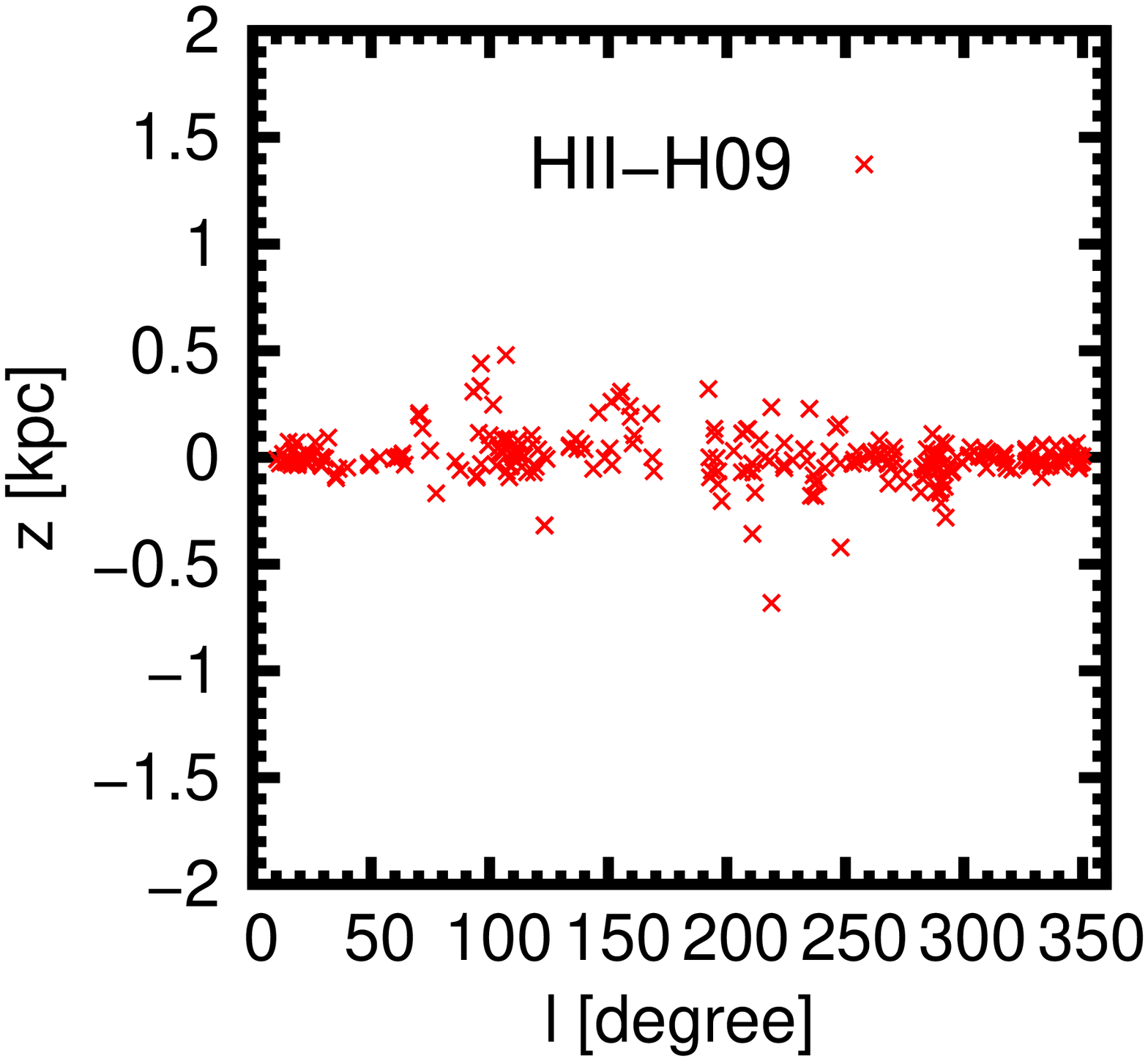,angle=270,width=2.5in}&
\hskip -1.7cm
\epsfig{file=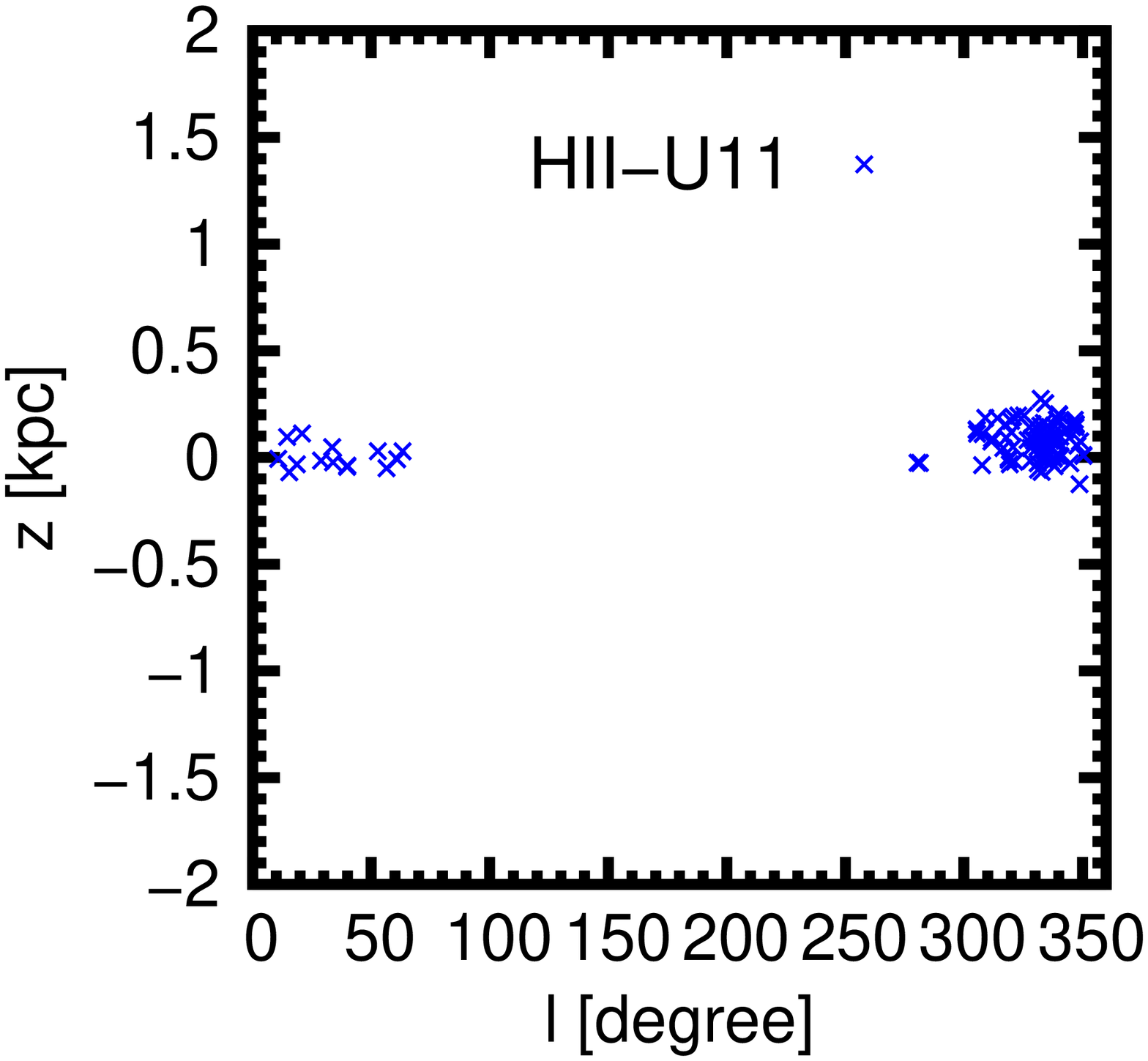,angle=270,width=2.5in}&
\hskip -1.7cm
\epsfig{file=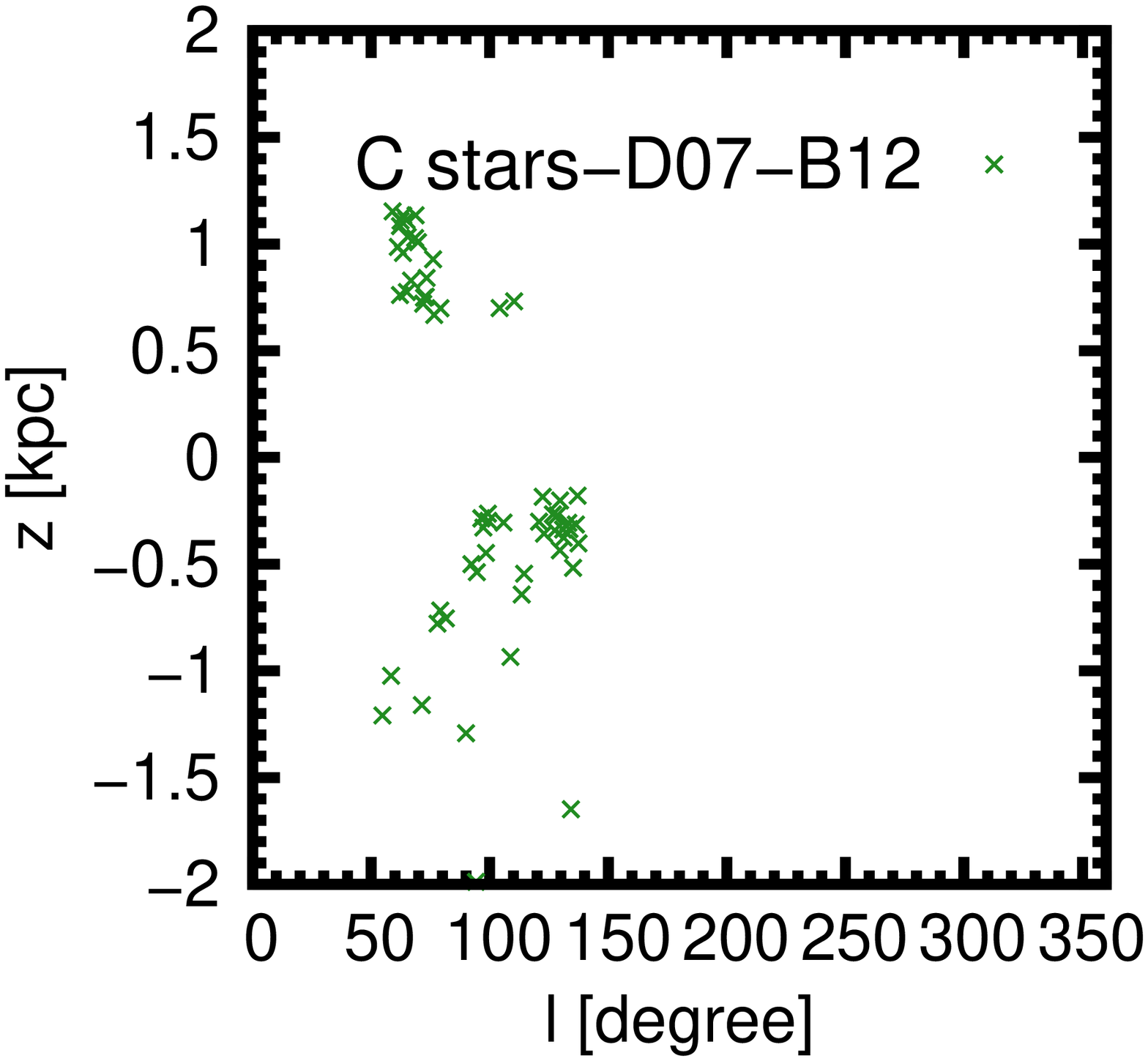,angle=270,width=2.5in}
\end{tabular}
\caption{Galactic longitude, $l$, versus height from Galactic
mid-plane, $z$, for the different disk tracer samples listed 
in Table \ref{Table:Disk_samples}, for the case $\Rsun=8.3\kpc$.}
\label{Fig:Fig_disk_lz_scatter_clr}
\end{figure*}

\vfill\eject

\begin{figure*}[!htb]
 \begin{tabular}{ccc}
\epsfig{file=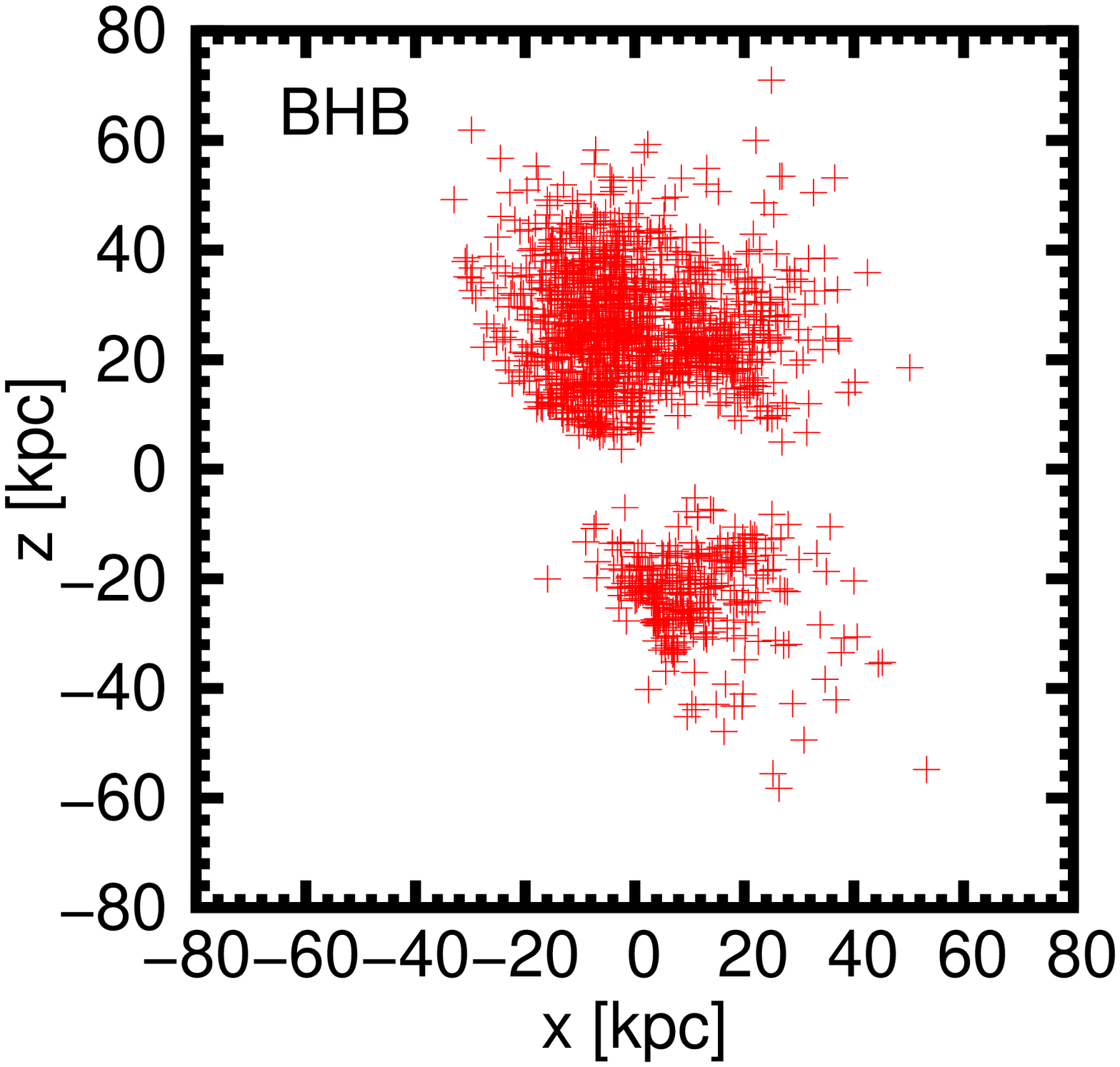,angle=270,width=2.7in}&
\hskip -2 cm
\epsfig{file=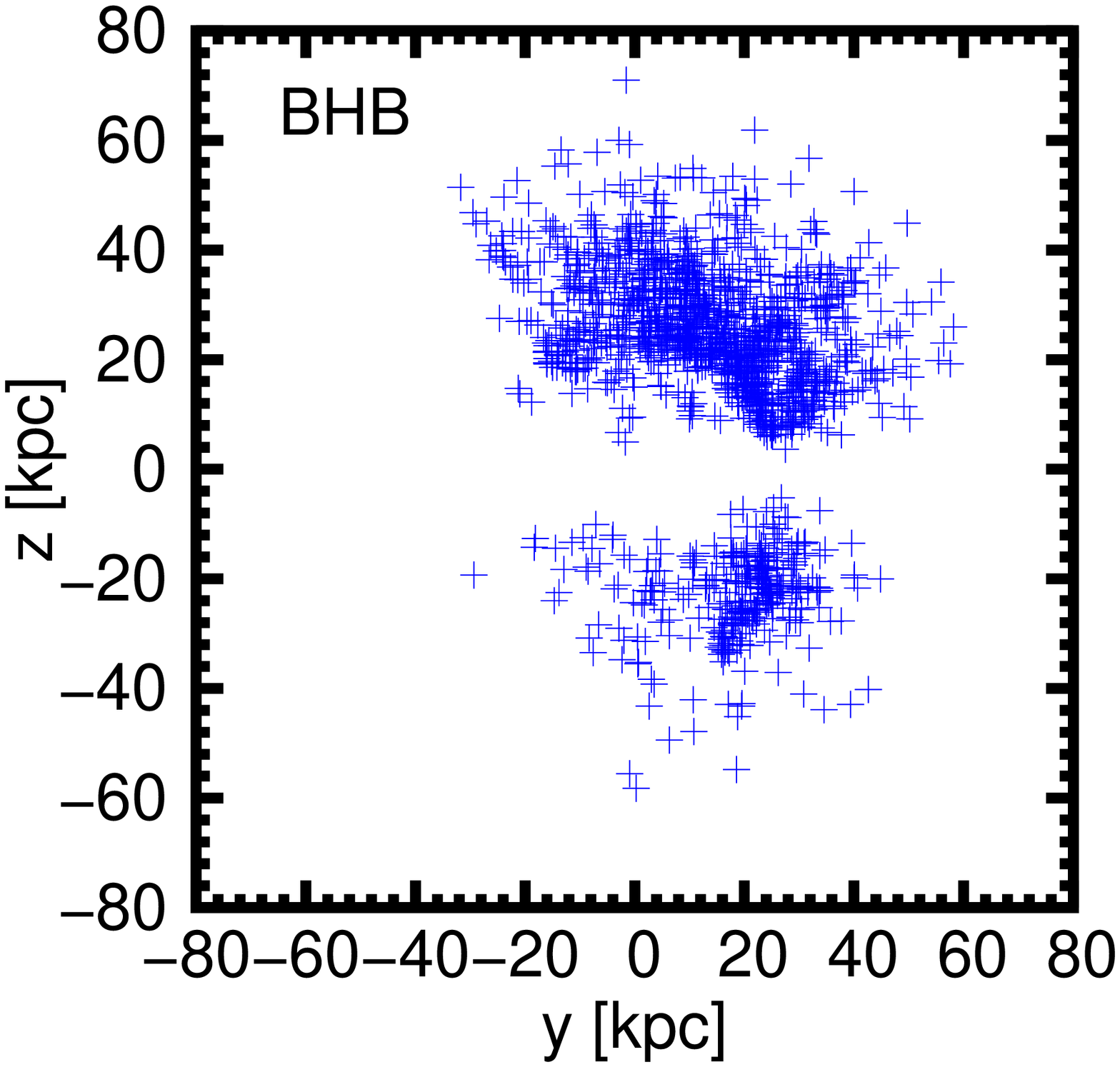,angle=270,width=2.7in}&
\hskip -2 cm
\epsfig{file=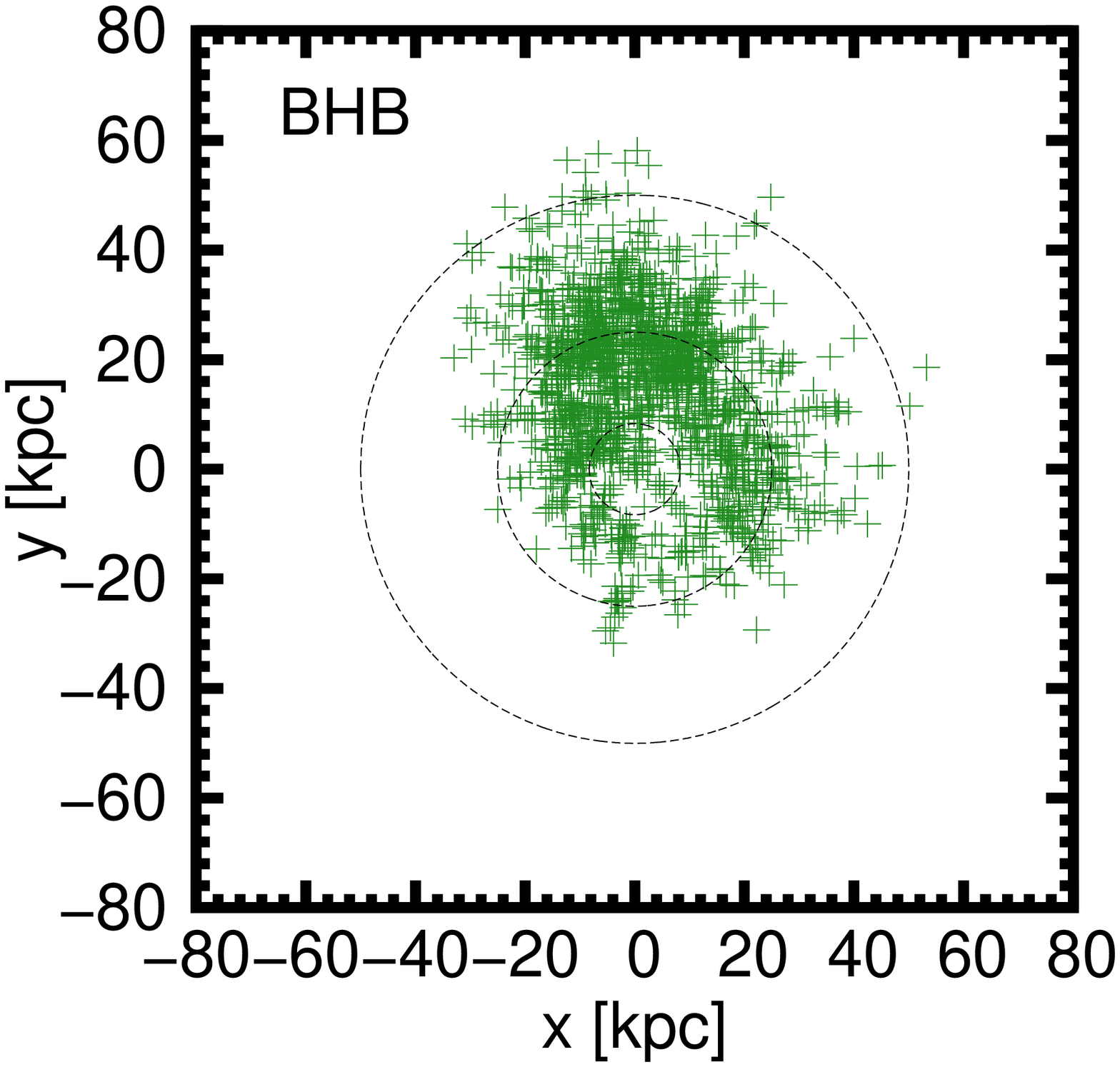,angle=270,width=2.7in}\\
%
\epsfig{file=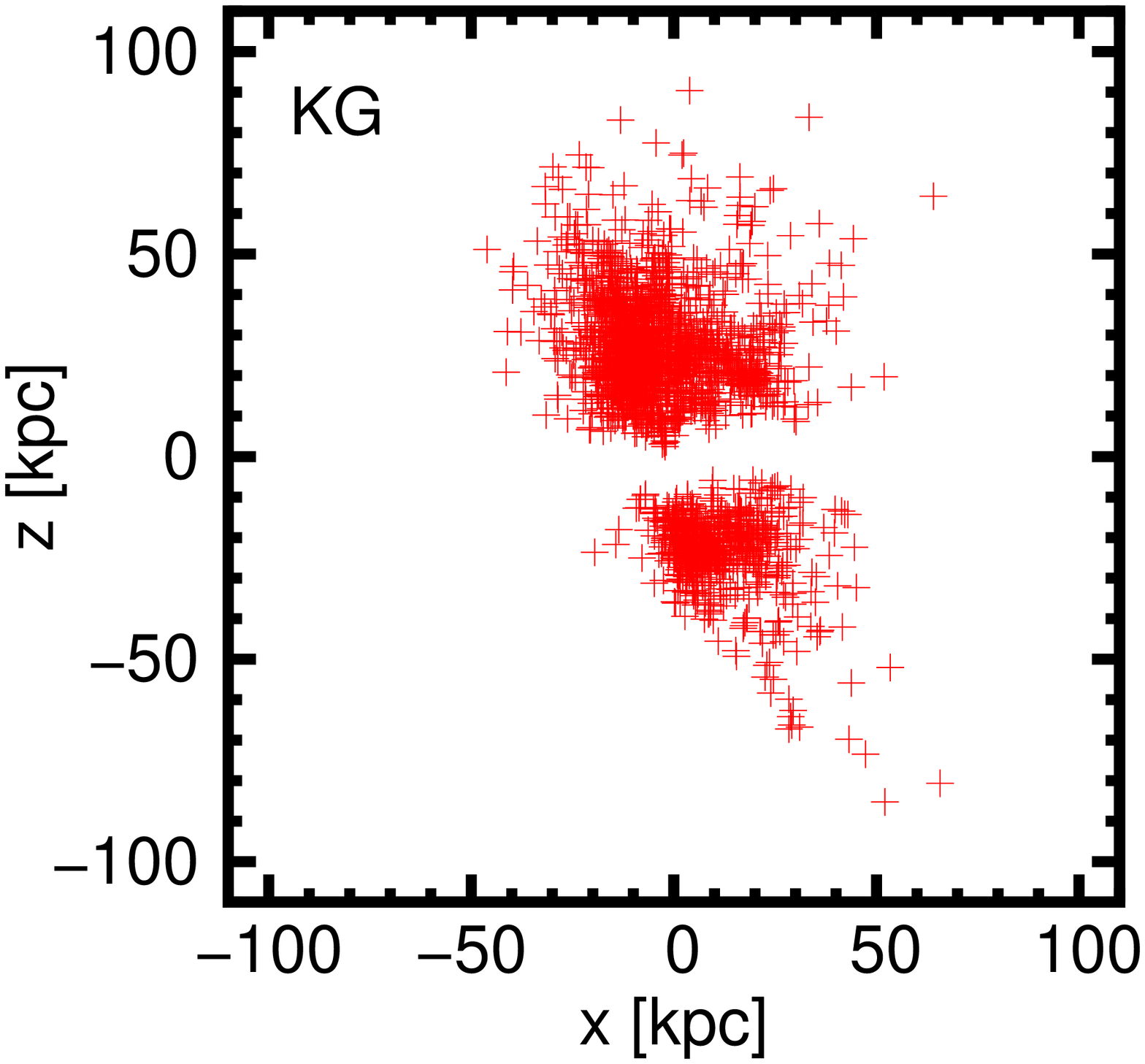,angle=270,width=2.7in}&
\hskip -2 cm
\epsfig{file=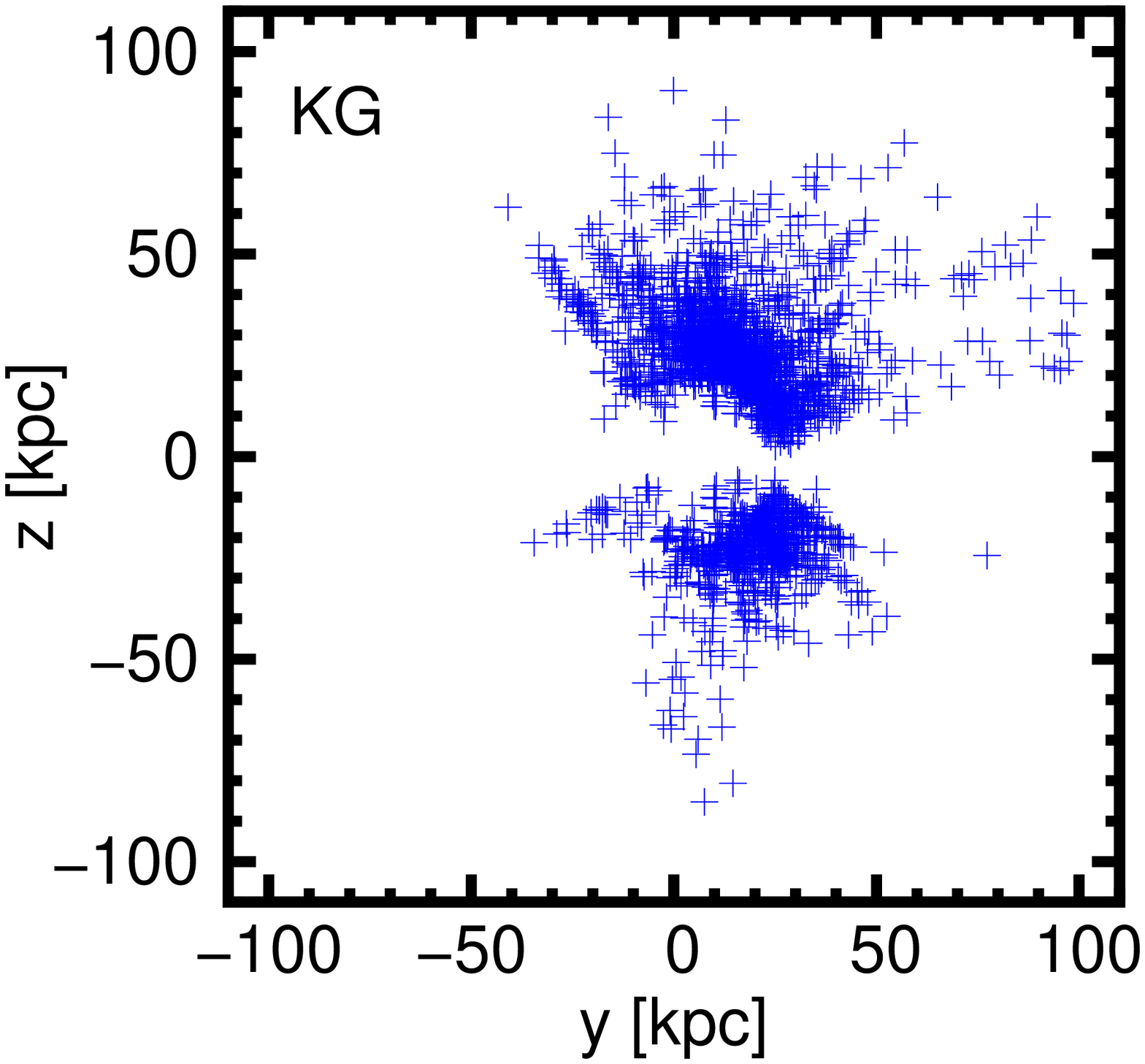,angle=270,width=2.7in}&
\hskip -2 cm
\epsfig{file=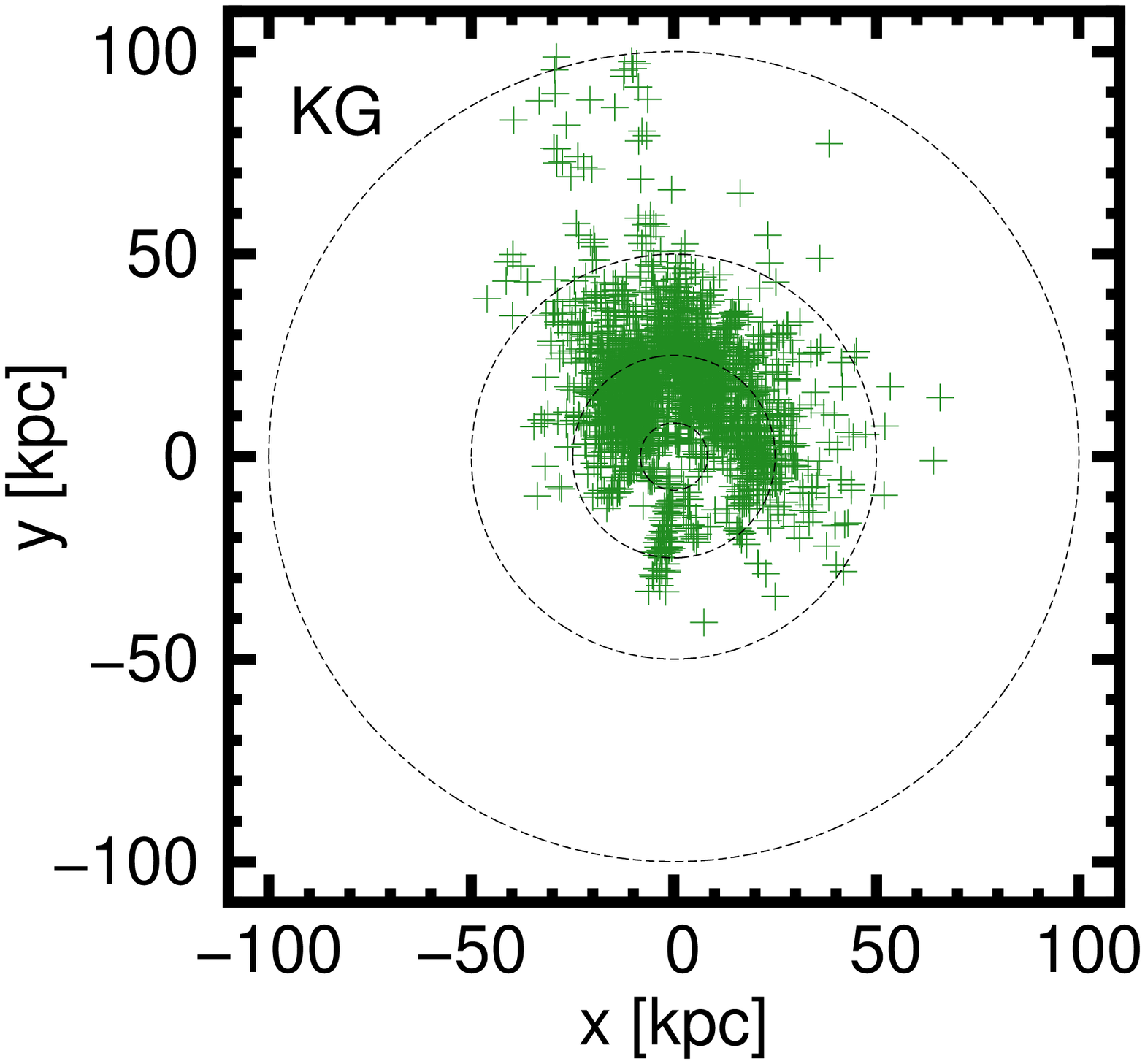,angle=270,width=2.7in}\\
%
\epsfig{file=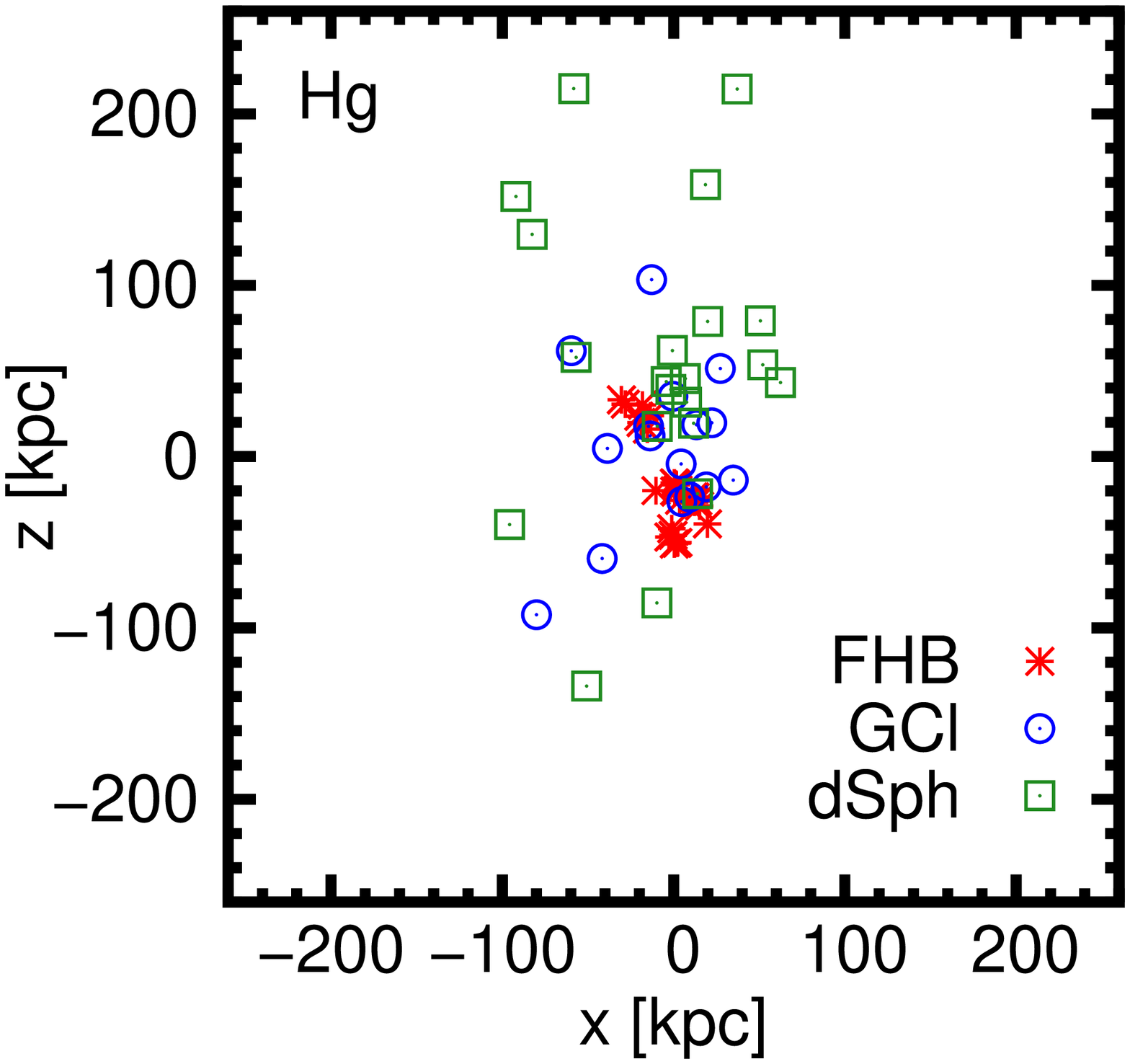,angle=270,width=2.7in}&
\hskip -2 cm
\epsfig{file=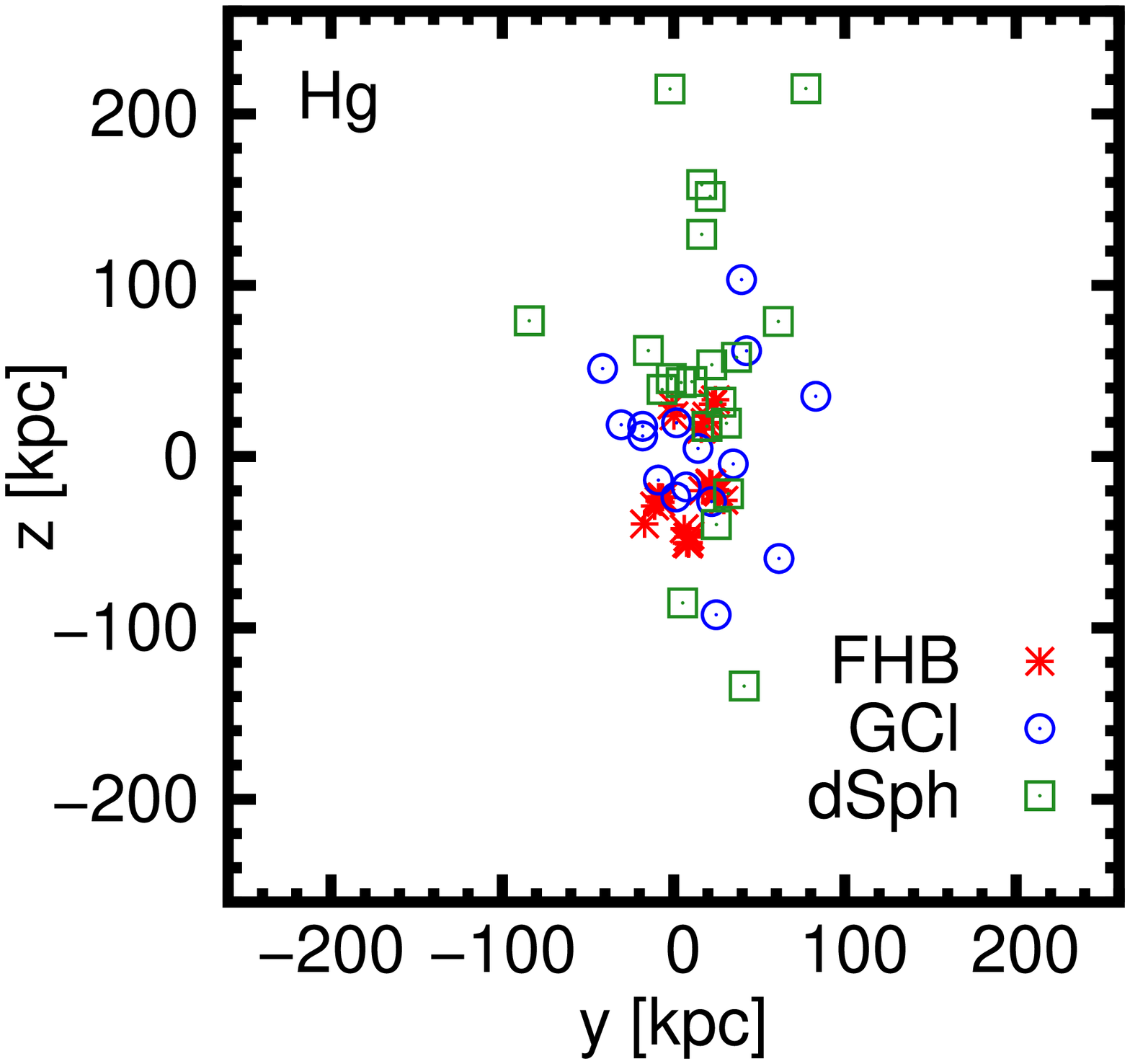,angle=270,width=2.7in}&
\hskip -2 cm
\epsfig{file=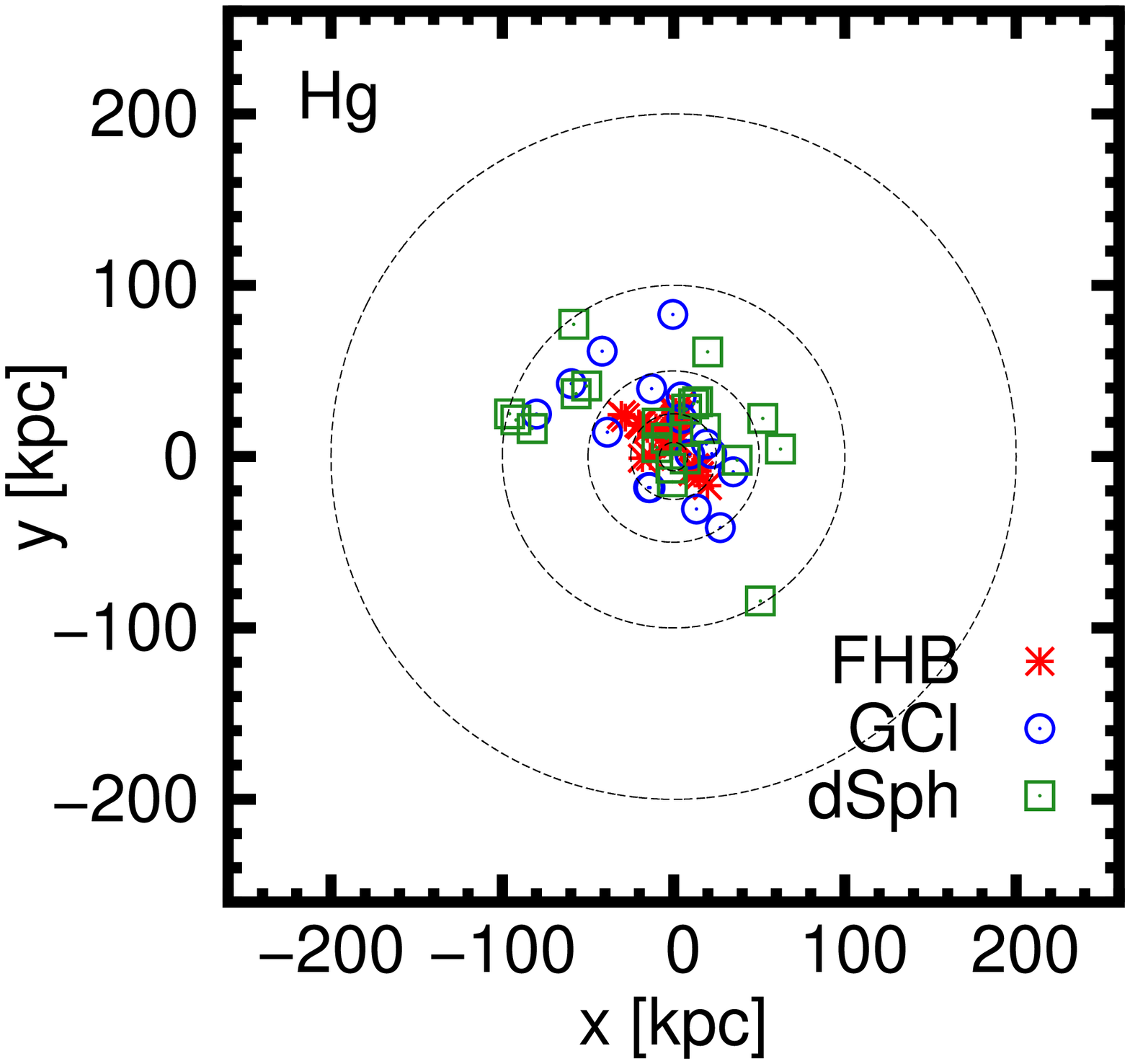,angle=270,width=2.7in}
\end{tabular}
\caption{$x$-$z$, $y$-$z$ and $x$-$y$ scatter plots (after 
removing objects with $r<25\kpc$; see text) for the three 
samples of non-disk tracer objects considered in this paper, 
namely, (1) the ``BHB" sample, a set of 1457 blue horizontal branch 
stars from the compilation of \citet{Xue_etal_BHB_SDSS_DR8_2011}, (2) 
the ``KG" sample, a set of 2227 K-Giant stars from the compilation of
\citet{Xue_etal_KGiants_SDSS_DR9_2012}, and (3) the ``Hg" sample, a 
heterogeneous set of 65 objects comprising of 16 Globular Clusters (GCl)
from \citet{Harris_GCl_2010_1996}, 28 field blue horizontal branch (FHB) 
stars from \citet{Clewley_etal_2004}, and 21 dwarf spheroidals (dSph) 
from \citet{McConnachie_DG_2012}, for $\rsun=8.3\kpc$ with 
the sun located at $(x=0,y=\rsun,z=0)$. 
}
\label{Fig:Fig_nondisk_scatter_BHB_KG_HS_clr}
\end{figure*}

\vfill\eject

\end{document}